\documentclass[11pt,a4paper]{article}
\pdfoutput=1
\usepackage{jcappub}

\usepackage{graphicx,xcolor}
\usepackage{epsfig}
\usepackage{amssymb}
\usepackage{amsmath,amssymb,textcomp,array}

\newcommand{\apj}{{Astroph.\@ J.\ }}
\newcommand{\prl}{{Phys.\@ Rev.\@ Lett.\ }}
\newcommand{\prd}{{Phys.\@ Rev.\@ D\ }}

\newcommand{\mn}{{Mon.\@ Not.\@ Roy.\@ Ast.\@ Soc.\ }}
\newcommand{\asta}{{Astron.\@ Astrophys.\ }}
\newcommand{\aj}{{Astron.\@ J.\ }}

\newcommand{\etal}{{et al.,~}}
\newcommand{\ie}{{i.e.,~}}

\newcommand{\eg}{{e.g.,~}}

\newcommand{\beq}{\begin{equation}}
\newcommand{\eeq}{\end{equation}}
\newcommand{\ber}{\begin{eqnarray}}
\newcommand{\eer}{\end{eqnarray}}
\newcommand{\lleq}{\lower0.9ex\hbox{ $\buildrel < \over \sim$} ~}
\newcommand{\ggeq}{\lower0.9ex\hbox{ $\buildrel > \over \sim$} ~}
\newcommand{\lsim}{\ \lower-1.5pt\vbox{\hbox{\rlap{$<$}\lower5.3pt\vbox{\hbox{$\sim$}}}}\ }
\newcommand{\gsim}{\ \lower-1.5pt\vbox{\hbox{\rlap{$>$}\lower5.3pt\vbox{\hbox{$\sim$}}}}\ }

\newcommand{\omt}{\Omega_{0 m}}
\newcommand{\omr}{\Omega_{0 r}}

\begin{document}

\title{An Exploration of Heterogeneity in Supernova Type Ia Samples}

\author[a, b]{Ujjaini Alam}
\author[b]{ Jeremie Lasue}
\affiliation[a]{Physics \& Applied Mathematics Unit, Indian Statistical Institute, Kolkata India}
\affiliation[b]{IRAP, Universit\'e de Toulouse, CNRS, UPS, CNES, Toulouse, France}
\emailAdd{ujjaini.alam@gmail.com}
\emailAdd{jeremie.lasue@irap.omp.eu}

\sloppy

\abstract{\small{   
We examine three SNe Type Ia datasets: Union2.1, JLA and Panstarrs to
check their consistency using cosmology blind statistical analyses as
well as cosmological parameter fitting.  We find that the Panstarrs
dataset is the most stable of the three to changes in the data,
although it does not, at the moment, go to high enough redshifts to
tightly constrain the equation of state of dark energy, $w$.  The
Union2.1, drawn from several different sources, appears to be somewhat
susceptible to changes within the dataset.  The JLA reconstructs well
for a smaller number of cosmological parameters. At higher degrees of
freedom, the dependence of its errors on redshift can lead to varying
results between subsets.  Panstarrs is inconsistent with the other two
datasets at about $2\sigma$ confidence level, and JLA and Union2.1 are
about $1\sigma$ away from each other. For the $\omt-w$ cosmological
reconstruction, with no additional data, the $1\sigma$ range of values
in $w$ for selected subsets of each dataset is two times larger for
JLA and Union2.1 as compared to Panstarrs. The range in $\omt$ for the
same subsets remains approximately similar for all three datasets.  We
find that although there are differences in the fitting and correction
techniques used in the different samples, the most important criterion
is the selection of the SNe, a slightly different SNe selection can
lead to noticeably different results both in the purely statistical
analysis and in cosmological reconstruction. We note that a single,
high quality low redshift sample could help decrease the uncertainties
in the result. We also note that lack of homogeneity in the magnitude
errors may bias the results and should either be modeled, or its
effect neutralized by using other, complementary datasets. A supernova
sample with high quality data at both high and low redshifts,
constructed from a few surveys to avoid heterogeneity in the sample,
and with homogeneous errors, would result in a more robust
cosmological reconstruction.}}

\maketitle

\thispagestyle{empty}

\section{Introduction}\label{sec:intro}

Observations of distant type Ia supernovae by the Supernova Cosmology
Project (SCP) and the High Redshift Search Team (HZT) \cite{hzt, scp},
in the late 1990s, uncovered one of the most mystifying cosmological
phenomena today-- the accelerated expansion of the Universe. One way
to explain this observational result is to theorize the existence of a
new form of energy, with negative pressure, often called ``dark
energy''. Another is to postulate a modification of the rules of
gravity. Many different models have been suggested for these, some of
which are reviewed and analyzed in \cite{de_rev1, de_rev2, de_rev3,
  de_rev4, de_rev5, de_rev6, de_rev7, de_rev8, de_rev9, de_rev10,
  de_rev11, de_rev12, de_rev13, de_rev14, de_rev15, ss06,
  ua16}. Current cosmological observations are commensurate with the
cosmological constant \cite{planck, planckde}, where the dark energy
equation of state is $-1$ and its energy density is constant. However,
other dark energy models are by no means ruled out \cite{sss14}, and
the search for the true nature of dark energy is a continuing process.

Different observations have provided indirect evidence for the
presence of dark energy, such as the Cosmic Microwave Background
Radiation (CMB) \cite{planck}, and Baryon Acoustic Oscillations (BAO)
\cite{bao,baoh, sdss12}. Nevertheless, the most direct evidence for
the accelerated expansion of the Universe remains the type Ia
SNe. These observations are not without their share of
controversy. Type Ia SNe are considered to be ``calibrated candles'',
\ie although their peak brightness is not identical to each other,
after being subjected to a stretch factor, they fall on the same
light-curve with remarkable uniformity. The evidence for this is purely
empirical and although the SNe type Ia as a group follow the stretch
relations extremely well, there remain some outliers. There is no
strong theoretical proof yet as to why the type Ia SNe should be
``calibrated candles'', nor as to why there should not be any
magnitude evolution between near and distant SNe. Indeed, it is
possible to study potential magnitude evolution in type Ia SNe to
examine whether sufficient magnitude evolution could negate or weaken
the result of accelerated expansion of the universe. Some groups have
studied the techniques for correcting for stretch and color in SNe
\cite{mag_sne}. Others have studied the possibility of non Gaussian
distribution of SNe \cite{saini_sne} or the effect of cosmic dust on
SNe magnitudes \cite{coras_sne}. The impact of metallicity on the SNe
rates has been examined in \cite{metal_sne}. Host selection for Type
Ia SNe has also been studied recently \cite{host_sne}. The effect of
interstellar dust on the luminosity of distant SNe, the exact
mechanism behind type Ia SNe explosions, all these are as yet ill
understood. As such, results from the type Ia SNe have met with some
scepticism from the community. Despite the many unknowns, the SNe
results are however strongly corroborated by many other observations,
and over the years, as the SNe data have grown qualitatively and
quantitatively, the proof for accelerated expansion of the universe
has only grown stronger.

Currently, the focus is on constraining dark energy parameters such as
the dark energy equation of state today, $w_0$, and its evolution with
the scale factor, $w_a$. While SNe cosmology is entering its precision
era, the apparent heterogeneity in the SNe Ia available to us remains
a cause for concern. There have been attempts to quantify the effects
of this heterogeneity on the cosmological reconstruction
\cite{stat_sne, stat_sne2}, as yet there are no conclusive results.
In this work we study the three most used and comprehensive currently
available SNe Ia datasets to understand the consistency within and
across the different samples and to assess how this may affect the
cosmological analysis. The paper is arranged as follows: section
\ref{sec:data} contains a description of the SNe data used, section
\ref{sec:stat} describes the results obtained by applying
cosmology-independent statistical techniques to SNe Ia data, in
section \ref{sec:cosmo} we obtain constraints on cosmological
parameters, and in section \ref{sec:concl} we present our conclusions.
 
\section{Supernova Data }\label{sec:data}

The first type Ia SNe datasets hinting at the accelerated expansion of
the universe were published in the late 1990s by \cite{hzt,scp}. At
this time, only 36 SNe were available from SCP, and 42 from HZT with
fairly high magnitude errors. Naturally these few observations could
not constrain dark energy parameters to appreciable levels. Since
then, various SNe teams have augmented the type Ia SNe dataset and
also utilized improved techniques for data reduction. In 2003 the HZT
team published the luminosity-redshift relation of 230 type Ia SNe
\cite{tonry03}. This paper constrained the equation of state of dark
energy at present to $-1.48 < w < -0.72$ at $2\sigma$ for a flat
$w$CDM model. In the same year, \cite{barris04} added a further 23
high redshift SNe to this set, doubling the high redshift SNe sample
at $z > 0.7$. In 2004, \cite{hst04} added 16 SNe from the Hubble Space
Telescope (HST) to the previous samples, and defined a subset of 157
SNe with well defined spectroscopic features, host redshift,
well-sampled colour and light curves as the Gold dataset. With added
priors on $\omt$ from WMAP and the Two-Degree Field Galaxy Redshift
Survey (2dfGRS), this constrained the equation of state of dark energy
at present to $w = -1.08^{+0.20}_{-0.18}$. All these analyses used the
MLCS2k2 light-curve fitter. The first results from the SuperNova
Legacy Survey (SNLS) were reported in 2005 \cite{snls06} with 71 new
high redshift SNe, leading to $\omt = 0.271 \pm 0.021 \ (stat) \ \pm
0.007 \ (syst), w = -1.023 \pm 0.090 \ (stat) \ \pm 0.054 \ (syst)$
with added constraints from BAO and assumed flatness. This group
introduced the SALT-I fitter for light-curves. The 2007 results from
the ESSENCE survey \cite{essence07} were commensurate with the SNLS
results, with $w = -1.05^{+0.13}_{-0.12}$ and $\omt =
0.274^{+0.033}_{-0.020}$ in a similar analysis. The 2008 Union2 dataset
\cite{union08} comprised of 307 SNe from ESSENCE, SNLS, HST and older
datasets, using the SALT-I fitter. These, with BAO and CMB
observations, led to constraints of $w = -0.969^{+0.059}_{-0.063}
\ (stat) ^{+0.063}_{-0.066} \ (sys)$. In 2009, magnitude-redshift
relations for 185 type Ia SNe observed at the F. L. Whipple
Observatory of the Harvard-Smithsonian Center for Astrophysics (CfA)
were reported in \cite{cfa09}. This resulted in the creation of the
Constitution dataset, which added the Union2 dataset to the CfA3 data
and BAO to report $w = -1.013^{+0.066}_{-0.068} \ (stat) \ \pm 0.11
\ (syst)$. This paper also carried out a comparison of the various
light-curve fitting methods: MLCS2k2, SALT-I and SALT-II.

The first season of the Sloan Digital Sky Survey-II (SDSS-II)
Supernova Survey filled the ``redshift desert'' between low and high
redshift SNe \cite{sdss09}, and combining these with the ESSENCE, SNLS
and HST SNe, obtained $\omt = 0.307 \pm 0.019 \ (stat) \ \pm 0.023
\ (syst), w = -0.76 \pm 0.07 \ (stat) \ \pm 0.11 \ (syst)$ for the
MLSC2k2 fitter, and $\omt = 0.265 \pm 0.016 \ (stat) \ \pm 0.025
\ (syst), w = -0.96 \pm 0.06 \ (stat) \ \pm 0.12 \ (syst)$ for the
SALT-II fitter. This paper highlighted the interesting fact that
different empirical light curve fitters could give significantly
different cosmological results for the same SNe. A further 20 high
redshift SNe from the HST Cluster Supernova Survey added to the
existing Union2 collection, and analyzed using the SALT-II fitter,
forms the Union2.1 set with 580 SNe. This results in constraints of $w
= -1.013^{+0.068}_{-0.073}$ \cite{union11}. The three year SNLS
program along with other previous SNe yielded a total of 472 high
quality SNe (with 242 SNe from SNLS alone), resulting in constraints
of $w = -0.91^{+0.17}_{-0.24} \ (stat+sys)$, $\omt = 0.18\pm 0.10
\ (stat+sys)$ \cite{snls11}. The 146 SNe at redshifts $0.03 < z <
0.65$ discovered during the first 1.5 years of the Pan-STARRS1 Medium
Deep Survey \cite{panstarrs14}, added to existing low redshift data,
as well as BAO, CMB and $H_0$ data, provide a constraint of $\omt =
0.280^{+0.013}_{-0.012}, w = -1.166^{+0.072}_{-0.069}$. A joint
analysis of SNe obtained by the SDSS-II and SNLS collaborations using
the SALT-II fitter, totals 740 spectroscopically confirmed type Ia
supernovae with high quality light curves. This ``JLA'' sample yields
$w =-1.027 \pm 0.055 \ (stat+sys)$ with added BAO and CMB constraints
\cite{jla14}.

Several points are to be noted about these SNe surveys:
\begin{itemize}
\item
Firstly, the SNe data has gone from less than 100 SNe with errors
$\sigma_{\mu} \sim 1-10$ \cite{hzt,scp} to $\ggeq 500$ SNe with errors
$\sigma_{\mu} \sim 0.1-1.0$. These observations have seen more than an
order of magnitude improvement both in numbers and accuracy.
\item
Many different light curve fitters are now available for analysis, \eg
MLCS2K2 \cite{mlcs2k2}; Stretch \cite{stretch}; SALT-I, SALT-II
\cite{salt}; SiFTO \cite{sifto}; BayeSN \cite{bayesn, bayesn1}. These
have improved the accuracy of individual SNe light curves to $\sim 5
\%$, but there remain inconsistencies between the different fitters,
and cosmological results may vary depending on the fitter
used. Currently, the most commonly used fitter is SALT-II.
\item  
Although many different surveys have reported high redshift type Ia
SNe, the low redshift range $z \lleq 0.1$ does not have many new
observations. This range, however, is crucial for the normalization of
the data. Thus, even the new and high quality high redshift data have
to use the older low redshift data, which is a fairly heterogeneous
dataset.
\item
Supernova data, by itself, can only constrain the total energy
density. Thus, in order to extract information on dark energy from it,
it is imperative to include information on the curvature of the
Universe (\eg from CMB) and the matter density (\eg from BAO). Even
with these information, it may not be possible to obtain reasonable
constraints on both the equation of state at present and its evolution
over time. Thus, it is often the practice to quote results for a
constant equation of state dark energy. This is not theoretically a
very viable dark energy model. The data should therefore reach an
accuracy where we are able to constrain both the present value and
rate of evolution of the dark energy equation of state. This could be
achieved by better quality SNe data, augmenting SNe data with other,
complementary observations, or by using more sophisticated statistical
analysis techniques.
\end{itemize}
In this work we assess the consistency of the SNe databases and the
effect any inconsistencies may have on the cosmological parameters
constrained from them.

\section{Statistical analysis of Supernovae data}\label{sec:stat}

The SALT-II model transforms the light-curve fit parameters into
distances as:
\beq
\mu = m_B + \alpha \centerdot X - \beta \centerdot C - M_B
\eeq
where $\mu$ is the distance modulus which contains the cosmological
information, $m_B$ is the peak B-band brightness, $X$ is a light-curve
shape parameter, and $C$ is a colour parameter.  $\alpha$ is
determined by the relation between luminosity and stretch while
$\beta$ is determined by the relation between luminosity and
color. $M_B$ is the absolute B-band magnitude of a fiducial SNe Ia
with $X = 0, C = 0$.  The parameters $\alpha, \beta, M_B$ are nuisance
parameters that are fitted simultaneously with the cosmological
parameters. Both the absolute magnitude $M_B$ and $\beta$ parameter
depend on host galaxy properties. However, the effect of this on
cosmological results is expected to be negligible, so we neglect this
dependence for our analysis as this is a sub-dominant systematic
effect \cite{nuisance}.

\begin{table*}
\caption{\footnotesize
Sample distribution of Union2.1 data}
\label{tab:union}      
\centering          
\begin{tabular}{|cccccccc|} 
\hline       
Sample&No.&Redshift&$\sigma_{m_B}$&$\sigma_X$&$\sigma_C$&$\sigma_{\mu_B}$&Reference \\ 
\hline                    
1&$18$&$0.02-0.10$&$0.04-0.19$&$0.10-0.93$&$0.01-0.09$&$0.16-0.22$&\cite{calantololo}\\
2&$6$&$0.02-0.03$&$0.07-0.14$&$0.12-0.21$&$0.01-0.03$&$0.09-0.15$&\cite{kris05}\\
3&$11$&$0.02-0.12$&$0.03-0.14$&$0.14-0.49$&$0.01-0.04$&$0.16-0.22$&\cite{cfa99}\\
4&$15$&$0.02-0.05$&$0.05-0.18$&$0.14-1.54$&$0.01-0.05$&$0.22-0.31$&\cite{cfa06}\\
5&$8$&$0.02-0.16$&$0.03-0.15$&$0.10-0.85$&$0.01-0.04$&$0.08-0.16$&\cite{union08}\\
6&$94$&$0.02-0.08$&$0.04-0.21$&$0.08-1.30$&$0.01-0.11$&$0.16-0.29$&\cite{cfa09}\\
7&$18$&$0.02-0.08$&$0.03-0.15$&$0.05-0.31$&$0.01-0.04$&$0.10-0.18$&\cite{csp10}\\
8&$129$&$0.04-0.42$&$0.02-0.11$&$0.12-1.56$&$0.01-0.09$&$0.11-0.25$&\cite{holtz08}\\
9&$11$&$0.24-0.97$&$0.05-0.15$&$0.30-1.85$&$0.04-0.27$&$0.31-0.81$&\cite{schmidt98}\\
10&$33$&$0.17-0.83$&$0.05-0.27$&$0.15-1.67$&$0.05-0.43$&$0.42-1.01$&\cite{scp}\\
11&$19$&$0.34-0.98$&$0.07-0.26$&$0.30-2.88$&$0.04-0.37$&$0.20-0.73$&\cite{barris04}\\
12&$5$&$0.18-0.27$&$0.03-0.12$&$0.19-0.51$&$0.02-0.09$&$0.19-0.26$&\cite{aman08}\\
13&$11$&$0.36-0.86$&$0.05-0.09$&$0.25-0.53$&$0.03-0.08$&$0.10-0.20$&\cite{knop03}\\
14&$72$&$0.25-1.01$&$0.03-0.14$&$0.13-1.01$&$0.01-0.14$&$0.13-0.43$&\cite{snls06}\\
15&$74$&$0.16-0.78$&$0.05-0.21$&$0.20-1.85$&$0.04-0.19$&$0.20-0.39$&\cite{miknaitis07}\\
16&$6$&$0.28-1.06$&$0.06-0.12$&$0.39-1.01$&$0.04-0.18$&$0.18-0.46$&\cite{tonry03}\\
17&$30$&$0.22-1.39$&$0.07-0.21$&$0.21-1.23$&$0.03-0.13$&$0.20-0.37$&\cite{hst07}\\
18&$6$&$0.51-1.12$&$0.06-0.15$&$0.26-0.85$&$0.03-0.36$&$0.09-0.95$&\cite{union10}\\
19&$14$&$0.62-1.41$&$0.09-0.17$&$0.31-1.53$&$0.05-0.20$&$0.17-0.56$&\cite{union11}\\
\hline                             
\end{tabular}
\end{table*}

We have at our disposal at present, three main SNe type Ia datasets
which have been processed using the SALT-II model. These have some
common SNe, and some SNe which are survey specific.
\begin{itemize}
\item
The Union2.1 dataset \cite{union11}: This dataset contains 580 SNe
gathered from 19 different sources. The break-up of the data in
samples is shown in Table~\ref{tab:union}. As we see, there are 19
different samples from which the data is taken, these include SNe from
the early 1990s up to 2011. Samples 1--7 contain low redshift data,
with error bars for the distance modulus ranging between $0.08 <
\sigma_{\mu} < 0.31$ (and that for the peak B-band brightness between
$0.02 < \sigma_{m_B} < 0.21$). These SNe cannot constrain the
cosmology very stringently since at low redshifts the universe follows
the simple linear Hubble law, but these are essential to provide a
low-$z$ anchor to the Hubble diagram. For the high redshift data in
samples 8--19, we have $0.09 < \sigma_{\mu} < 1.08$ and $0.03 <
\sigma_{m_B} < 0.31$. The highest redshift data is in sample 16 at $z
= 1.4$, the largest samples are sample 6 for low redshift, sample 8
for low to mid redshifts, and samples 14, 15 for high redshift
data. These samples might be expected to dominate the results.
\item
The JLA dataset \cite{jla14}: This dataset combines the SNLS and SDSS
SNe to create an extended sample of 740 SNe along with existing lower
redshift SNe. The sample distribution of the data is shown in
Table~\ref{tab:jla}. The two biggest samples are: the SDSS data at low
to mid redshift, with peak B-band brightness errors between $0.11 <
\sigma_{m_B} < 0.17$; and the SNLS three year data at high redshift
(up to $z = 1.06$) with $0.02 < \sigma_{m_B} < 0.21$. The remaining
data is from older samples. Various checks have been carried out for
consistency, bias and systematic uncertainties on this dataset and it
is expected to be a more homogeneous dataset than the Union2.1 data.
\item
The Panstarrs dataset \cite{panstarrs14}: This dataset contains $311$
SNe, with the high redshift SNe measured from the Pan-STARRS1 Medium
Deep Survey, and the low redshift SNe from various sources.  The
sample distribution is shown in Table~\ref{tab:panstarrs}. The
Panstarrs data only goes up to $z \leq 0.634$, but its high redshift
data is the most homogeneous, observed from a single survey. A few low
redshift SNe were also observed by this survey, but they are too few
in number to make a difference, and low redshift SNe from other
surveys are required for normalization. In the low redshift data,
JRK07 is a composite of the samples 1-4 of the Union2.1 sample. The
cosmological results so far obtained from this dataset are markedly
different from the results from the other two samples. This may
however be an effect of less data points and lower redshifts.
\end{itemize} 

\begin{table*}
\caption{\footnotesize
Sample distribution of JLA data}
\label{tab:jla}      
\centering          
\begin{tabular}{|ccccccc|} 
\hline       
Sample&No.&Redshift&$\sigma_{m_B}$&$\sigma_X$&$\sigma_C$&Reference\\ 
\hline                    
Calan/Tololo&$17$&$0.01-0.08$&$0.14-0.16$&$0.07-0.39$&$0.03-0.06$&\cite{calantololo}\\
CfAI&$7$&$0.02-0.05$&$0.14-0.16$&$0.12-0.27$&$~0.03$&\cite{cfa99}\\
CfAII&$15$&$0.01-0.06$&$0.14-0.18$&$0.04-0.35$&$0.02-0.04$&\cite{cfa06}\\
CfAIII&$55$&$0.01-0.07$&$0.14-0.17$&$0.03-0.47$&$0.02-0.04$&\cite{cfa09}\\
CSP&$13$&$0.01-0.08$&$0.14-0.17$&$0.02-0.11$&$~0.02$&\cite{csp10}\\
Other&$11$&$0.01-0.08$&$0.14-0.17$&$0.04-0.23$&$0.02-0.04$&\cite{kris05, union08}\\
SDSS&$374$&$0.04-0.40$&$0.11-0.17$&$0.03-1.38$&$0.01s-0.09$&\cite{sdss14}\\
SNLS&$239$&$0.13-1.06$&$0.09-0.14$&$0.05-1.64$&$0.02-0.09$&\cite{snls11}\\
HST&$9$&$0.84-1.29$&$0.11-0.13$&$0.29-0.69$&$0.04-0.11$&\cite{hst07}\\
\hline                             
\end{tabular}
\end{table*}

\begin{table*}
\caption{\footnotesize 
Sample distribution of Panstarrs data}
\label{tab:panstarrs}      
\centering          
\begin{tabular}{|cccccccc|} 
\hline       
Sample&No.&Redshift&$\sigma_{m_B}$&$\sigma_X$&$\sigma_C$&$\sigma_{\mu_B}$&Reference \\ 
\hline  
JRK07&$48$&$0.02-0.12$&$0.03-0.24$&$0.04-2.87$&$0.03-0.07$&$0.15-0.41$&\cite{calantololo, cfa99, cfa06, mlcs2k2}\\
CfA3&$84$&$0.01-0.08$&$0.03-0.19$&$0.06-1.08$&$0.03-0.07$&$0.14-0.31$&\cite{cfa09}\\
CfA4&$43$&$0.01-0.07$&$0.03-0.26$&$0.10-0.98$&$0.03-0.11$&$0.19-0.48$&\cite{cfa12}\\
CSP&$22$&$0.01-0.08$&$0.02-0.17$&$0.04-0.35$&$0.02-0.04$&$0.13-0.25$&\cite{csp10}\\
Panstarrs-1&$113$&$0.03-0.63$&$0.03-0.23$&$0.12-1.98$&$0.02-0.22$&$0.10-0.64$&\cite{panstarrs14}\\
\hline                             
\end{tabular}
\end{table*}

We note here that some data subsets, although given different names by
different groups, represent subgroups of the same observational sets,
and many of these are common to the 3 SNe samples. Examples of subsets
which are drawn from the same set of observations are: Sample 1 in
Union2.1 and the Calan-Tololo sample in JLA, Sample 3 of Union2.1 and
CfA2 of JLA, Sample 4 of Union2.1 and CfA2 of JLA, Sample 6 of
Union2.1 and CfA3 in both JLA and Panstarrs, Samples 1-4 of Union2.1
and JRK07 in Panstarrs, Sample 7 of Union2.1 and CSP in JLA and
Panstarrs, Sample 16 of Union2.1 and HST of JLA and Panstarrs. Most of
the low redshift data is drawn from the same observations while the
high redshift data is mostly different in the three datasets.

An additional effect to be considered is that of intrinsic
dispersion. This quantity can be estimated in different ways. For the
Panstarrs data, the intrinsic scatter attributed to color variation
and that attributed to luminosity variation lead to a fairly large
variation in the parameter $\beta$, in the Union2.1 analysis, the
median $\sigma_{sample}$ is used as a measure of intrinsic scatter,
while JLA uses color corrected intrinsic dispersion. Neither of the
latter two analyses reports a large variation in $\beta$. The accurate
measure of intrinsic dispersion continues to be a matter of some
debate, and the study of the variation of standardisation parameters
such as $\beta$ with intrinsic dispersion remains inconclusive at
present. Since the main motivation of this paper is to compare the
different datasets for statistical stability, rather than to extract
cosmological information, hence we avoid using the very different
values of $\sigma_{int}$ used in the different samples. We rather use
the simplest method and use the median $\sigma_{sample}$ to correct
for $\sigma_{int}$. Utilizing a bias correction with different
simulations of data using different values of $\beta$, as in
\cite{panstarrs14}, may result in as much as $4\%$ difference in the
distance moduli, and may change cosmological results, but we constrain
ourselves to the simplest form of intrinsic scatter to reduce
potential uncertainities pertaining to different systematics in the
different datasets.

In the rest of this section, we use the free source R software package
to develop the data exploration \cite{R16}.

\subsection{Exploratory data analysis}\label{sec:mva}

A first look at the data consistency may be obtained by exploratory
data analysis using multivariate projections of the different
parameters fitted for each SNe ($m_B$, $\sigma_{m_B}$, $X$,
$\sigma_X$, $C$,$\sigma_C$). This is especially useful to characterize
clustering and similarity within datasets and get insight into their
variations and consistency \cite{Las10, Las11}.

The data matrix is normalized by centering and scaling the columns by
their mean and variance.  Then principal components analysis is
performed to determine the directions of maximum variance of the data.
This is equivalent to projecting the data along the first two
eigenvectors of the data matrix \cite{Joll02}.  The cumulative
variance on the principal components follows approximately the same
distribution for every dataset, which indicates a similar spread of
the data in the multivariate space for Panstarrs, JLA and
Union2.1. The cumulative variance of the first two principal
components correspond to about 60\% of the total variance for every
set. This low value indicates a relatively even spread of the data
points in the different dimensions of the multivariate space, or
equivalently that no dimensional direction singly dominates the extent
of the set of data points, as expected from uncorrelated variables.

\begin{figure}
\centering
$\begin{array}{ccc}
\includegraphics[width=0.32\linewidth]{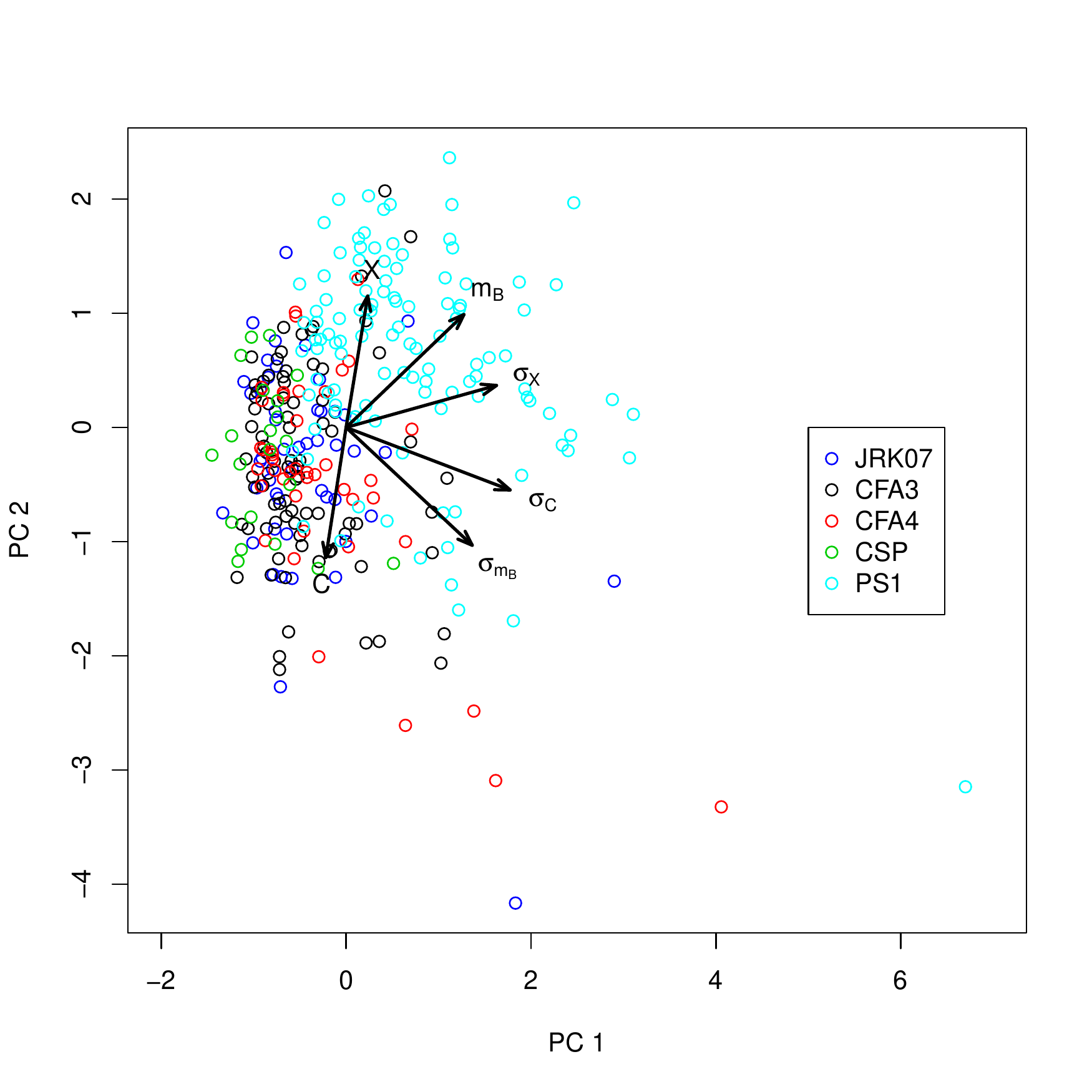} &
\includegraphics[width=0.32\linewidth]{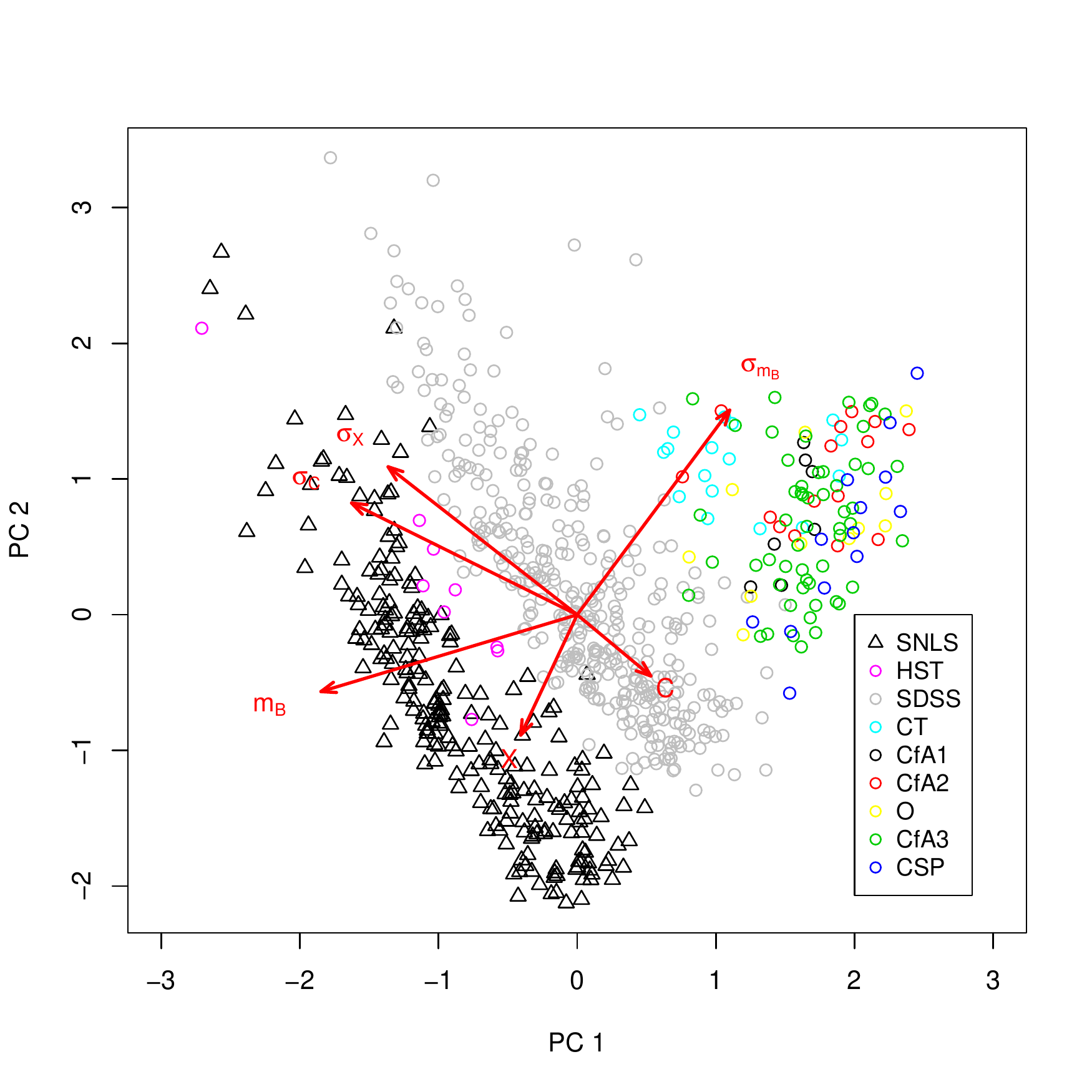} &
\includegraphics[width=0.32\linewidth]{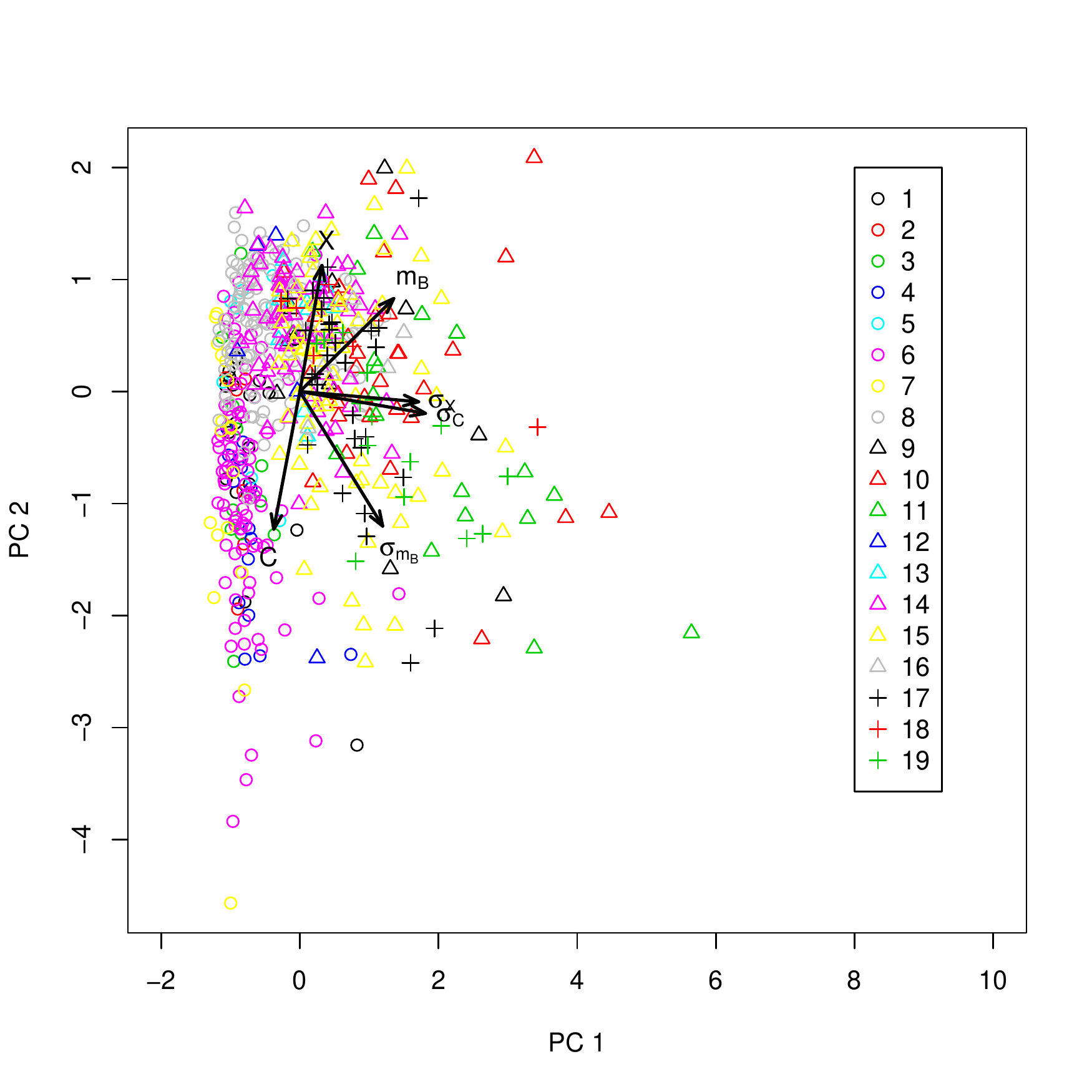} \\
\mbox{(a)} & \mbox{(b)}  & \mbox{(c)}
\end{array}$
\caption{\footnotesize
First 2 principal components projections of the parameter matrices for
(a) Panstarrs, (b) JLA and (c) Union2.1 datasets.  The scores of the
data points along the principal components are used to generate the
map, and the loadings of the different directions to the eigenvectors
are represented by the arrows, with the label indicating which
parameter direction is used. The colours and data point shapes
correspond to the different subsets of each dataset as explained in
the legend of the figures.}
\label{fig:pcproj}
\end{figure}

The projection on the first two principal components for each dataset
is plotted in Figure~\ref{fig:pcproj}, with the contribution of the
different loadings represented by the length and direction of the
arrows, with labels corresponding to the relevant parameters.  One
notes that the projection of Panstarrs and Union2.1 datasets are
consistent with each other. Their parameter loadings appear
distributed over a half-space in a similar manner, from the direction
of the arrows projected in the plane to the same parameters being
projected in the same directions.  As is expected, the $m_B$ loading
(corresponding also to the variation in redshift) separates the low
redshift data (JRK07, CFA3, CFA4, CSP) from the high redshift data
(PS1) of the Panstarrs dataset. This is also the case for the Union2.1
dataset, for which the $m_B$ direction indicates the transition from
the circles (low redshift) to the triangles (high redshift).  The
other loadings do not appear to play a significant role in separating
the data, and no other obvious clustering appears from the projection
on the first two principal components for Panstarrs or Union2.1,
indicating consistency in the spread of data.  One can notice though
that in the Union2.1 dataset, subset 8 appears to form a tighter
cluster of data points in this projection, due to lower errors of the
fitting parameters as compared to the other subsets, which can also be
noticed from Table~\ref{tab:union}.

The JLA projection presents some interesting structure and clustering,
which appears correlated with the data subsets. Here, one notices that
as compared to Panstarrs or Union2.1, the loading directions and
modules are different, showing a different contribution of the various
parameters to the principal components. Firstly, unlike in the other
datasets, one notices that both $m_B$ and $\sigma_{m_B}$ separate the
low-$z$ (SNLS, HST, SDSS, CT, CFA1, Others) data points from the
medium-$z$ (CFA3) and high-$z$ (CFA2, CSP) data points. Secondly, in
this dataset the clusters are more strongly separated than for the
other datasets owing to the influence of $\sigma_{m_B}$.  Furthermore,
additional clustering is visible in the low redshift cluster where
CFA1 and CT appear to separate from the main cluster of data points at
low-$z$ due to the influence of $\sigma_X$ and $\sigma_C$.  Overall,
the effect of clustering due to different errors for the fitting
parameters should not play a significant role in the retrieved
cosmology model parameters as these effects are rather small. But the
fact that they correlate with separate subsets of the JLA dataset
indicates specific differences due to the origin of the data and
incomplete consistency of the datasets.

\begin{figure}
\centering
$\begin{array}{ccc}
\includegraphics[width=0.32\linewidth]{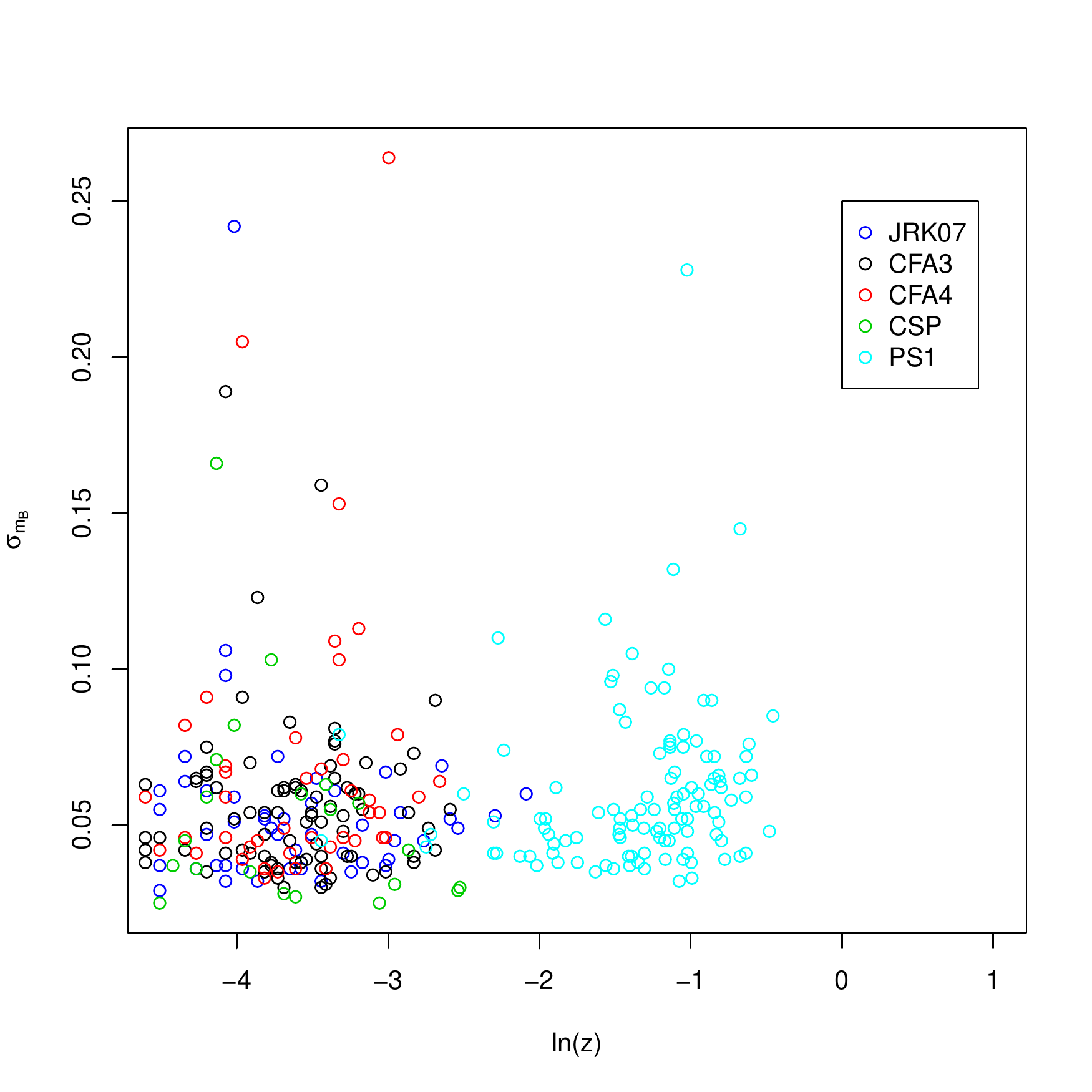} &
\includegraphics[width=0.32\linewidth]{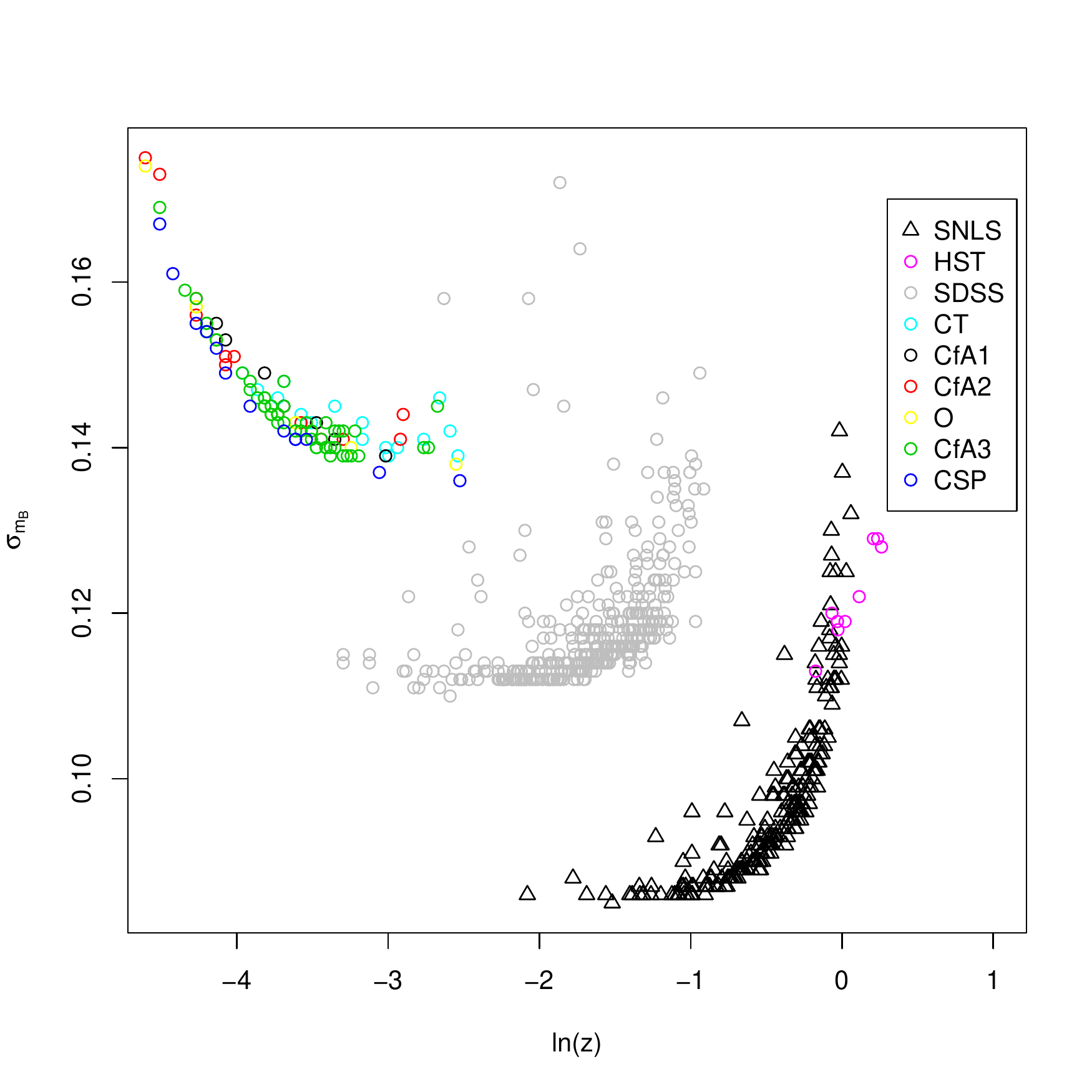} &
\includegraphics[width=0.32\linewidth]{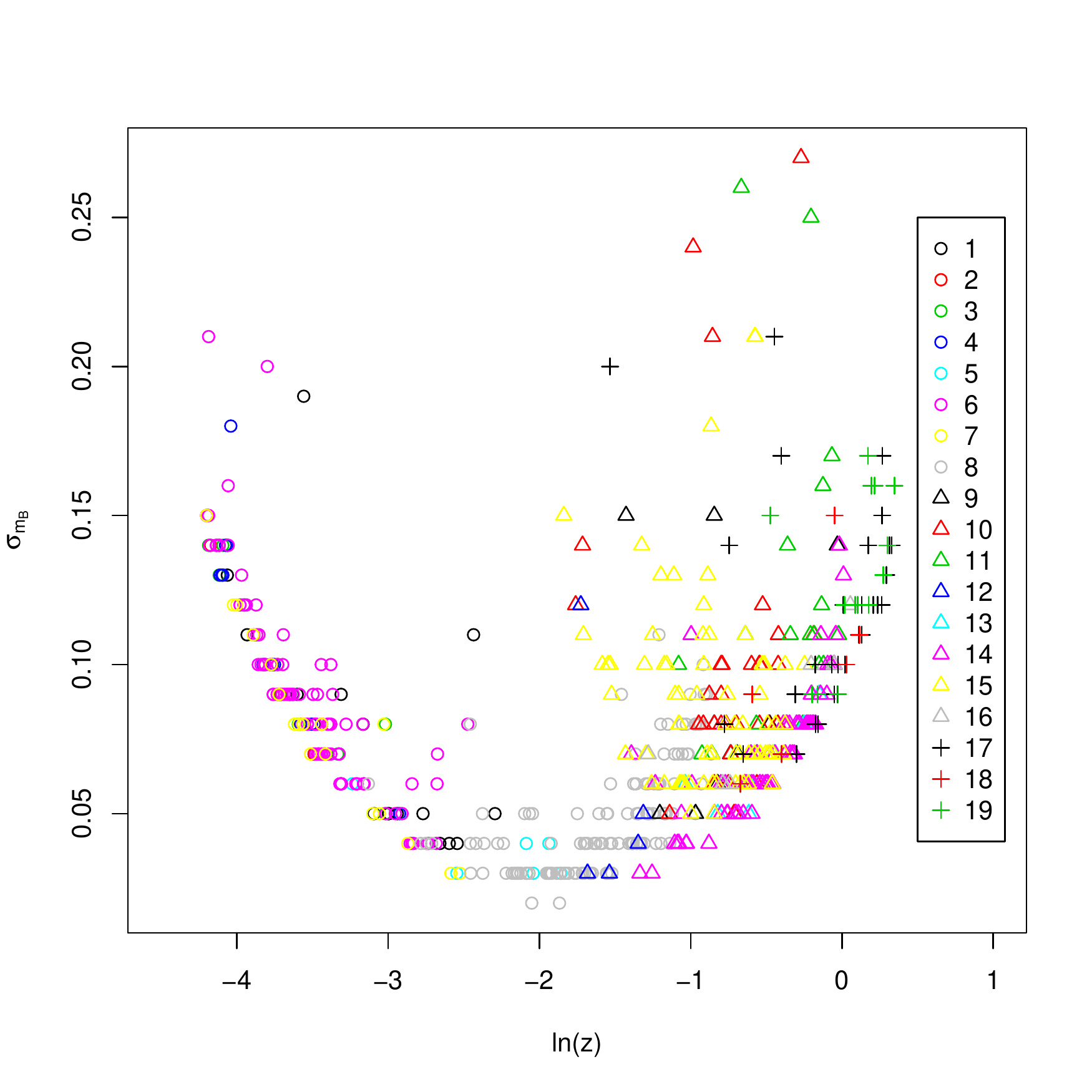} \\
\mbox{(a)} & \mbox{(b)}  & \mbox{(c)}
\end{array}$
\caption{\footnotesize
Residuals for the fitted magnitude, $\sigma_{m_B}$, as a function of
$\ln(z)$ for (a) Panstarrs, (b) JLA and (c) Union2.1
datasets. Differences in the behaviour of the residuals for the fitted
magnitude in each dataset are apparent. The colours and data point
shapes correspond to the different subsets of each dataset as
explained in the legend of the figures.}
\label{fig:sigmamb}
\end{figure}

In statistics, when the variability of an observation changes systematically as a function of a predictor variable, then the dataset is said to be heteroscedastic~\cite{Davidian90}. In this case, the usual assumption that the observations have a constant variance accross the response range is not satisfied and should be taken into account in the analysis. In figure~\ref{fig:sigmamb},
the plot of $\sigma_{m_B}$-- the fitted errors of the magnitude
$m_B$-- as a function of $\ln(z)$ for the three datasets shows
different variabilities in the errors as a function of the
redshift. The Panstarrs dataset appears to be fairly well behaved,
with a relatively constant distribution of the errors and few
outliers.  The error behaviour of Union2.1 is dependent on the subset
considered with the lowest errors for subset 8. One notices the
heteroscedasticity of the data with larger errors at low and high
redshifts. This would indicate that SNe with extreme redshifts may
present higher chances of being outliers to the magnitude-redshift
relationship.  Finally, the JLA $\sigma_{m_B}$ present a strong
dependence on the subset considered with clear clustering of the data
at low, mid and high redshifts, combined with an apparent
heteroscedasticity of the data at low and high redshifts.  It is
interesting to note the increase in errors at the low and high
redshifts, which may translate into increased uncertainties for the
SNe located at the extremes of the surveys.  In such cases, care needs
to be taken to make sure that the data points at low and high
redshifts do not strongly bias the parameter estimations for the
magnitude-redshift fit~\cite{Davidian90}.

Further multidimensional scaling analyses do not indicate that more
clustering of the data is present in any dataset.  In the next
sections, we focus our analysis on the relationship between the
magnitude and redshift, which is used to retrieve the cosmological
model parameters.

\subsection{Linear regression model analysis for low redshift supernovae}\label{sec:lin}

We first consider the low redshift SNe, since in these, $\ln(z)$ has a
linear, cosmology-independent relationship to $\mu$, and therefore to
$m_B$. According to the Hubble law, the recession velocity of the
system is given as $v_r=Hd$. In terms of redshift, $v_r=cz$. Thus for
nearby SNe, we may approximate the distance to $d_L=cz/H_0$ where
$H_0=$Hubble parameter at present. Thus the magnitude is related to
redshift as:
\beq
\mu=5{\rm log}_{10} d_L + 25 = 5{\rm log}_{10} z +\left(5{\rm log}_{10}\frac{c}{H_0} +25\right)\,\,,
\eeq 
and the peak B-band brightness as 
\beq\label{eq:lowz}
m_B=\frac{5}{\ln(10)}\ln(z) + \left(\frac{5}{\ln(10)}\ln\left(\frac{c}{H_0}\right) +25 - \alpha \centerdot X + \beta \centerdot C + M_B\right)\,\,.
\eeq

\subsubsection{Methodology}

We use $z \leq 0.1$ as our cut-off point. Although this is a slightly
arbitrary number, any cut-off up to $z < 0.15$ would be valid for the
linear relationship to hold. We use $z \leq 0.1$ as a fairly
conservative cut-off. This limit is taken to be the one where quasi
linear relationship between $\ln(z)$ and $\mu_B$ will hold (see
eq.~\ref{eq:lowz}). In order to study the consistencies in the data,
we use a simple regression law between the redshift and the supernovae
magnitude. We fit the $m_B$ v/s $\ln(z)$ correlation to a linear model
with weights corresponding to $\sigma_{m_B}$. As expected from
theoretical considerations outlined above, we obtain a relationship
following $m_B(z) = a +b \times \ln(z)$ and determine the confidence
levels on the intercept $a$ and slope $b$ of the regression. Should
any statistically significant difference exist from the background
cosmology at low redshift, this would be detected in the
inconsistencies of the fitted parameters of the linear regression
model (especially the slope) between the different datasets and in
comparison to theory at a level determined by their respective
confidence levels.

\begin{figure}
\centering
$\begin{array}{cc}
\includegraphics[width=0.45\linewidth]{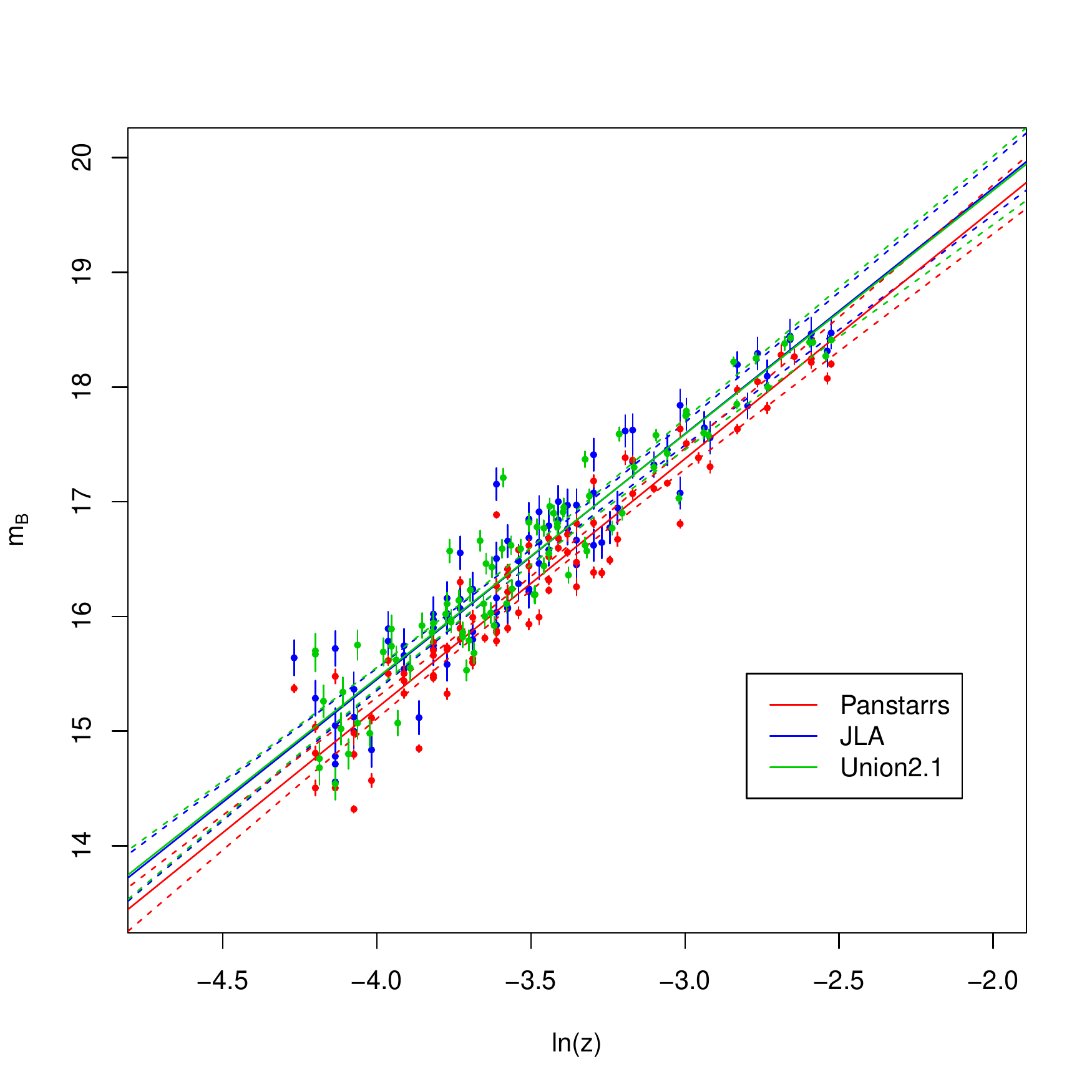} &
\includegraphics[width=0.45\linewidth]{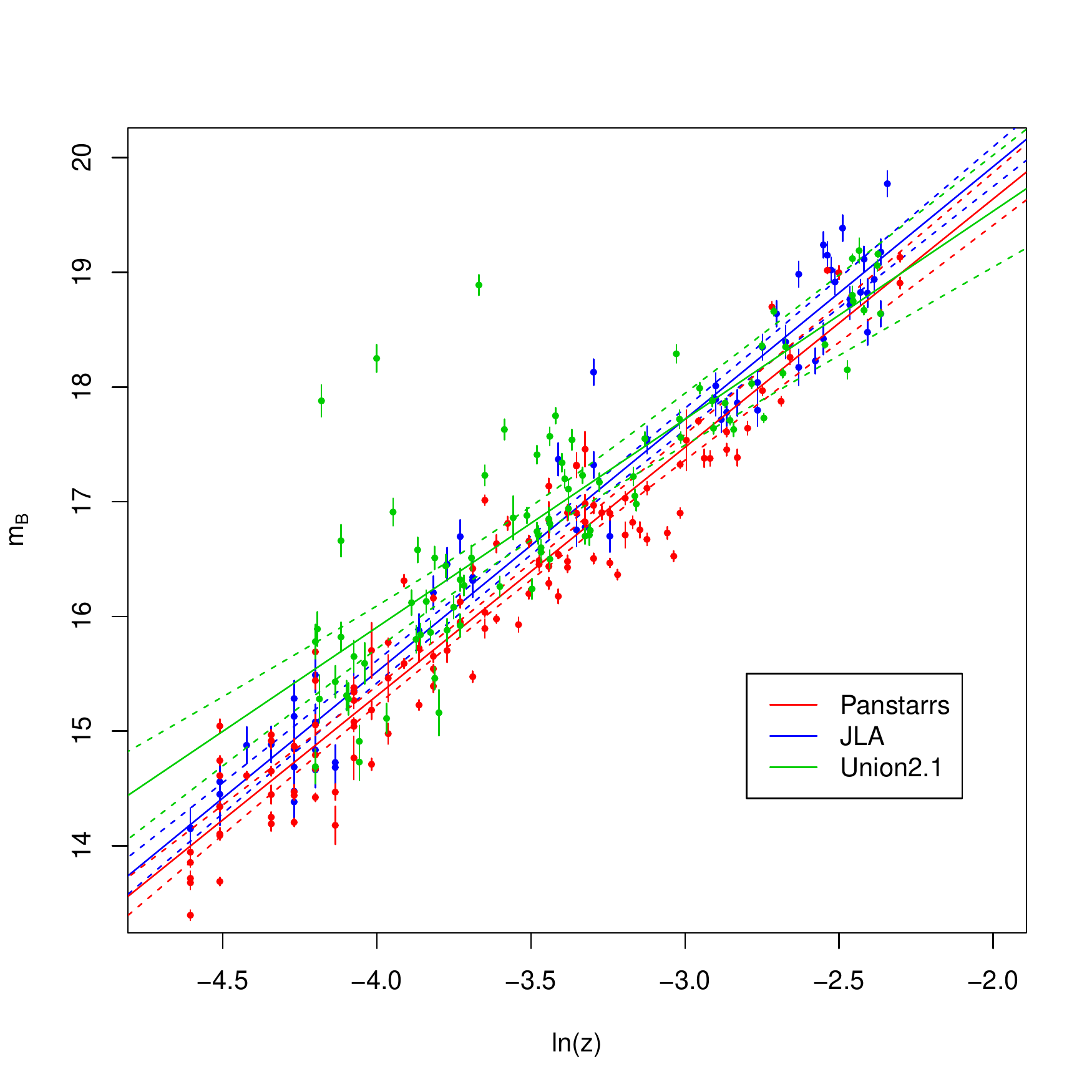} \\
\mbox{(a)} & \mbox{(b)}
\end{array}$
\includegraphics[width=0.45\linewidth]{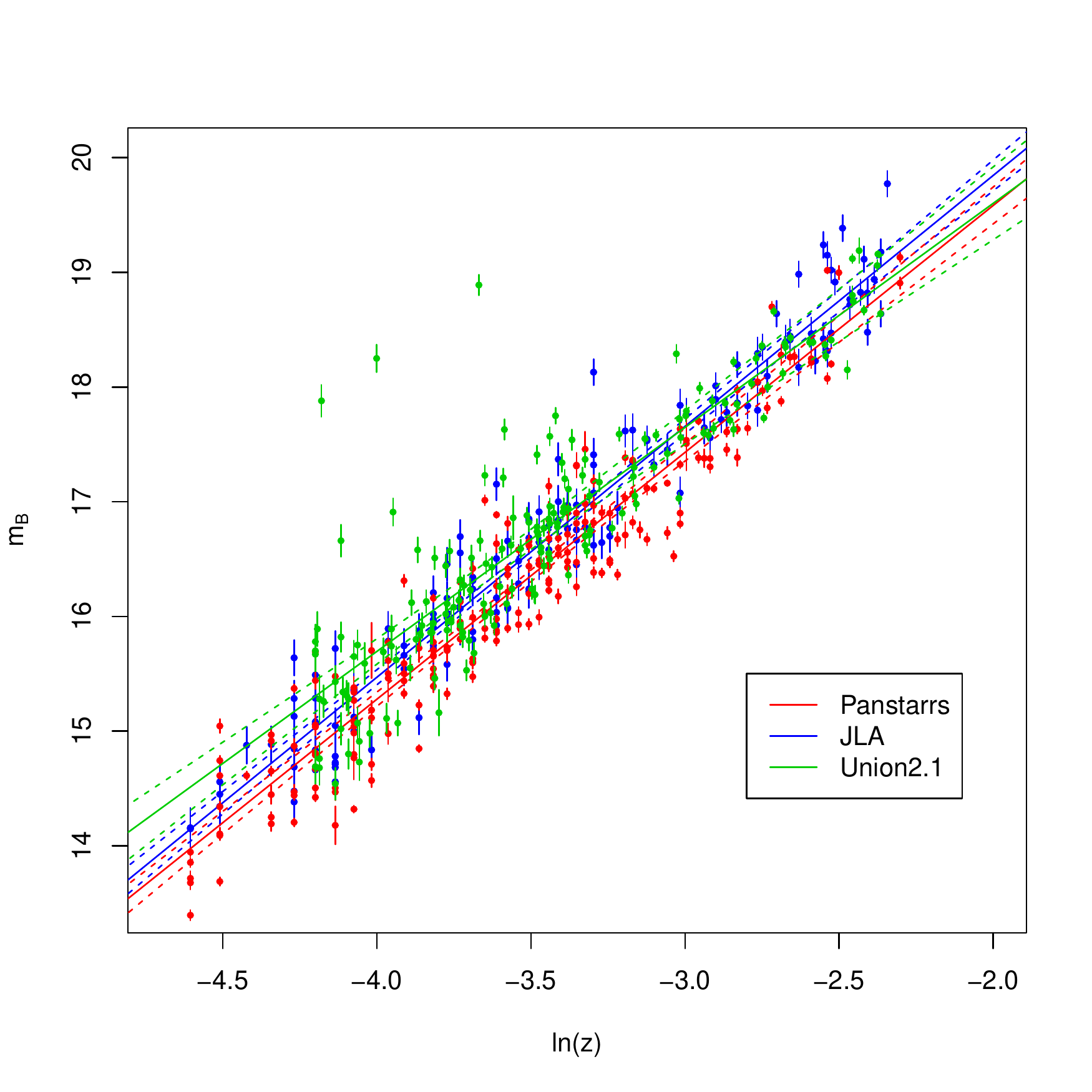} \\
$\mbox{(c)}$
\caption{\footnotesize
Linear regression plots of $m_B(z)$ as a function of $\ln(z)$ for the
three datasets (JLA in blue, Panstarrs in red and Union2.1 in green)
for (a) the supernovae common to all datasets at $z \leq 0.1$, (b) the
supernovae that are different between the datasets at $z \leq 0.1$,
and (c) all the supernovae at $z \leq 0.1$.  Regression lines are
calculated based on weighted least squares optimization. Continuous
lines are the regression line, dashed curves indicate the $2\sigma$
confidence levels of the regression parameters.  A clear difference at
the $2\sigma$ level appears in the regression lines when the extra
data points are added to the analysis.}
\label{fig:linreg}
\end{figure}

\begin{table*}
\caption{\footnotesize
Optimization parameters for the linear regression for the different
datasets at low $z$ ($z \leq 0.1$).}
\label{tab:linreg}      
\centering          
\begin{tabular}{|l|l|ccc|}     
\hline       
Data Subset & Dataset & Intercept & Slope & $R^2$ \\ 
\hline                    
Common & JLA & $24.02 \pm 0.26$ & $2.14 \pm 0.07$ & $0.91$ \\  
Common & Panstarrs & $23.90 \pm 0.24$ & $2.17 \pm 0.07$ & $0.92$\\  
Common & Union2.1 & $23.97 \pm 0.32$ & $2.13 \pm 0.09$ & $0.87$\\  
Extra & JLA & $24.33 \pm 0.19$ & $2.20 \pm 0.05$ & $0.96$ \\  
Extra & Panstarrs & $23.97 \pm 0.25$ & $2.17 \pm 0.07$ & $0.90$\\  
Extra & Union2.1 & $23.16 \pm 0.53$ & $1.81 \pm 0.15$ & $0.63$\\  
All & JLA & $24.22 \pm 0.15$ & $2.19 \pm 0.04$ & $0.95$ \\  
All & Panstarrs & $23.89 \pm 0.18$ & $2.15 \pm 0.05$ & $0.91$\\  
All & Union2.1 & $23.51 \pm 0.35$ & $1.95 \pm 0.10$ & $0.71$\\  
\hline                  
All & Union2.1 without outliers  & $23.97 \pm 0.23$ & $2.11 \pm 0.06$ & $0.87$\\  
\hline                  
\end{tabular}
\end{table*}

\subsubsection{Results}

As described in section~\ref{sec:data}, we consider three different
supernovae datasets that are typically used to constrain the
cosmological models.  Amongst those datasets some 86 supernovae were
found to be common to all datasets at low-$z$, although each has been
processed separately between the different groups, and sometimes also
observed by different telescopes.  To assess whether different
processing techniques or observations introduce bias between the
datasets, we first look at the properties of the common data points
between the different sets.

Figure~\ref{fig:linreg} shows the regression plots of the magnitude
$m_B$ for the three datasets as a function of $\ln(z)$ for (a) the
supernovae common to all datasets, (b) the supernovae that are
different between the datasets (called 'Extra') and (c) all the
supernovae in each dataset.  The regression lines are calculated based
on weighted least squares optimization. The values obtained for the
linear regression optimization are given in Table~\ref{tab:linreg}.
The parameter of interest to compare with theory is the slope of the
regression line. Therefore we do not highlight the results obtained on
the value of the intercept, though it may play a role for higher
redshifts.

From fig~\ref{fig:linreg}(a), it superficially appears that when
considering the 86 data points common to the three datasets, the
magnitude values obtained by the different processes do differ
significantly. Indeed, the magnitudes determined for the Panstarrs
data is consistently lower by 0.19 from the Union2.1 data, and by 0.23
from the JLA data. However, typically simple differences in the
intercept can be absorbed by the nuisance parameters ($\alpha, \beta,
M_B$) in cosmological analysis, and do not show up appreciably in the
cosmological parameters. If the slope of the regression line differs
considerably between datasets, this would be cause for concern. In
this case, as seen in table~\ref{tab:linreg}, the analysis shows that
the regression slope parameters are very close from one dataset to the
next, and within $1\sigma$ of the calculated regression confidence
levels. We also note that the slopes for all three datasets are quite
close to the expected theoretical value of $5/\ln(10)=2.17$.  This
indicates that the processing followed by the different surveys are
consistent and that the cosmological parameters probably would not
depend on the processing of the data done by each group or on the
observer of the supernova, at least at low redshift.

However, when looking at the data points that have been selected
separately in each dataset -- those that are not part of the common
pool of supernovae -- differences in the regression parameters at the
$1\sigma$ level or higher arise, as seen in fig~\ref{fig:linreg}(b).
These differences are sufficiently large that they translate to a
similar level difference in the regression slope between each dataset
when all the low redshift SNe are considered (see
fig~\ref{fig:linreg}(c)). For example, there is a $1\sigma$ level
difference in the slope obtained for the regression of the data points
between JLA and Panstarrs, and a $2\sigma$ difference between the
slopes calculated for Union2.1 and JLA or Panstarrs data.  This
discrepancy is also seen in Table~\ref{tab:linreg}.  It appears that
more than the choice of the technique used for light curve analysis of
supernovae data, the choice of which supernovae to fit is critical in
order to get consistent constraints on the cosmological parameters.

The values determined in Table~\ref{tab:linreg} also indicate
differences in the quality of the fit between the different
datasets. The quantity $R^2$ is the coefficient of determination of
the regression line, here it represents the square of the correlation
between $m_B$ and $\ln(z)$. While the common data points for all
datasets appear to be equivalently fitted amongst all datasets ($R^2
\approx 0.9$), the extra data points added to the Union2.1 dataset
drive the regression parameters away from the ones of JLA and
Panstarrs, and lead to a worse regression fit ($R^2 \approx
0.7$). This effect is seen in about ten outlier data points that are
clearly located away from the regression line in the upper part of
figures~\ref{fig:linreg}(b) and (c) between $-4 \lleq \ln(z) \lleq
-3.5$ or equivalently $0.018 \lleq z \lleq 0.03$.  The Panstarrs fit
for the data, whether using common points or extra appears consistent,
with $R^2$ values uniformly around 0.9.  The difference in slope
between Panstarrs and JLA appears driven by the added extra Panstarrs
points located at very low redshifts and redshifts close to 0.1.  The
JLA dataset behaves somewhat better than either Union2.1 and Panstarrs
because the fit quality improves as extra data points are added to the
common data points, with the $R^2$ value increasing to 0.96 and
reduced residuals.

\begin{figure}
\centering
\includegraphics[width=0.45\linewidth]{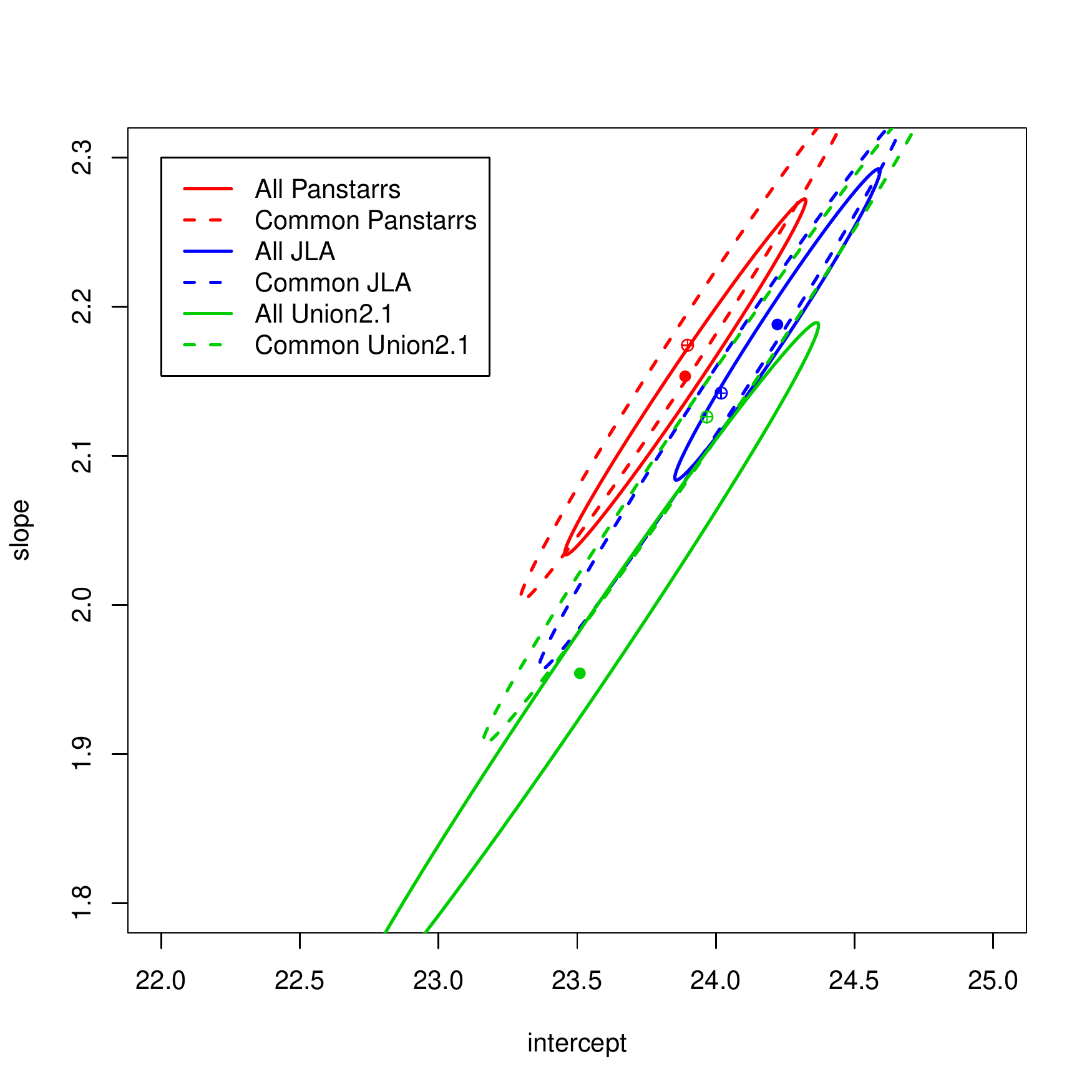} 
\caption{\footnotesize
$2\sigma$ confidence ellipses plotted for the linear regression
parameters in the slope v/s intercept space for common data points and
all the low-$z$ data ($z \leq 0.1$) for the three datasets. Best-fit
points are marked as crossed circles for the common datasets and
filled circles for the full datasets.}
\label{fig:linreg_cl}
\end{figure}

Figure~\ref{fig:linreg_cl} summarizes the linear regression results by
plotting the 95\% confidence ellipses of the linear regression
parameters in the slope v/s intercept space for the SNe common to all
the datasets and for the entire available data at $z \leq 0.1$. It can
be seen from this plot that Panstarrs is $2\sigma$ away from JLA and
Union2.1 whatever the data set considered. However, JLA and Union2.1
are consistent when considering their common data points, but not
anymore if all the data points are considered. JLA and Panstarrs are
well behaved with respect to the inclusion of additional data, as
their $2\sigma= 95\%$ confidence ellipses are consistent with each
other. This is not the case for the Union2.1 dataset, indicating a
significant change to the behaviour of the dataset as additional
points are included. We also see that the slopes for all common data
are consistent with each other and the theoretical value, while for
Union2.1, the full data, due to the effect of the extra data points, is
inconsistent with the others and the expected theoretical value.

\begin{figure}
\centering
\includegraphics[width=0.9\linewidth]{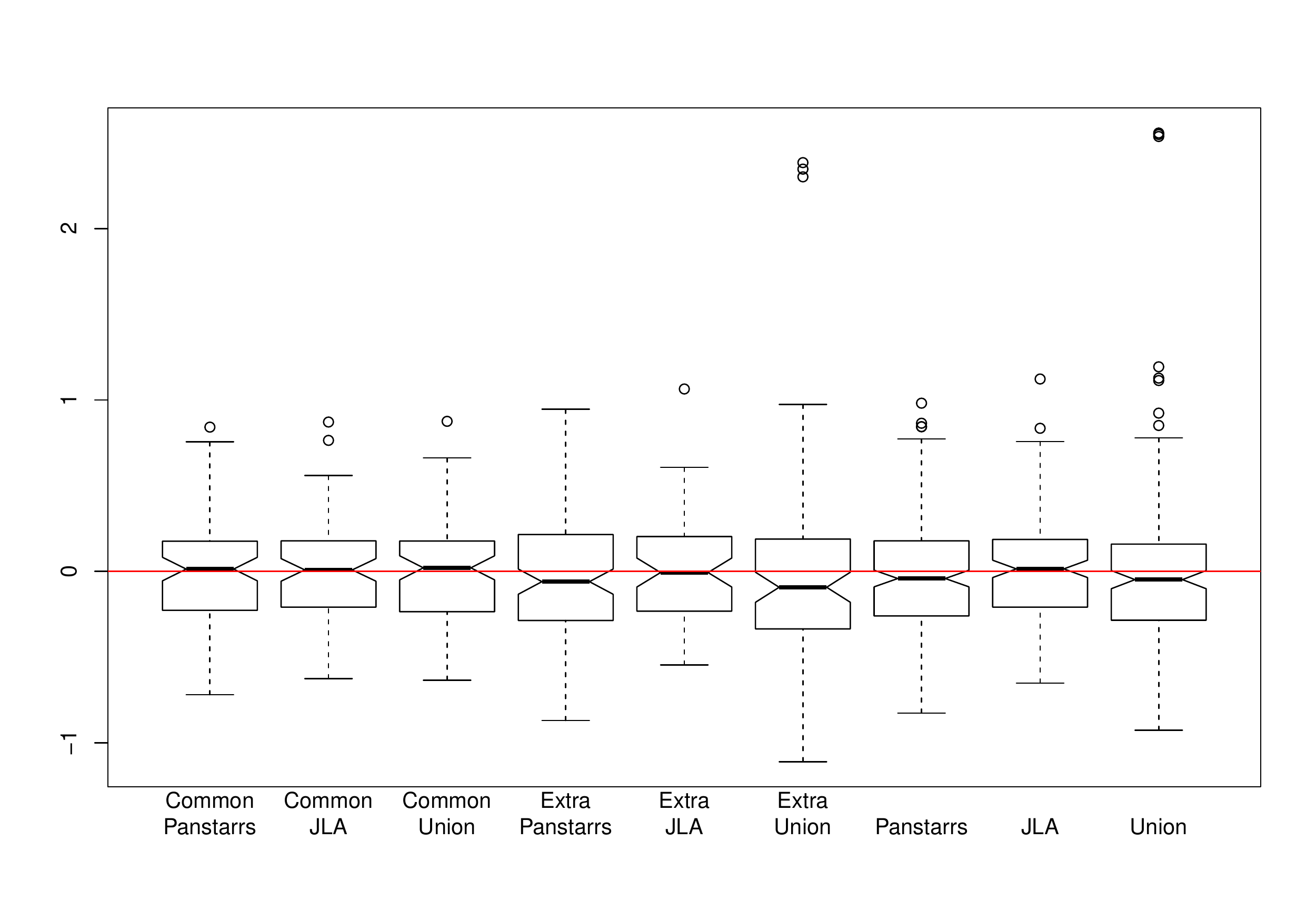} 
\caption{\footnotesize
Boxplot of the residual distribution of each fit for the three
datasets (Panstarrs, JLA, Union2.1).  The residual distribution
obtained for the data common to the three sets is consistent and
relatively tight.  Three strong outlier data points are seen in the
Union2.1 Extra subset, while all distributions present a tail of high
residual values when considering full datasets.}
\label{fig:lin_box}
\end{figure}

\begin{table*}
\caption{\footnotesize  
Optimization parameters for the linear regression for the different
subsets of Panstarrs.}
\label{tab:sub_ps}      
\centering          
\begin{tabular}{|l|ccc|}    
\hline       
Subset & Intercept & Slope & $R^2$ \\
\hline                    
All data & $23.89 \pm 0.18$ & $2.16 \pm 0.05$ & $0.91$ \\  
\hline                    
JRK07  (47) & $23.46 \pm 0.37$ & $2.04 \pm 0.10$ & $0.90$ \\  
CFA3  (84) & $23.87 \pm 0.28$ & $2.15 \pm 0.08$ & $0.91$ \\  
CFA4 (43) & $23.59 \pm 0.41$ & $2.06 \pm 0.12$ & $0.88$ \\  
CSP (21) & $24.41 \pm 0.72$ & $2.32 \pm 0.19$ & $0.88$ \\  
PS1 (7) & $24.36 \pm 0.71$ & $2.24 \pm 0.25$ & $0.93$ \\  
\hline                  
\end{tabular}
\end{table*}

Boxplots are a way to summarize the distribution of values by drawing
the limits of the extreme data and the first and third quartiles
surrounding the median value. Figure~\ref{fig:lin_box} presents
boxplots of the residual distribution obtained for each linear
regression model.  The residual distributions for the common
supernovae of the three sets are consistent and relatively tight.
Three strong outlier data points are seen in the Union2.1 subset of
extra supernovae data, while all distributions present a tail of high
residual values when considering all data points of the sets. This
justifies the lower fit quality determined for the Union2.1 data set,
and suggests that several outliers are present in this data set.  The
data points which have residuals above 95\% of the mean of the residual
distribution of the fitting model can be considered as potential
outliers. Six such points are found (1999gd, 2002hw, 2006br, 2006cm,
2006gj, 2005a) and if they are removed from the analysis, then both
the quality of the fit ($R^2=0.87$) and the values for the intercept
and slope of the regression lines are much closer to the ones obtained
from JLA and Panstarrs analyses as shown in Table~\ref{tab:linreg}.
However, the slope value still remains more than $1\sigma$ away from
the ones calculated for the JLA dataset, indicating that the
differences are not just driven by a few outlier data points, but are
present systematically between the sets of supernovae chosen for
analysis by the different groups.

As a further check, we obtain the same result by comparing the
variance (Fisher-Student test) of the residual populations between the
different datasets and for the different linear models. The comparison
between the variance of residuals for the linear regressions based on
the common data points indicates no statistical difference between the
residuals distributions (p value $> 0.05$).  This is not the case when
comparing the distributions of the residuals for the Union2.1 data set
consisting of extra points as well as for the whole data set of
low-$z$ values as compared to JLA and Panstarrs. The Fisher-Student
test does not show a statistical difference between JLA and Panstarrs
but the hypothesis of identical variance between the residuals of
Union2.1 and the other datasets should be rejected due to very small
p-values (around $10^{-6}$).  When the 6 outlier data points are
removed from the Union2.1 data set, then the Fisher-Student test
remains non-significant.

We next separate the datasets into subsets based on the divisions in
tables~\ref{tab:union},~\ref{tab:jla},~\ref{tab:panstarrs}. We use
these subsets to check if these separations could be significantly
driving the results with specific biases linked to a single
subsample. Each of the JLA, Panstarrs and Union2.1 groups have used
their own consistency checks to put together these datasets. Our
analyses would serve as a further test of the consistencies within and
between these datasets.

\begin{figure}
\centering
$\begin{array}{cc}
\includegraphics[width=0.45\linewidth]{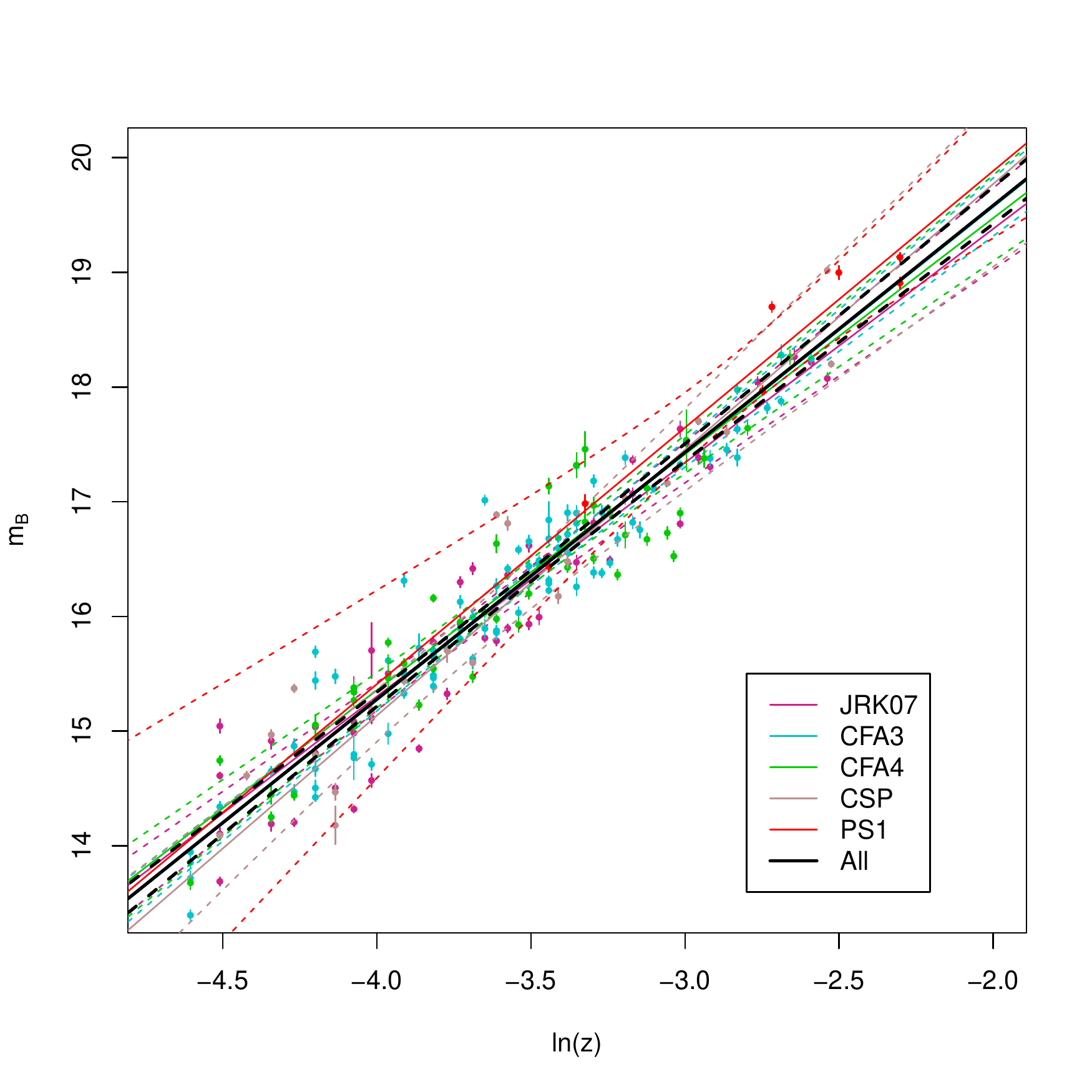} &
\includegraphics[width=0.45\linewidth]{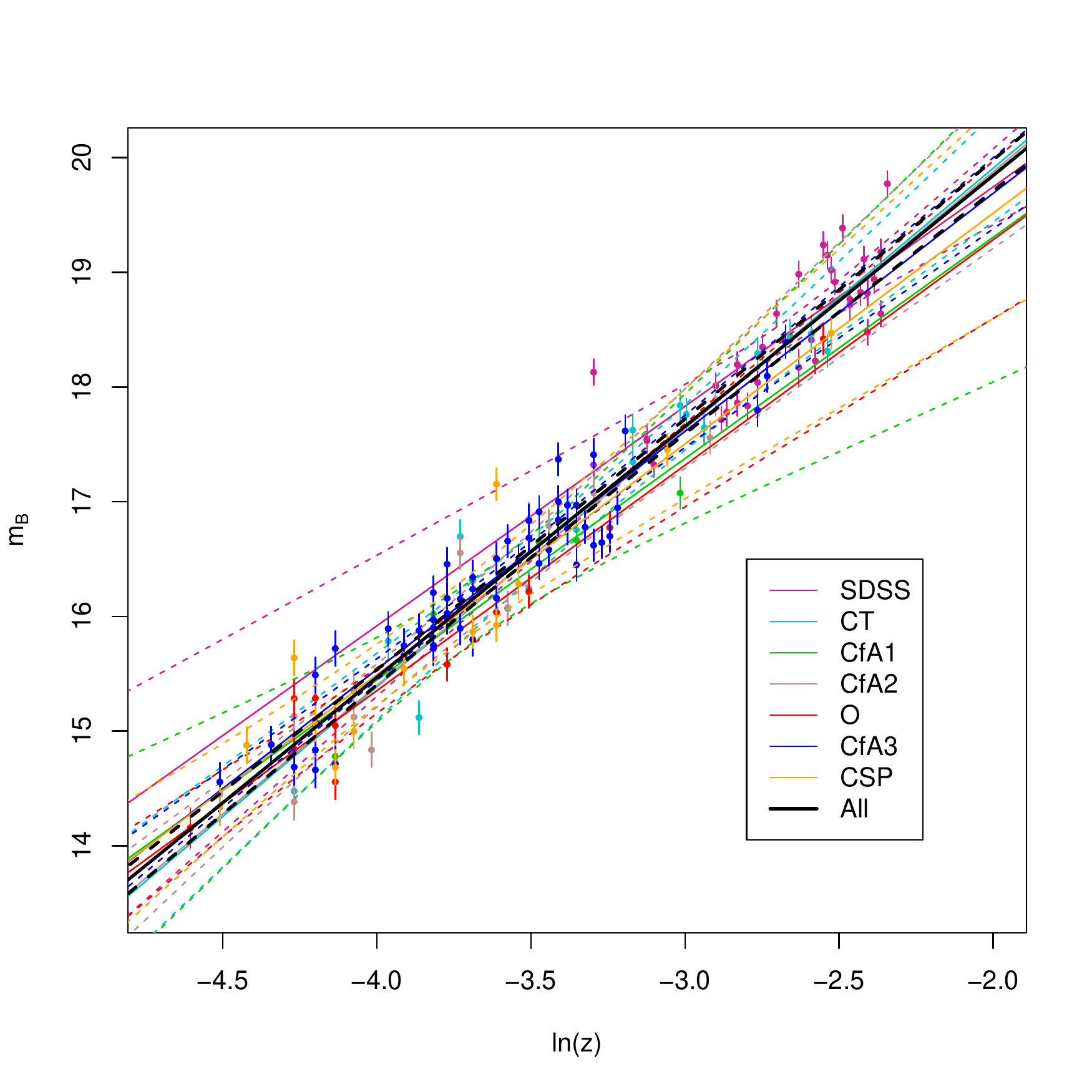} \\
\mbox{(a)} & \mbox{(b)} \\
\includegraphics[width=0.45\linewidth]{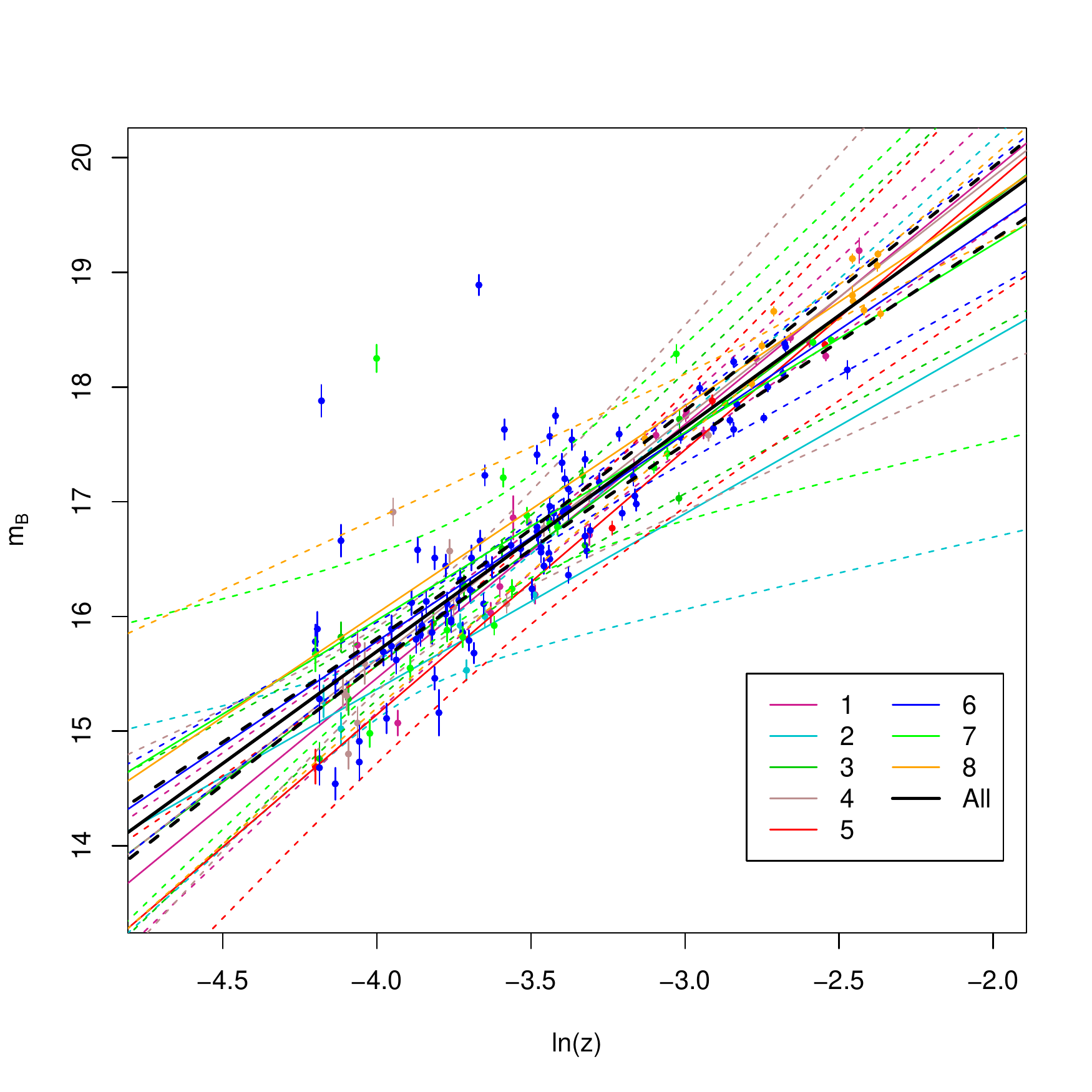} &
\includegraphics[width=0.45\linewidth]{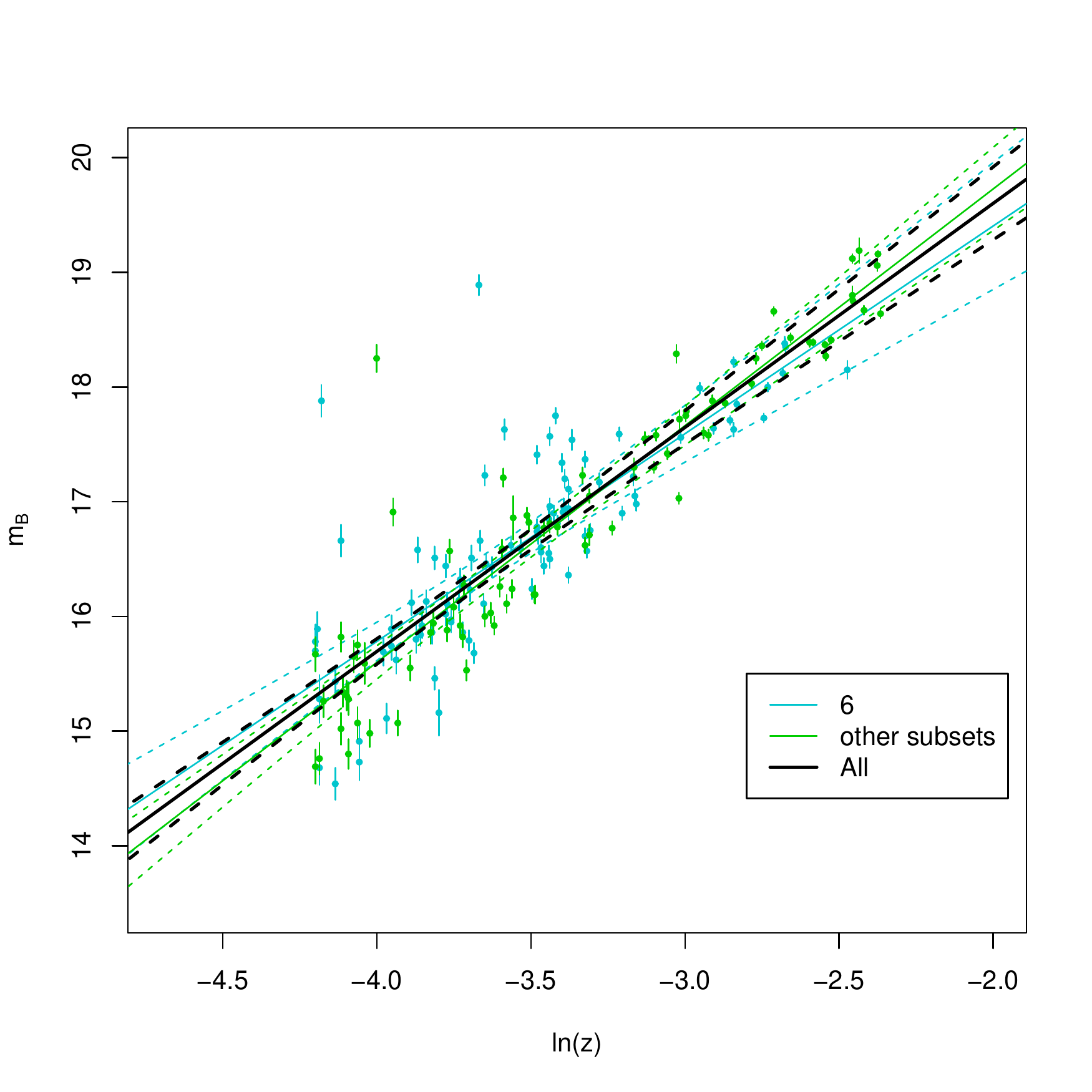} \\
\mbox{(c)} & \mbox{(d)} \\
\end{array}$
\caption{\footnotesize
Regression plots for (a) Panstarrs, (b) JLA (`O' stands for Others),
and (c) Union2.1 datasets using separation between the different
surveys.  Each plot compares the regression lines obtained from each
subset of the dataset and shows that no subset is more than $2 \sigma$
away from the others.  The black lines represent the regression line
and confidence levels for the linear model including all data points.
It can be clearly seen that no subset diverges more than $2 \sigma$
away from the total regression line shown in black.  Plot (d) compares
the regression obtained for subsample 6 with that for the rest of the
subsamples of the Union2.1 dataset.  While the lines are within $2
\sigma$ of each other, there is a clear trend driven by the subset 6
for a lower slope of the regression line.  }
\label{fig:subset}
\end{figure}

For example, the Panstarrs data are separated into 5 subsets (JRK07,
CFA3, CFA4, CSP and PS1). Of these, JRK07 is comprised of several
datasets put together, as mentioned in section~\ref{sec:data}.  Each
subset was used to generate a linear regression which could be
compared to the other ones. We note here that the low-$z$ PS1 group
has only 7 points (since the PS1 telescope mainly observes higher
redshift SNe) and should not be considered statistically relevant,
although we present results from it for the sake of completeness.  The
numerical results are shown in Table~\ref{tab:sub_ps} and the plots
presented in Figure~\ref{fig:subset}(a).  This panel of the figure
compares the regression lines obtained from each subset of the
Panstarrs dataset and demonstrates together with
Table~\ref{tab:sub_ps} that no subset is more than $2 \sigma$ away
from the other subsets.  Figure~\ref{fig:subset}(a) also compares the
regressions obtained for each subset with the one obtained for all the
data.  It can be clearly seen that no subset diverges more than $2
\sigma$ away from the total regression line shown in black and that
apart from CSP and PS1 which contain a relatively low number of data
points, the errors and confidence levels of the regression are
comparable.

The JLA dataset is separated in 7 subsets (SDSS, CT, CfA1, CfA2, O,
CfA3, CSP).  Similar to the analysis we did for the Panstarrs subsets,
we compare the linear regression parameters fitted for each subset.
The numerical results are shown in Table~\ref{tab:sub_jla} and the
plots presented in Figure~\ref{fig:subset}(b).  Here we compare the
regression lines obtained from each subset of the JLA dataset and
demonstrate together with Table~\ref{tab:sub_jla} that no subset is
more than $2 \sigma$ away from the other subsets, though the subsets
containing few data points appear to have larger confidence
levels. Subset SDSS is specific in the fact that it contains a
relatively large number of data points (34) but they are all located
at comparatively high $z$ ($z>0.04$), so that their linear regression
is not particularly tight, and presents large confidence levels. As in
the case of PS1 in Panstarrs, most of the SDSS data is actually for
the medium redshift range, so the lower redshift SNe from this set may
not be particularly relevant to the low redshift analysis.  Again,
from the figure~\ref{fig:subset}(b), it can be seen that no subset
diverges more than $2 \sigma$ away from the total regression line
shown in black.

\begin{table*}
\caption{\footnotesize 
Optimization parameters for the linear regression of the different
subsets of JLA.}
\label{tab:sub_jla}      
\centering          
\begin{tabular}{|l|ccc|}     
\hline\hline       
Subset & Intercept & Slope & $R^2$ \\
\hline                    
All data & $24.22 \pm 0.15$ & $2.19 \pm 0.04$ & $0.95$ \\  
\hline                    
SDSS  (34) & $23.57 \pm 0.60$ & $1.91 \pm 0.22$ & $0.67$ \\  
CT  (17) & $24.43 \pm 0.52$ & $2.26 \pm 0.16$ & $0.93$ \\  
CFA1 (7) & $23.17 \pm 1.06$ & $1.93 \pm 0.29$ & $0.88$ \\  
CFA2 (15) & $24.36 \pm 0.62$ & $2.24 \pm 0.16$ & $0.93$ \\  
CFA3 (55) & $23.84 \pm 0.33$ & $2.08 \pm 0.09$ & $0.91$ \\  
CSP (13) & $23.54 \pm 0.85$ & $2.01 \pm 0.22$ & $0.88$ \\  
Others (11) & $23.22 \pm 0.61$ & $1.97 \pm 0.16$ & $0.94$ \\  
\hline                  
\end{tabular}
\end{table*}

Finally, 8 subsets are present in the Union2.1 dataset given by the
different data releases from various observatories that have performed
the supernovae observations over the years. The subsets have a very
heterogeneous distribution of data points, one subset containing just
4 data points while the largest subset contains 94 of them.  We have
assessed whether any of the subsets was biasing the data as a whole.
The values obtained for the regression models in this case are shown
in Table~\ref{tab:sub_union}. The regression slopes are all within $2
\sigma$ of the value for the model that includes all the data
points. For the two subsets that contain the least number of data
points (6 and 4 data points respectively), we note that the best-fits
are somewhat different, but as these have large confidence levels,
they cannot be expected to drive the trend of the dataset.

Figure~\ref{fig:subset}(c) compares the regression lines obtained from
each subset of Union2.1 and shows that no subset is more than $2
\sigma$ away from the other subsets. Though it can be noted
immediately that the subsets with the lowest number of data points
have very large confidence levels, they are still consistent with the
other more constrained regression lines. Again, no subset diverges
more than $2 \sigma$ away from the total regression line shown in
black.  So the behaviour of the data selected within each survey
appears consistent.

\begin{table*}
\caption{\footnotesize
Optimization parameters for the linear regression of the different
subsets of Union2.1. !6 corresponds to the set of all subsamples apart
from subsample 6.}
\label{tab:sub_union}      
\centering          
\begin{tabular}{|l|ccc|}   
\hline\hline       
Subset & Intercept & Slope & $R^2$ \\
\hline                    
All data & $23.51 \pm 0.35$ & $1.95 \pm 0.10$ & $0.71$ \\  
\hline                    
1  (17) & $24.31 \pm 0.56$ & $2.21 \pm 0.17$ & $0.92$ \\  
2  (6) & $21.49 \pm 1.28$ & $1.53 \pm 0.33$ & $0.80$ \\  
3 (10) & $23.69 \pm 1.02$ & $2.03 \pm 0.27$ & $0.86$ \\  
4 (15) & $24.05 \pm 1.59$ & $2.10 \pm 0.41$ & $0.64$ \\  
5 (4) & $24.38 \pm 0.48$ & $2.31 \pm 0.13$ & $0.99$ \\  
6 (94) & $23.02 \pm 0.60$ & $1.81 \pm 0.16$ & $0.57$ \\  
7 (18) & $22.52 \pm 1.77$ & $1.64 \pm 0.48$ & $0.38$ \\  
8 (11) & $23.27 \pm 0.66$ & $1.81 \pm 0.25$ & $0.83$ \\  
\hline
!6 (81) & $23.86 \pm 0.40$ & $2.06 \pm 0.11$ & $0.82$ \\  
\hline                  
\end{tabular}
\end{table*}

However, one can see that the subset 6 of Union2.1, which contains the
most data points, has a low value for the correlation ($R^2 \approx
0.6$) and the lowest slope for a large subset for any of the
regression models calculated so far, with a value of 1.81.  If one
compares the regression obtained from the subset 6 with that from the
rest of the data points (as shown in Table~\ref{tab:sub_union}), one
notices that the regression model obtained for the remaining data
points of the Union2.1 dataset is consistent with the model values
obtained by JLA and Panstarrs as shown in Table~\ref{tab:linreg}. It
seems that subset 6 is the most inconsistent set of data with respect
to all the other supernovae surveys and is driving the difference
between Union, JLA and Panstarrs at low redshift. The respective plots
of the models are shown in Figure~\ref{fig:subset}(d) using separation
between the survey number 6 and the rest of the surveys.  The two
plots compare the regression lines obtained from the subset 6 of
Union2.1 with the regression line obtained by considering all the
other subsets except for 6.  The lines are about $2 \sigma$ away from
each other, in addition, there is a clear trend driven by the subset 6
for a lower slope of the regression line, while the rest of the data
from the Union2.1 dataset is much more consistent with both JLA and
Panstarrs.

Thus we find that, while the subsets in the Panstarrs and JLA datasets
are quite consistent with each other and the full datasets, the set 6
of the Union2.1 dataset is almost $2\sigma$ inconsistent with the
other Union2.1 subsets and with the JLA and Panstarrs datasets. It is
also inconsistent with the expected theoretical value of $2.17$. We
note here that subset 6 of Union2.1 is drawn from the same dataset as
the CfA3 subset of both JLA and Panstarrs. However, the results from
it do not match those from either the JLA subset of CfA3 or the
Panstarrs one (as seen from
tables~\ref{tab:sub_ps},~\ref{tab:sub_jla},~\ref{tab:sub_union}).

Therefore, from the above analysis we may conclude that the
significant differences obtained when comparing the three low-$z$
datasets most likely correspond to a bias introduced by the selection
of the supernovae within the different surveys, as it does not appear
to stem from the processing techniques. The difference may be driven
by the selected data points and could be traced to some extent to the
data subset 6 of Union2.1 (drawn from CfA3) which appears to be
driving the trends in the regression seen in the Union2.1 dataset. It
is not clear by how much, if at all, these differences in the low
redshift datasets translate to cosmological results. This will be
studied in section~\ref{sec:cosmo}.

\subsection{Nonlinear regression model analysis for all supernovae data}\label{sec:nonln}

We now consider the entire datasets for each of Panstarrs, JLA and
Union2.1. We note that while low redshift data is essential for
calibrating the datasets, the cosmological information in the
supernovae Type Ia is really available in the mid to high-$z$
data. The high-$z$ data for Union2.1 is a composite of several
different surveys whereas for JLA it is mostly dependent on SNLS and
SDSS data, while for Panstarrs it is entirely drawn from the Panstarrs
telescope survey.

\subsubsection{Methodology}

When considering redshift values larger than 0.1, the linear
relationship of the previous section cannot be used, and needs to be
adapted to a non-linear variation of the data. The initial
relationship we considered for low-$z$ data was $m_B(z) = a + b \times
\ln(z)$. To take into account nonlinear regression by adding minimal
parameters in a cosmology-blind analysis, we now introduce a third
parameter that corresponds to a quadratic term in $\ln(z)$ so that the
magnitude to redshift relationship is expressed as $m_B(z) = a + b
\times \ln(z) + c \times \ln(z)^2$. As the initial magnitude to
redshift correlation at low-$z$ is almost linear, the parameter $c$ is
a minor correction to the relationship and we find that it gives good
results for the model fitting in SNe data. We obtain the best fit by
using a generalized nonlinear least square fitting algorithm
\cite{R16,Bates88}.  It is important to remember here that this fit
does not take into consideration any cosmological models. It is simply
a statistical fit to the data which has no physical or cosmological
significance, and thus has no bias towards any specific cosmology.

\begin{table*}
\caption{\footnotesize
Optimization parameters for the non-linear regression of the different
datasets. Clow refers to the common low redshift data subset, H to the
high redshift dataset.}
\label{tab:nonlin}      
\centering          
\begin{tabular}{|l|ccccc|}     
\hline       
subset & N & $a$  & $b$ & $c$ & RSD \\
\hline
All data &&&&&\\           
\hline      
JLA  & 740 & $24.82 \pm 0.03$ & $2.59 \pm 0.03$ & $0.06 \pm 0.007$ & $0.097$ \\  
Panstarrs  & 310 & $24.32 \pm 0.09$ & $2.36 \pm 0.09$ & $0.02 \pm 0.02$ & $0.082$ \\  
Union2.1  & 580 & $24.90 \pm 0.04$ & $2.61 \pm 0.05$ & $0.08 \pm 0.01$ & $0.128$  \\  
\hline 
Clow + H &&&&&\\   
\hline                 
JLA  & 674 & $24.83 \pm 0.03$ & $2.61 \pm 0.03$ & $0.06 \pm 0.008$ & $0.094$  \\  
Panstarrs & 194 & $24.29 \pm 0.10$ & $2.31 \pm 0.11$ & $ 0.007 \pm 0.02$ & $0.070$ \\  
Union2.1 & 491 & $24.88 \pm 0.03$ & $2.54 \pm 0.05$ & $0.04 \pm 0.01$ & $0.103$ \\  
\hline       
Clow + H ($<0.64$) &&&&&\\    
\hline         
JLA & 546 & $24.85 \pm 0.06$ & $2.63 \pm 0.06$ & $0.07 \pm 0.01$ & $0.099$  \\  
Panstarrs & 194 & $24.29 \pm 0.10$ & $2.31 \pm 0.11$ & $0.007 \pm 0.02$ & $0.070$  \\  
Union2.1 & 383 & $24.67 \pm 0.08$ & $2.30 \pm 0.1$ & $-0.005 \pm 0.02$ & $0.105$  \\  
\hline 
H&&&&&\\   
\hline                   
JLA & 588 & $24.88 \pm 0.03$ & $2.78 \pm 0.06$ & $0.15 \pm 0.03$ & $0.090$ \\  
Panstarrs & 108 & $24.28 \pm 0.24$ & $2.34 \pm 0.37$ & $0.04 \pm 0.13$ & $0.076$  \\  
Union2.1 & 405 &  $24.90 \pm 0.03$ & $2.67 \pm 0.08$ & $0.15 \pm 0.05$ & $0.104$  \\  
\hline 
H ($<0.64$)&&&&&\\   
\hline                    
JLA & 460 & $25.18 \pm 0.1$ & $3.22 \pm 0.15$ & $0.3 \pm 0.06$ & $0.095$  \\  
Panstarrs & 108 & $24.28 \pm 0.24$ & $2.34 \pm 0.37$ & $0.04 \pm 0.13$ & $0.076$  \\  
Union2.1 & 297 &  $24.4 \pm 0.16$ & $1.75 \pm 0.28$ & $-0.21 \pm 0.11$ & $0.106$  \\  
\hline 
Large subsets&&&&&\\   
\hline                    
JLA & 668 & $24.81 \pm 0.03$ & $2.60 \pm 0.03$ & $0.07 \pm 0.008$ & $0.093$ \\  
Panstarrs & 197 & $24.24 \pm 0.1$ & $2.25 \pm 0.11$ & $0.001 \pm 0.02$ & $0.076$  \\  
Union2.1 & 369 &  $24.8 \pm 0.07$ & $2.52 \pm 0.09$ & $0.06 \pm 0.02$ & $0.124$ \\  
\hline                  
\end{tabular}
\end{table*}

\begin{figure}
\centering
$\begin{array}{cc}
\includegraphics[width=0.45\linewidth]{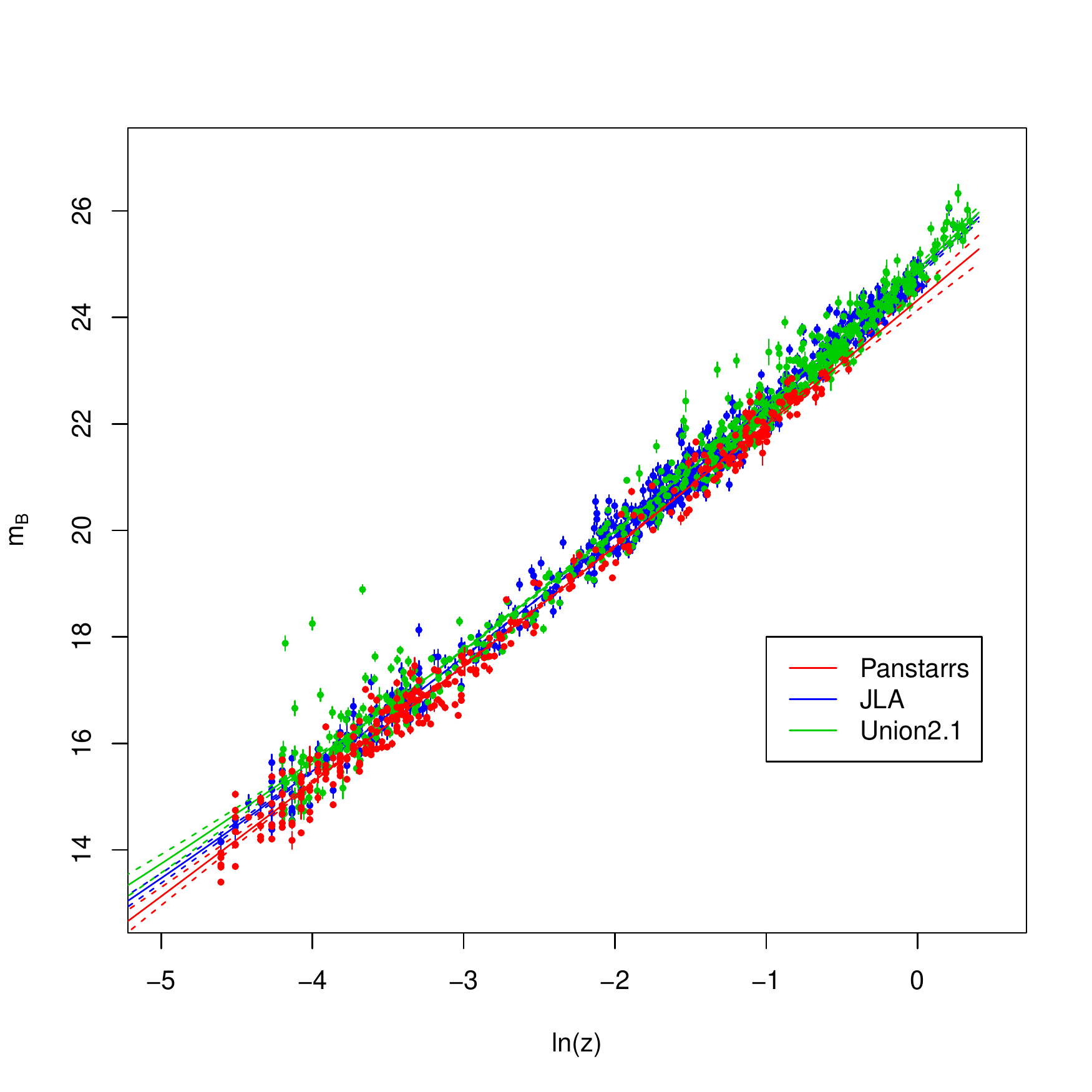} &
\includegraphics[width=0.45\linewidth]{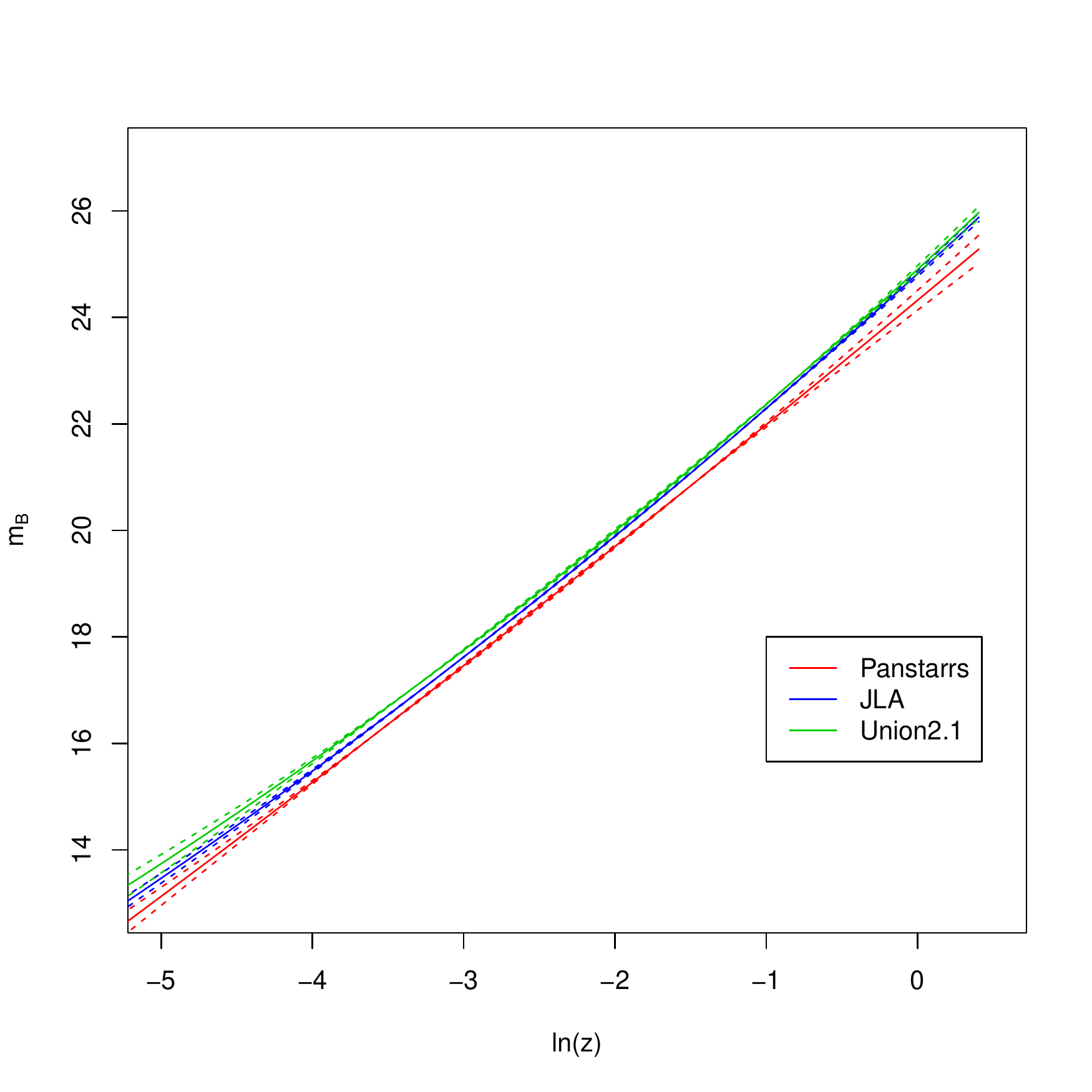} \\
\mbox{(a)} & \mbox{(b)} \\
\end{array}$
\caption{\footnotesize
Nonlinear regression plots of the datasets using all the data from the
three data sets.  (a) With and (b) without the data points to allow
visualization of the regression curves.  }
\label{fig:nonlin}
\end{figure}

\subsubsection{Results}

Table~\ref{tab:nonlin} presents a summary of all the results for the
nonlinear regression fitted parameters.  We have separated the data
into various subsets to compare models amongst the datasets. The first
nonlinear regression analysis was done on the entire data sets for
Panstarrs, JLA and Union2.1 without any further processing.  From the
results of the table, it can be seen that the regression model is
relatively good.  This can also be seen in Figure~\ref{fig:nonlin}.

In the table, it can be seen that the parameters calculated for the
Panstarrs data are very different from the JLA and Union2.1 data by
more than $2 \sigma$. While JLA and Union2.1 parameters are closer,
they still are around $1 \sigma$ different and thus not as consistent
as they could be. This difference could be driven by the low-$z$ data
difference that was discussed in the previous section as it is clearly
noticed that for high-$z$ data, the curves appear to be superposed for
the three datasets, \eg in the Figure~\ref{fig:nonlin}(b).
 
It is to be noted that on all the fits calculated for the Panstarrs
data, the parameter $c$ added for the nonlinear regression is never
contributing significantly to the fit (as can be seen from its
estimated value which is typically much smaller than the other two
parameters), and therefore although it is reported here for
consistency, we do not attach much meaning to it.  This shows that the
behaviour of Panstarrs data is different from JLA and Union2.1 as it
is not different from a linear regression even including the high-$z$
data points, whereas a better fit is obtained for JLA and Union2.1
when including the non linear term.  This difference could be due to
the lower number of data points at high-$z$ for Panstarrs, since the
current dataset presents a cut-off around $z=0.64$. Thus, as the
Panstarrs data does not go to very high redshifts, the quadratic term
containing $c$ may not be very strongly required for fitting this
dataset. However, one can see from Table~\ref{tab:nonlin} that $c$
remains a significant parameter in the non-linear fits of especially
the JLA dataset even when the high-$z$ ($>0.64$) data points are
removed.  If this tendency to linearity is confirmed by future, higher
redshift Panstarrs measurements, it would have strong implications on
the cosmology constraints.  At the moment we are not well-placed to
present any clear judgement on this.

Interestingly, among the three datasets, the one which presents the
least weighted residual standard deviation, RSD, is Panstarrs.  For a
regression fit, the RSD is the root mean square weighted residual sum
of squares for the model, $\sqrt{\sum \frac{1}{n-p }
  (m_B-m_{B_{fit}})^2}$, where $n$ is the number of data points and
$p$ the number of free parameters.  Thus, small RSD indicates a good
fit to the data and tighter residual distribution.  The Panstarrs
data, even though it is the smallest dataset, or perhaps because of
it, appears to be the most consistent within itself.

\begin{figure}
\centering
$\begin{array}{cc}
\includegraphics[width=0.45\linewidth]{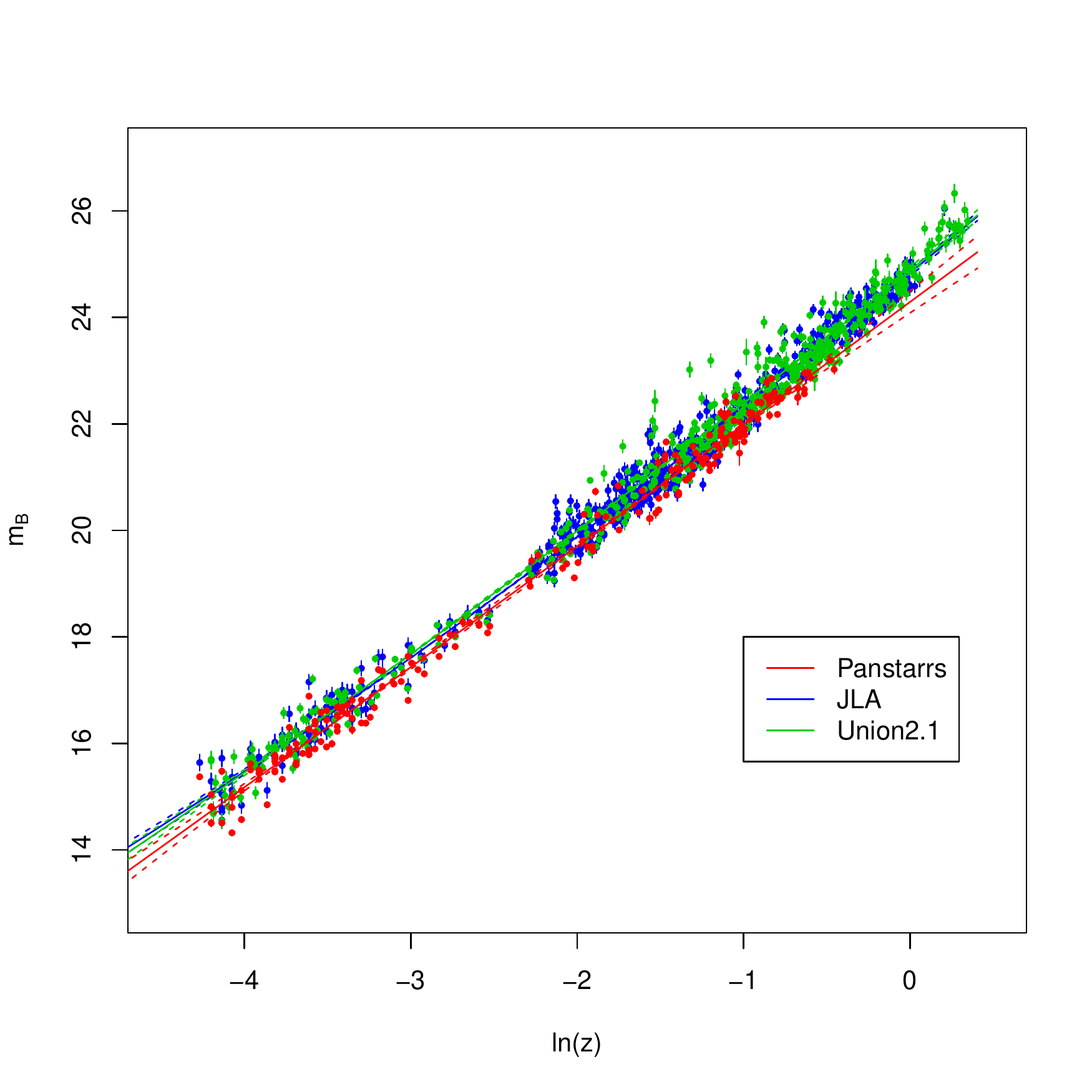} &
\includegraphics[width=0.45\linewidth]{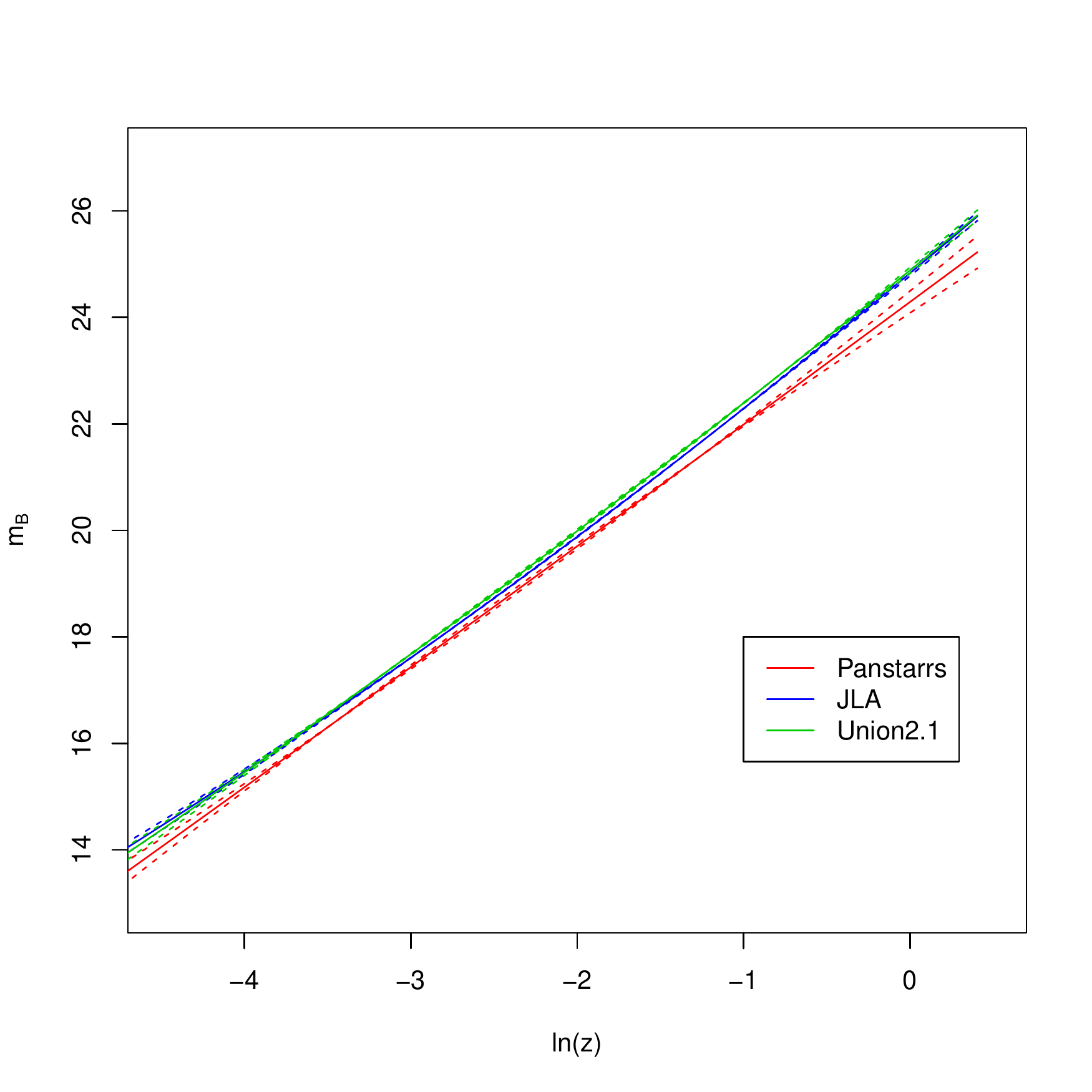} \\
\mbox{(a)} & \mbox{(b)} \\
\end{array}$
\caption{\footnotesize
Nonlinear regression plots of the datasets using common low-$z$ and
all high-$z$ data from the three data sets.  (a) With and (b) without
the data points to allow visualization of the regression curves.  }
\label{fig:nonlin_comm}
\end{figure}

\begin{figure}
\centering
$\begin{array}{cc}
\includegraphics[width=0.45\linewidth]{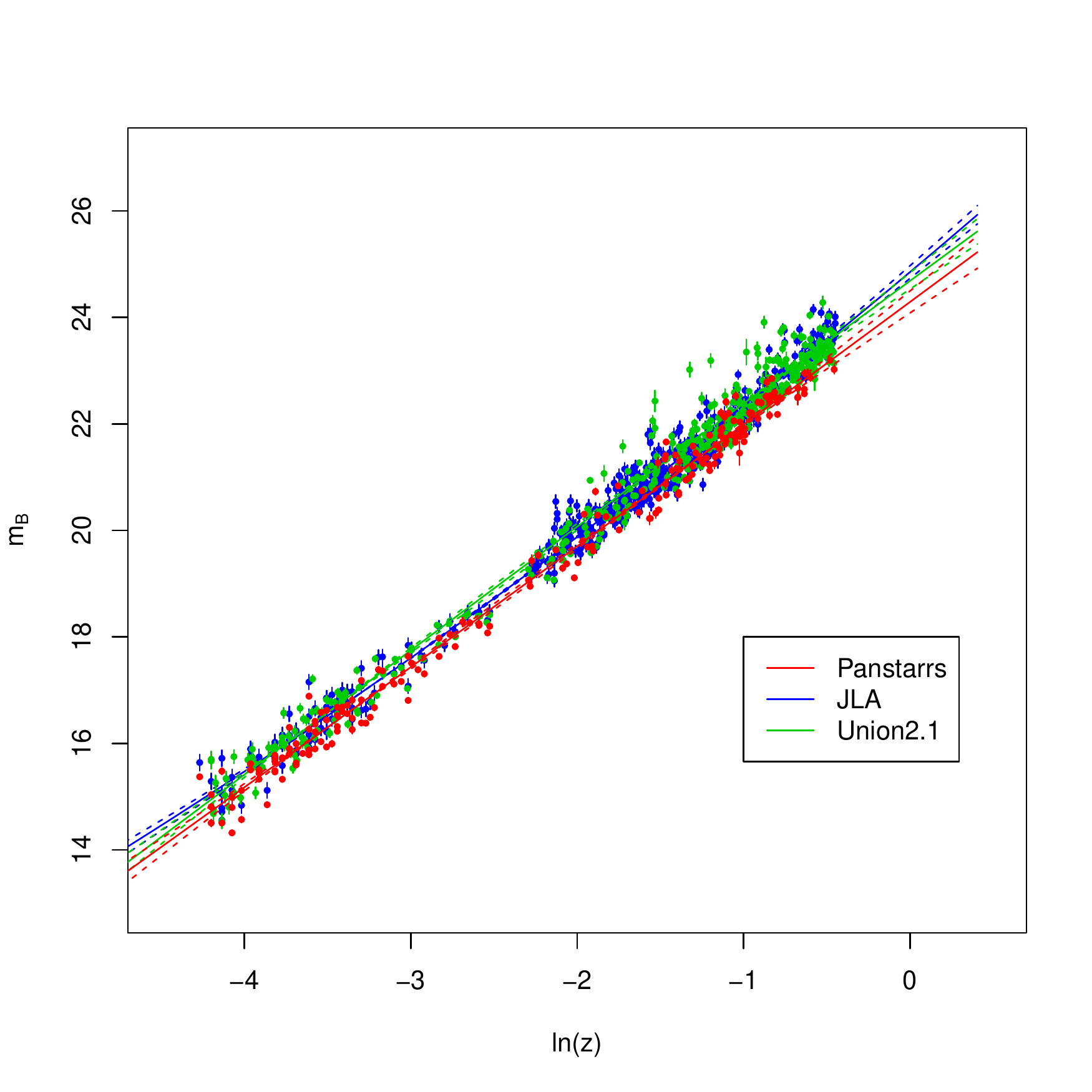} &
\includegraphics[width=0.45\linewidth]{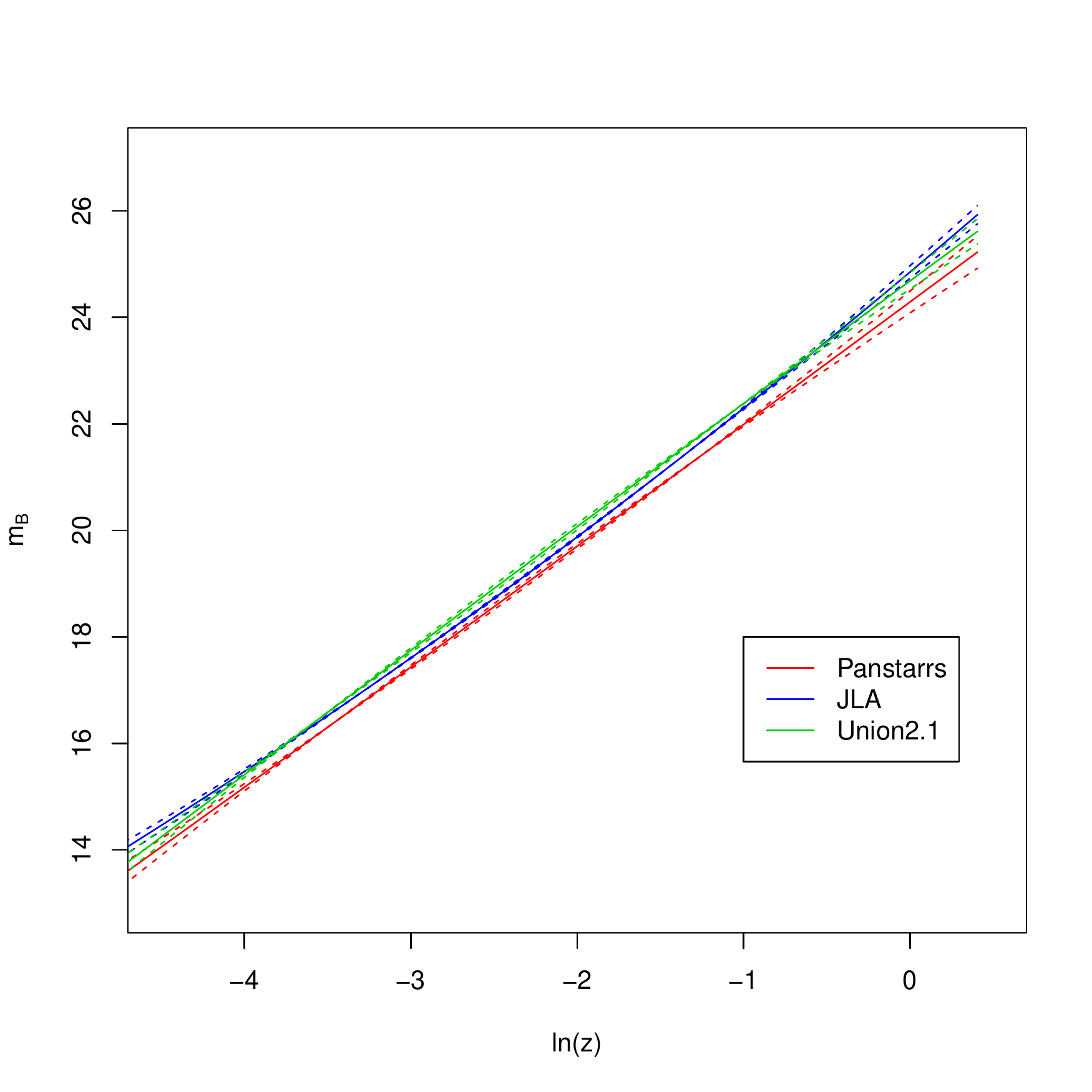} \\
\mbox{(a)} & \mbox{(b)} \\
\end{array}$
\caption{\footnotesize
Nonlinear regression plots of the datasets using common low-$z$ and
high-$z$ data ($<0.64$) from the three data sets.  (a) With and (b)
without the data points to allow visualization of the regression
curves.  }
\label{fig:nonlin_cut}
\end{figure}

\begin{figure}
\centering
$\begin{array}{cc}
\includegraphics[width=0.45\linewidth]{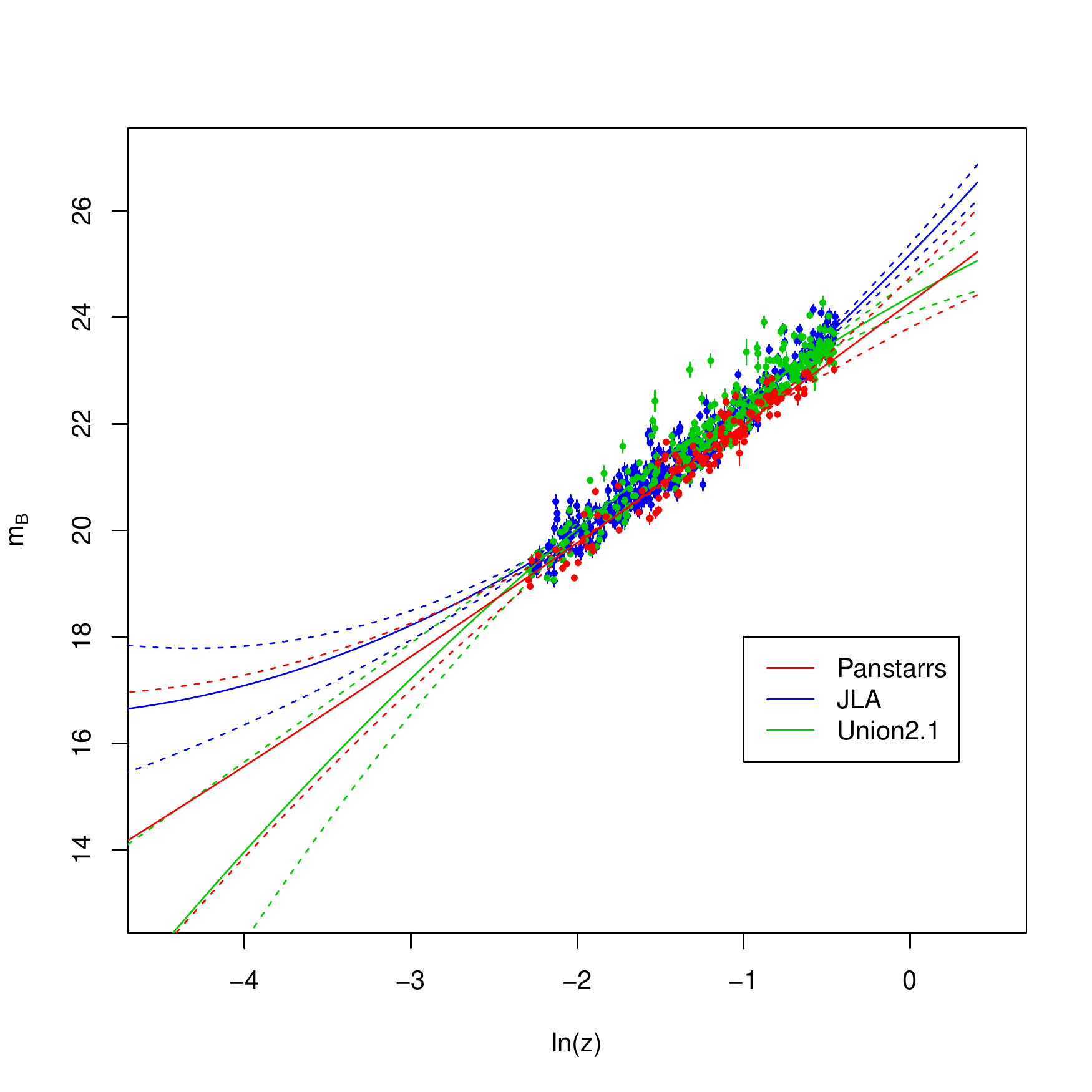} &
\includegraphics[width=0.45\linewidth]{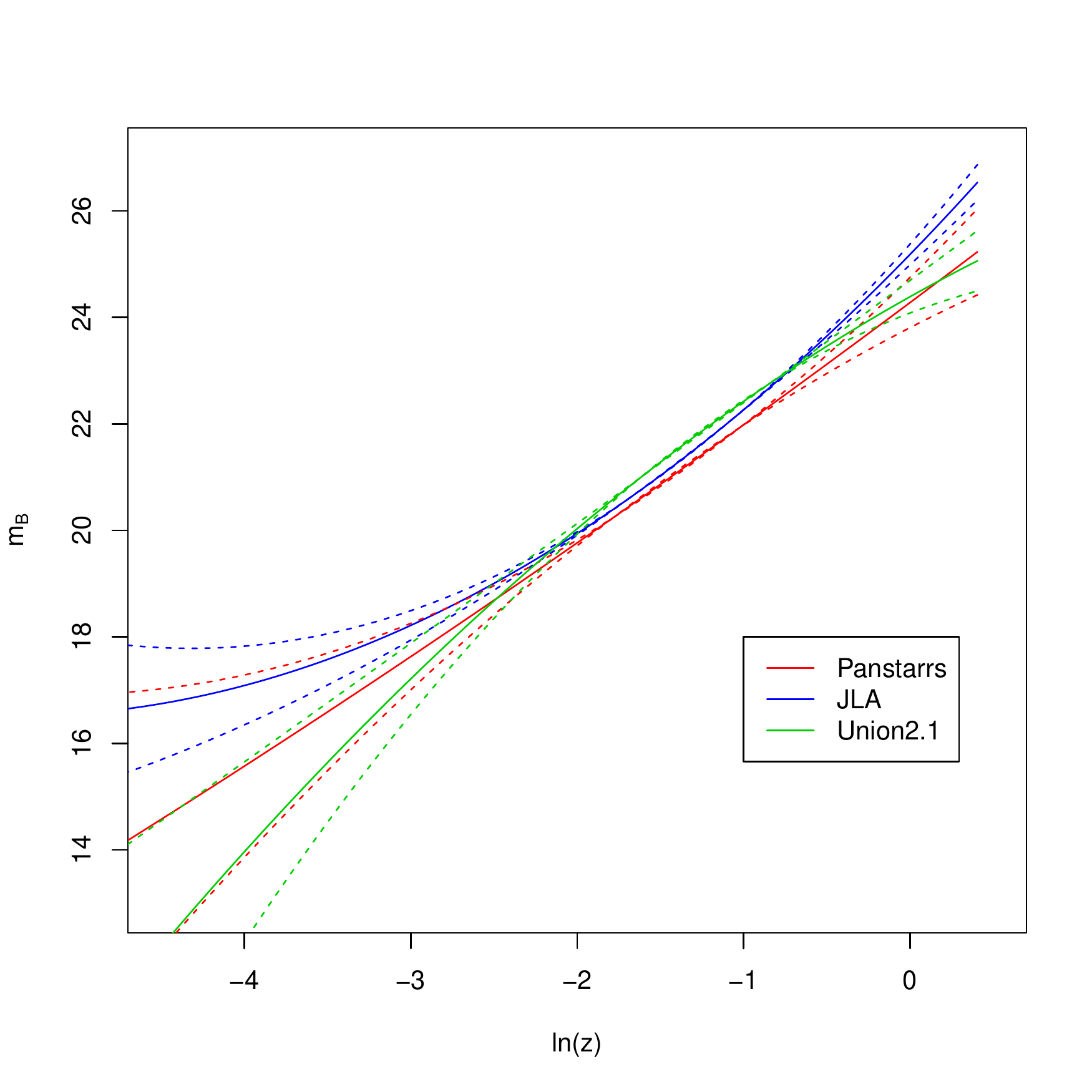} \\
\mbox{(a)} & \mbox{(b)} \\
\end{array}$
\caption{\footnotesize
Nonlinear regression plots of the datasets using only high-$z$ data
($<0.64$) from the three data sets.  (a) With and (b) without the data
points.  }
\label{fig:highz}
\end{figure}

\begin{figure}
\centering
$\begin{array}{cc}
\includegraphics[width=0.45\linewidth]{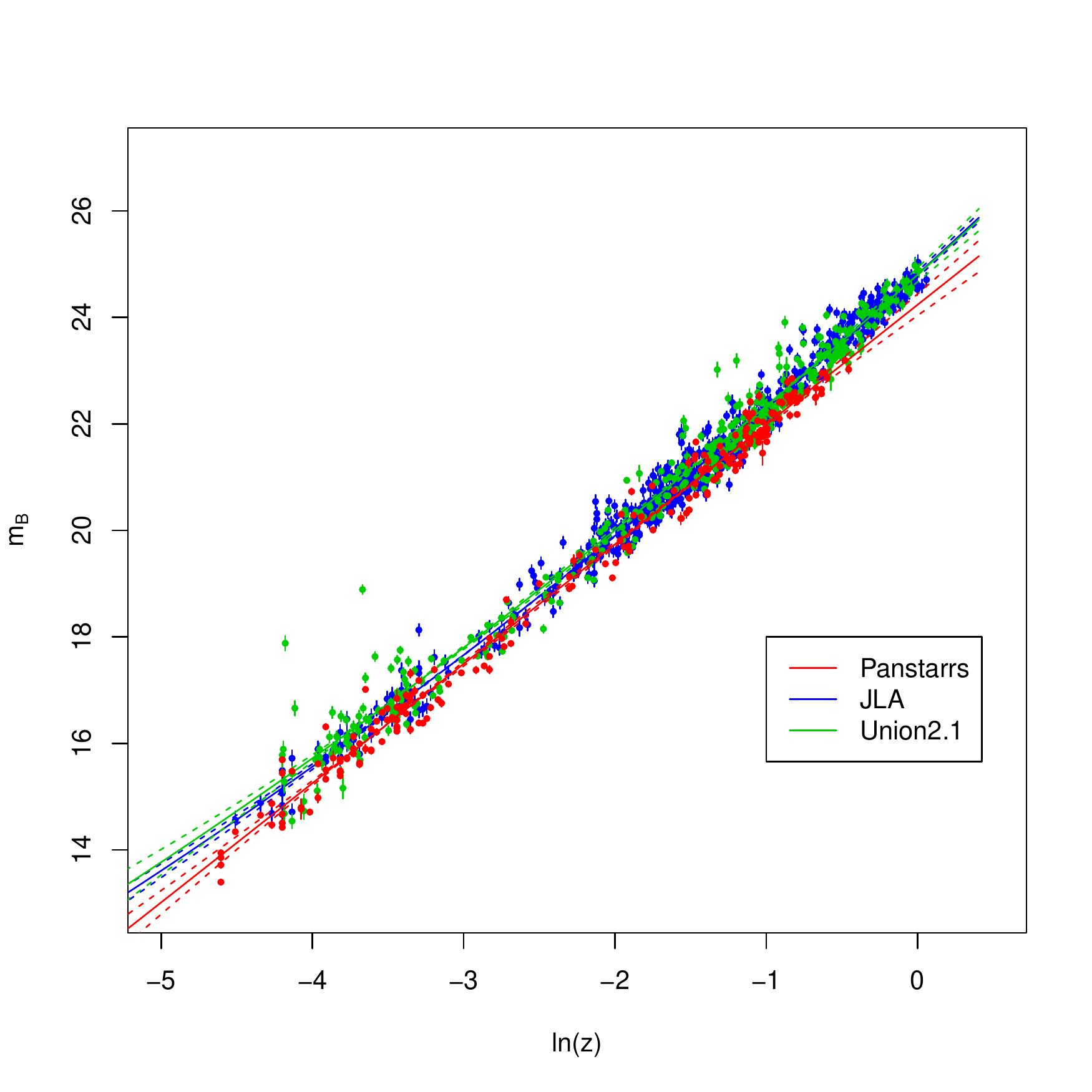} &
\includegraphics[width=0.45\linewidth]{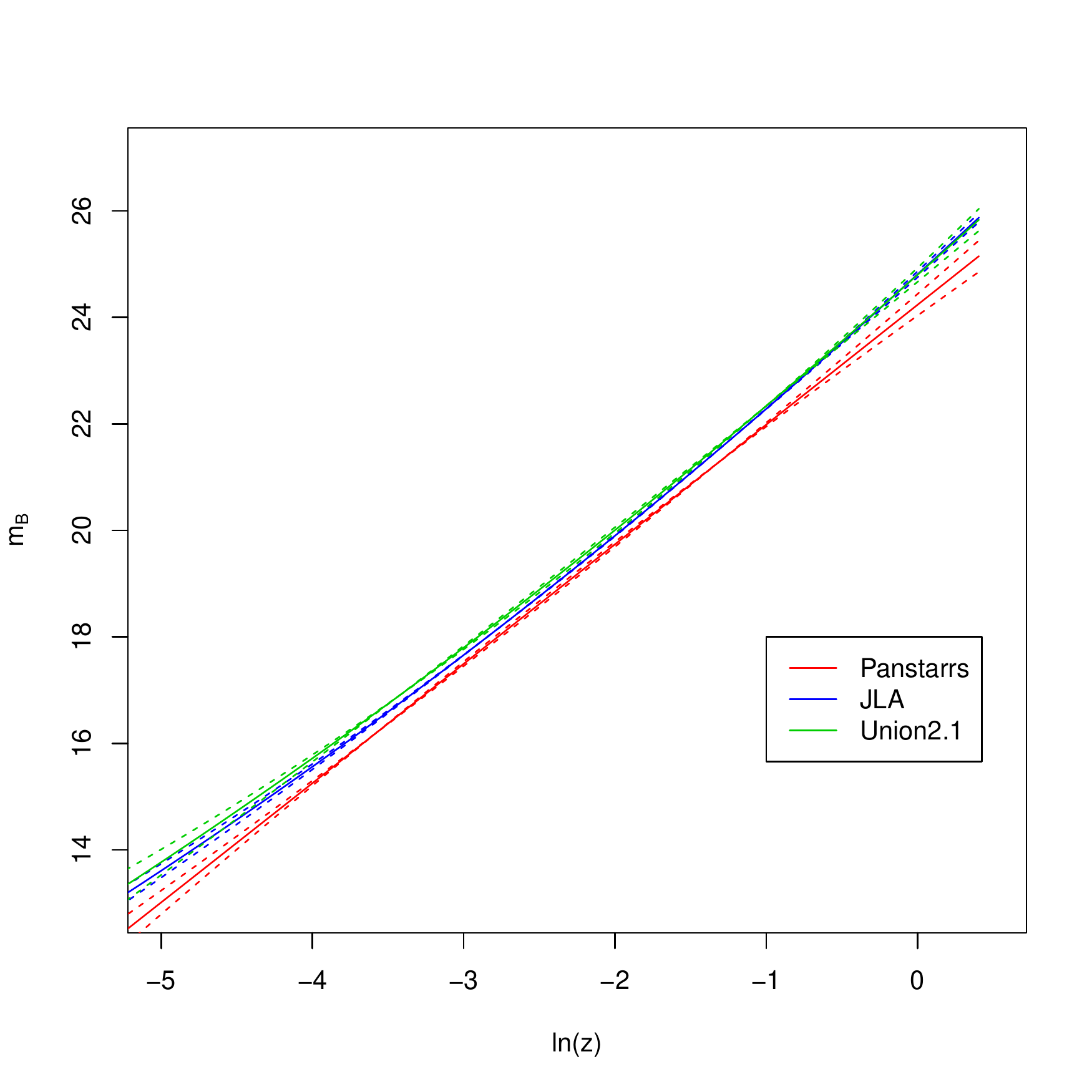} \\
\mbox{(a)} & \mbox{(b)} \\
\end{array}$
\caption{\footnotesize
Nonlinear regression plots of the datasets using the largest data sets
within Panstarrs, JLA and Union2.1. (a) With and (b) without the data
points.  }
\label{fig:LargeSet}
\end{figure}

We can now assess the influence of the high-$z$ data by considering
the common low-$z$ data points for every data set, since we
demonstrated in Section 3.2 that the fitted models are consistent for
the common low-$z$ data points.  In this case, as we see in
fig~\ref{fig:nonlin_comm}, the values for the parameters do not change
much, but the errors are tightened as this common subset of data is
more consistent amongst all data sets. However, the slopes for JLA and
Union2.1 are now a bit further away from each other, showing that
consistency in low-$z$ data does not enhance the consistency of the
global nonlinear regression, probably due to a smaller number of
supernovae data points below $z=0.1$ as compared to higher redshift
data.

The Panstarrs data only goes up to a redshift of $z<0.634$. If we now
remove the high-$z$ data above the Panstarrs upper limit by
introducing a cut-off at $z=0.64$ for all datasets, this will give us
an idea if the results from Panstarrs are different from those of JLA
and Union2.1 mainly due to it having a smaller redshift range or
not. From figure~\ref{fig:nonlin_cut}, we see that apparently the
difference between Panstarrs and the other two remains. However, from
table~\ref{tab:nonlin}, we see that the linear parameter $b$ is now
very consistent between Panstarrs and Union2.1, and the quadratic term
$c$ also is closer, while JLA remains different from the other two
datasets.  It is tempting to assume that, since it is mainly the
gradient $b$ and possibly the quadratic term $c$ which represent the
changes in cosmology (rather than the intercept $a$), the Union2.1 and
Panstarrs data, when cut-off at $z<0.64$, might give cosmological
results consistent with each other. However, we should remember that
in a quadratic expression, the first and second terms do not simply
represent the intercept and gradient of the equivalent linear
regression, and the three parameters $a, b, c$ are degenerate with
each other. So two datasets can be said to behave equivalently only
when all three parameters $a, b, c$ have similar values.

It is singular to note from Table~\ref{tab:nonlin} that when
considering only the high-$z$ datasets, JLA and Union2.1 appear to be
more consistent, while strong differences arise between them when
introducing the cutoff at $z<0.64$. This further supports the
conclusion that the choice of data points, especially within the
Union2.1 dataset, could be the main driver for determining the
constraints for cosmological parameter fitting. We also note that the
Panstarrs dataset probably remains to this date too small to draw
strong conclusions from its cosmological constraints.

We also look at just the high redshift data at $0.1<z<0.64$ for three
datasets in fig~\ref{fig:highz}. These regressions are extremely
different from each other, and without the low redshift data acting as
an anchor, the JLA data in particular shows a highly non-linear
behaviour. The Panstarrs data remains the one with the lowest
deviance, but the values of the parameters for all three cases are
very different, and the interesting consistency between Union2.1 and
Panstarrs when using common low-$z$ and high-$z (< 0.64)$ data is now
gone. This illustrates the importance of the low redshift data in
normalizing the entire dataset.

Finally, the regression based on the largest subsets of each dataset
is also considered. This allows us to see if the largest subsets are
consistent or not with the general behaviour of each dataset. For
Panstarrs, we select CfA3 and Panstarrs-1 samples for a total of 197
data points; for JLA, we select CfA3, SDSS and SNLS for a total of 668
data points, and for Union2.1, we use subsets number 6, 8, 14, and 15
for a total of 369 data points. Using these subsets that correspond to
the majority of the data points in each dataset, we see in
figure~\ref{fig:LargeSet} that the fit parameters are practically
identical to the ones that were obtained using the whole dataset for
JLA, and reasonably close to the whole dataset results for Panstarrs
and Union2.1.

\begin{figure}
\centering
$\begin{array}{cc}
\includegraphics[width=0.45\linewidth]{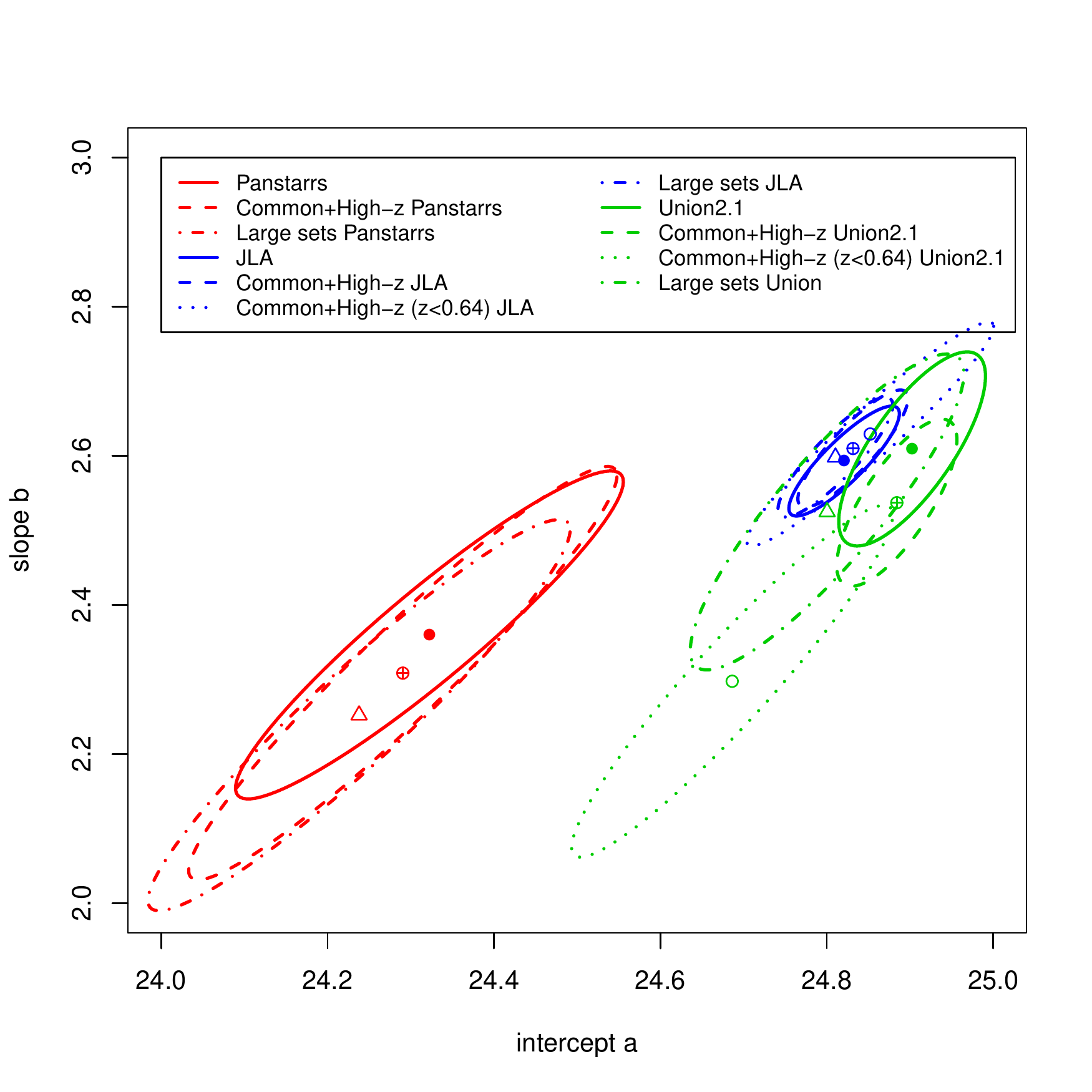} &
\includegraphics[width=0.45\linewidth]{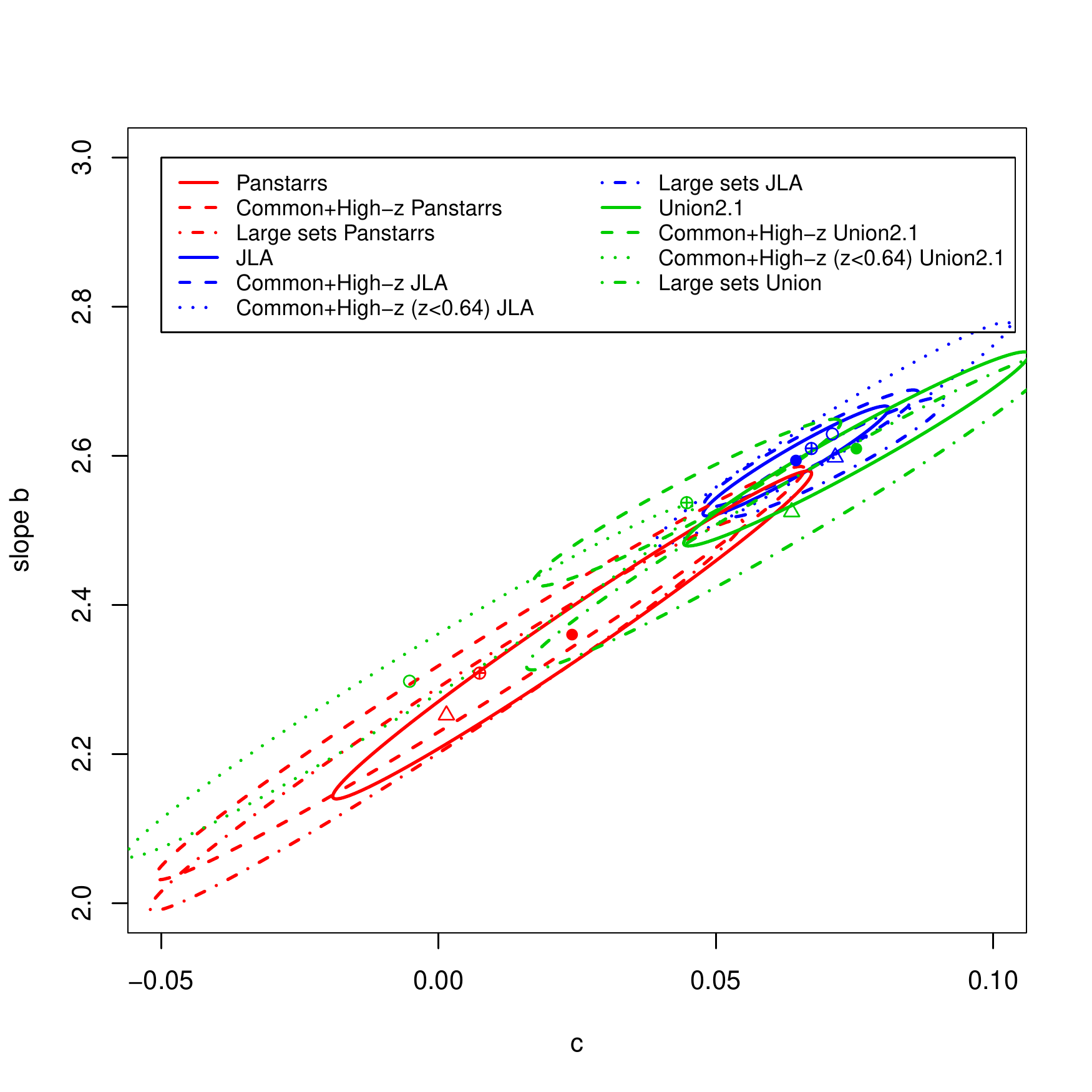} \\
\mbox{(a)} & \mbox{(b)} \\
\end{array}$ 
\caption{\footnotesize
$2\sigma$ confidence ellipses plotted for the non-linear regression
parameters in (a) $b$ v/s $a$ space and (b) $b$ v/s $c$ space, for the
entire dataset (filled circle for best-fit), common low-$z$ + high-$z$
data (crossed circle), common low-$z$ + high-$z$ ($<0.64$) (empty
circle), and for the largest subsets (triangle) for the three
datasets. }
\label{fig:nonlin_cl}
\end{figure}

Figure~\ref{fig:nonlin_cl} summarizes the regression results by
plotting the $2\sigma$ confidence ellipses of the quadratic regression
parameters in the $b$ v/s $a$ space, and in the $b$ v/s $c$ space for
the entire datasets, for the sets with common low-$z$ data, for the
sets with common low-$z$ and high-$z$ cut-off, and for the large
subsets. It can be seen from this plot that Panstarrs is typically
more than $2\sigma$ away from JLA and Union2.1 whatever the data
subset considered. However, JLA and Union2.1 are consistent at
$2\sigma$ when considering all the data. One may note that when
considering the entire redshift range, the influence of the common and
extraneous data points at low-$z$ plays a minimal role.  However,
choosing an upper limit cut-off of $z<0.64$ (the highest redshift for
Panstarrs) changes completely the confidence levels for
Union2.1. Although the intercept remains different from Panstarrs,
both the slope and the quadratic term come very close to the Panstarrs
value. This points out once more the strong sensitivity of this
dataset to the choice of the data points included. This also indicates
the possibility that the cosmological constraints from Union2.1 and
Panstarrs may be slightly more consistent in the limit $z<0.64$, while
the JLA should retain its discrepancy with Panstarrs irrespective of
the subsets considered.

\begin{figure}
\centering
\includegraphics[width=0.9\linewidth]{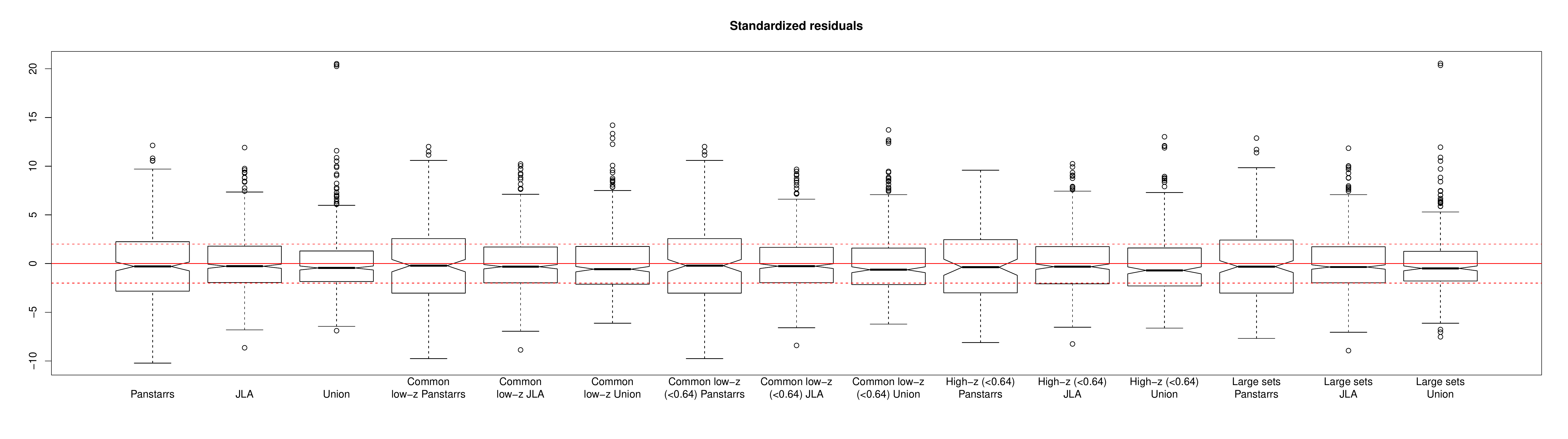} 
\caption{\footnotesize
Boxplots of the standardized residual distributions for each nonlinear
model of the supernovae data. Horizontal lines indicate the $2\sigma$
levels. }
\label{fig:nl_box}
\end{figure}

Figure~\ref{fig:nl_box} represents the boxplot distribution of the
standardized residuals for each subset. One can see that there is a
systematic presence of potential outliers on the upper part of the
supernovae curves.  This corresponds to supernovae for which the
magnitude is overestimated with respect to their redshift, indicating
a population of supernovae that appear to have a lower emission flux
than expected. This probably does not reflect an observational bias
since brighter supernovae should be detected at the same rate.  It
could be due to a difference in the absorbing medium between the
source and the observer or other effects changing the brightness of
the supernovae considered, like different properties of their host
galaxies (see \eg \cite{sne_bias}, \cite{sne_bias2}).

\begin{figure}
\centering
\includegraphics[width=0.9\linewidth]{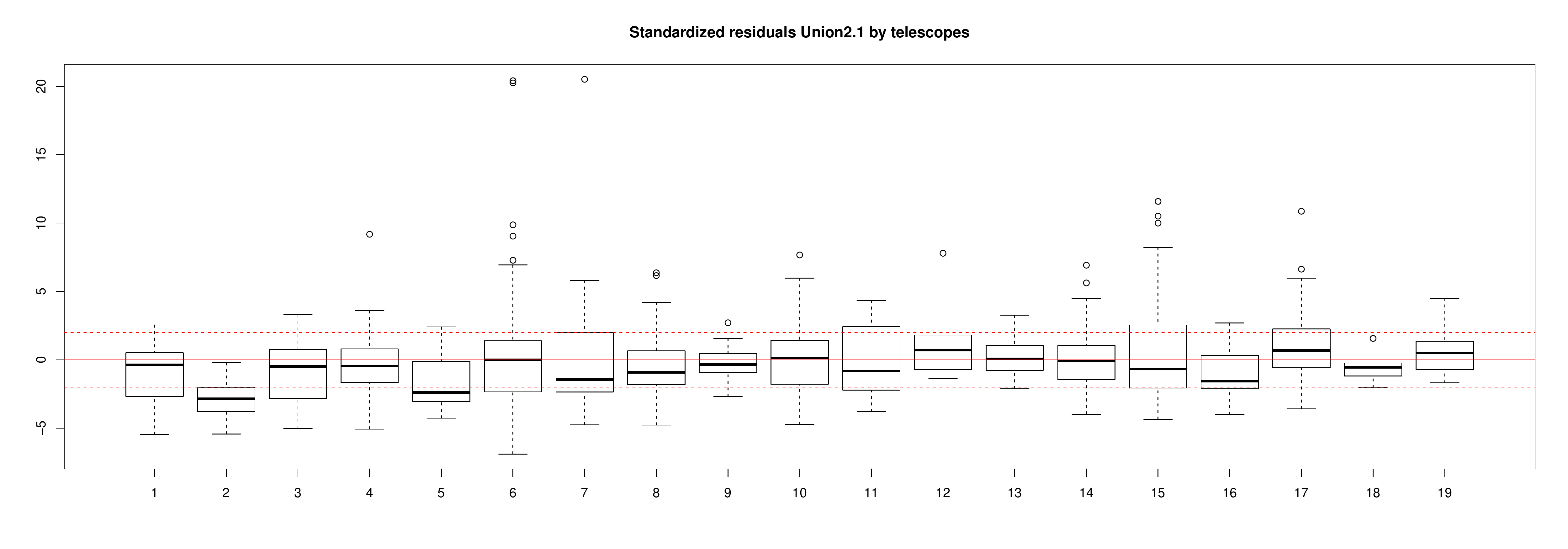} 
\caption{\footnotesize
Boxplots of the residual distributions for each subset of Union2.1
data.  Horizontal lines indicate the $2\sigma$ levels.}
\label{fig:union_box}
\end{figure}

\begin{figure}
\centering
$\begin{array}{ccc}
\includegraphics[width=0.32\linewidth]{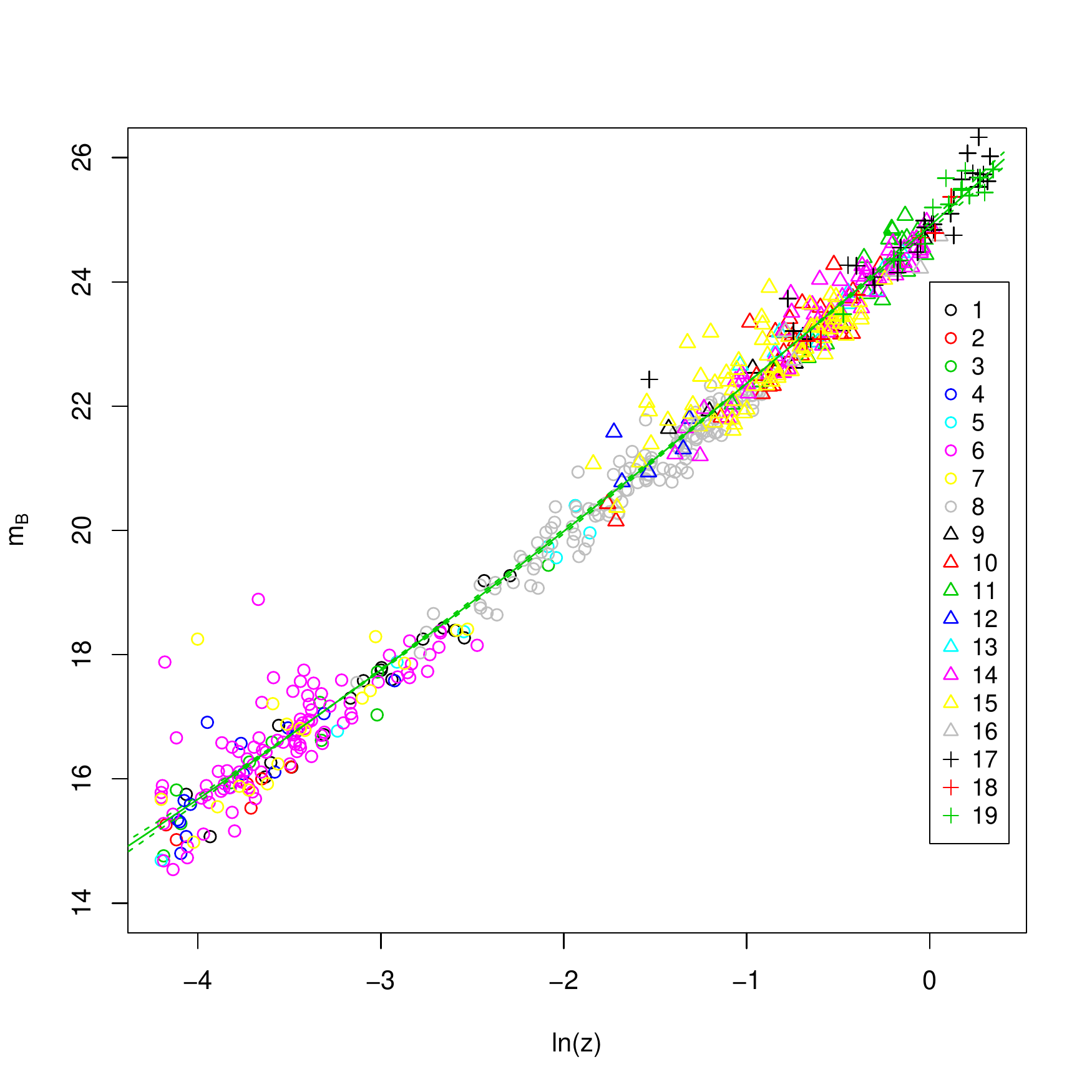} &
\includegraphics[width=0.32\linewidth]{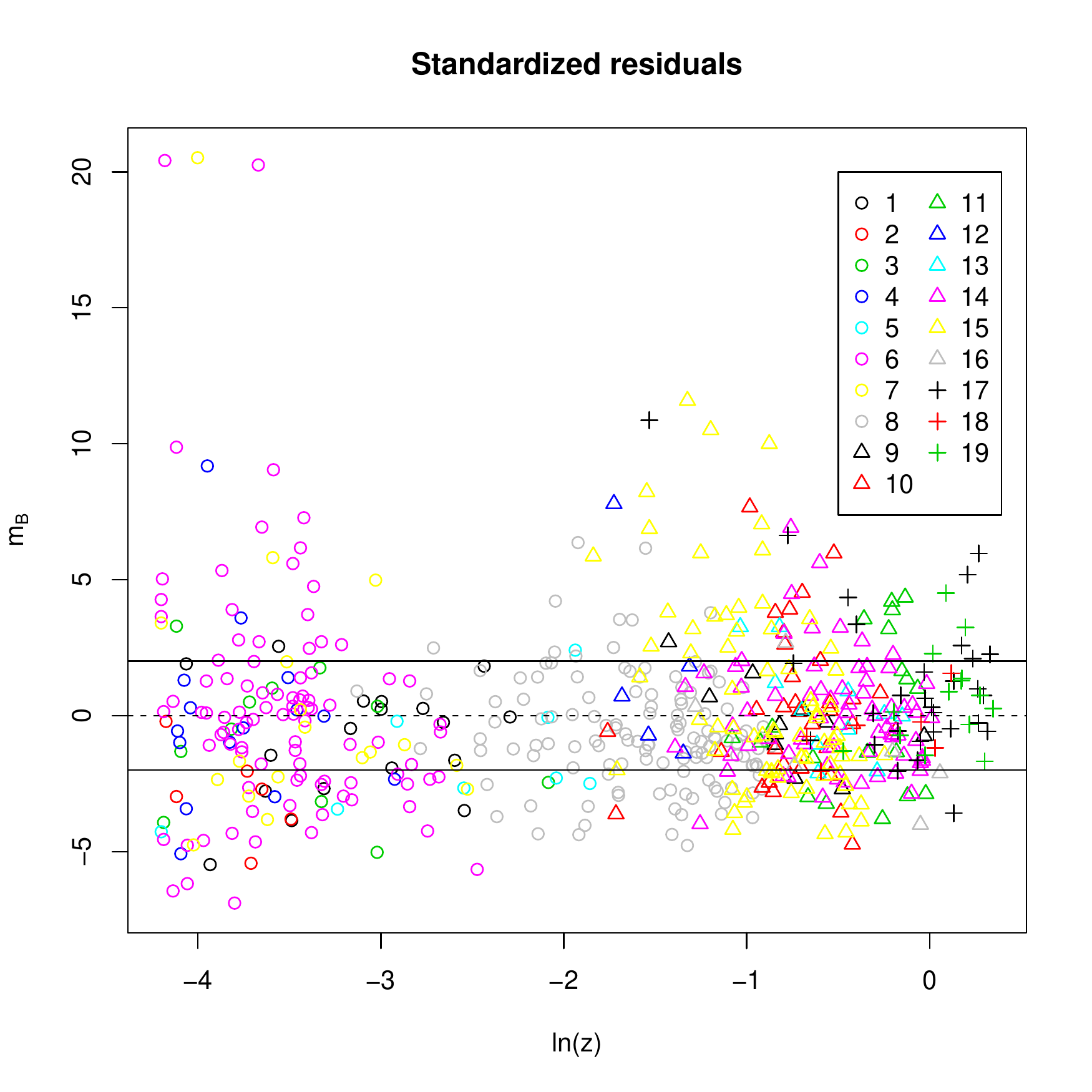} & 
\includegraphics[width=0.32\linewidth]{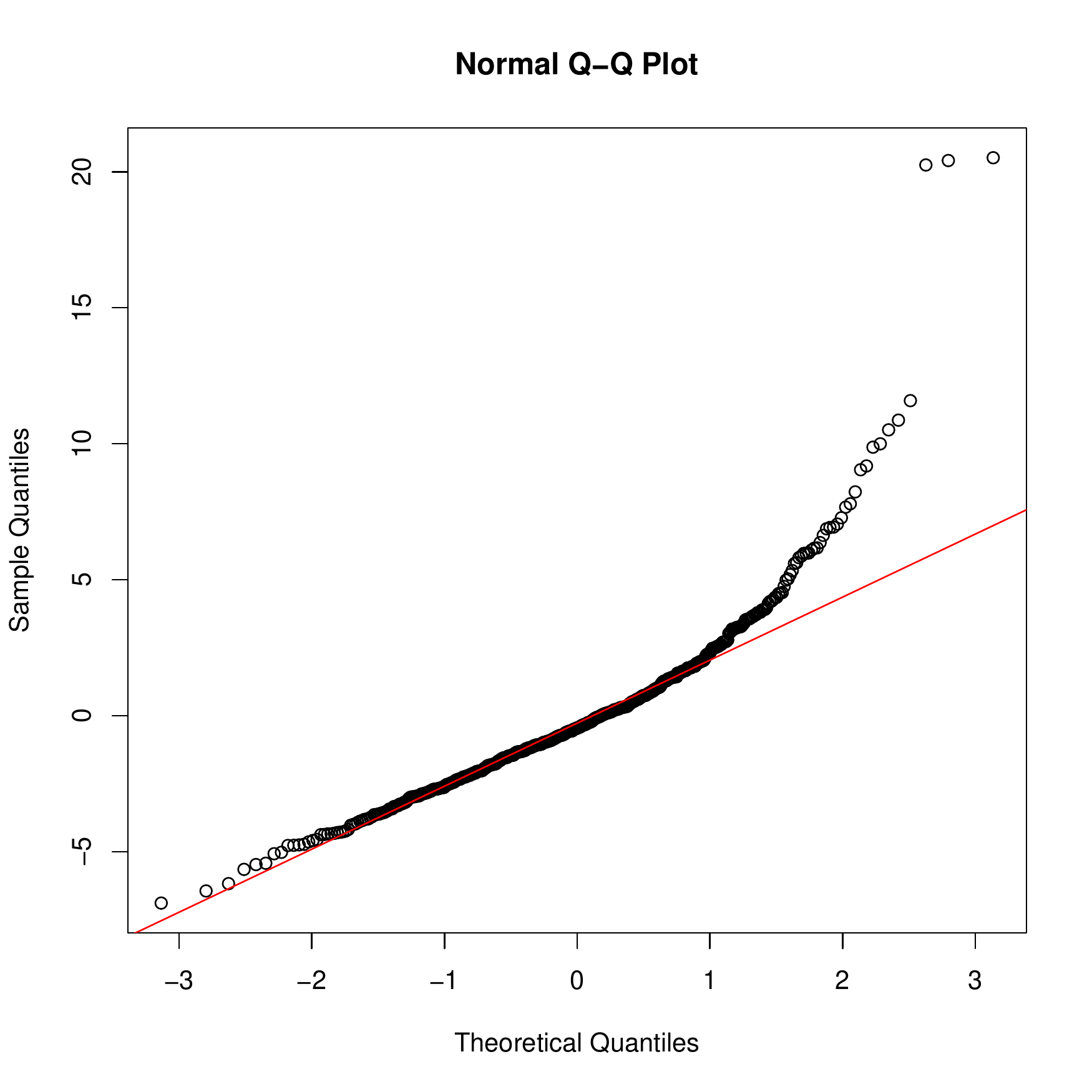} \\
\mbox{(a)} & \mbox{(b)} & \mbox{(c)} \\
\end{array}$
\caption{\footnotesize
Study of subset behaviour of Union2.1. (a) Non-linear fit, (b)
standardized residuals, and (c) associated quantile-quantile plot
comparing the residuals and normal distributions.  The curves and
lines indicate the $2\sigma$ levels.  }
\label{fig:union_res}
\end{figure}

Similar to the low redshift data, we may now attempt to check the
effect of the different subsets in the data. In the case of JLA and
Panstarrs, the high redshift data is typically taken from a few
telescopes (only one in the case of Panstarrs), but Union2.1 has
several subsets at high redshift, with various quantities of SNe in
them. In the figure~\ref{fig:union_box}, we show the residuals for the
different subsets in Union2.1. We note here that the largest high
redshift datasets (14, 15) are actually quite well-behaved being
centred around the mean with few outliers. The effect of the largest
low redshift dataset (\ie subset 6) is somewhat mitigated by better
behaviour from the high redshift data. So we do not expect the results
to be drastically different between the data subsets. The
figure~\ref{fig:union_res} shows the standardized residuals for each
subset of Union2.1. Except for a few outliers the different subsets
appear to be reasonably consistent with each other. Thus, although
this dataset is quite diverse, the different subsets are fairly
well-chosen. The analysis thus does not show a specific behaviour by
subset for the Union2.1 data.  In the figure~\ref{fig:union_res}(b),
one can appreciate the variations in residual that are visible for low
and high redshifts, presenting a visible non-normal distribution and
heteroscedasticity as noted earlier in section 3.1. This is further
illustrated by the quantile-quantile comparison between the residuals
and the normal distribution shown in~\ref{fig:union_res}(c), which
indicates the over-representation of large positive residuals. These
variations may play a role in the cosmological parameter fitting.  A
similar analysis was done using the Panstarrs and JLA data subsets,
for which just a couple of different observatories have measured the
high-$z$ supernovae. None of these analyses show an indication that a
specific set of data points particularly drives the resulting
models. This demonstrates that the multiple data subsets appear
broadly consistent with one another, even though over smaller redshift
ranges, the Union2.1 sample in particular shows some subset-dependence
(\eg subset 6 drives the data at low-$z$).

From the above analyses in linear and non-linear regimes, several
points of interest should be noted.
\begin{itemize}
\item
Firstly, as has been noted elsewhere from cosmological reconstruction,
the Union2.1 and JLA data appear statistically consistent with each
other, while the Panstarrs is $2\sigma$ away from either. However, if
the data is cut-off at $z<0.64$ which is the upper limit of Panstarrs
data, Union2.1 and Panstarrs now appear to be possibly more consistent
with each other, while JLA retains its difference. It is not clear if
this discrepancy is simply due to the fact that Panstarrs has less
data, and is at lower redshifts than the others.
\item 
In the low redshift data, the data subset 6 of Union2.1 appears to be
rather discrepant from the other subsets of Union2.1 and with JLA and
Panstarrs data, as well as with the expected theoretical
behaviour. This behaviour may or may not result in drastic differences
in cosmology since low redshift data acts mainly as an anchor, but it
is worth noting especially since the other data subsets appear fairly
consistent with each other.
\item 
The Union2.1 dataset appears to be especially sensitive to which data
subsets are chosen. We find the most variance within this dataset and
the statistical results appear to change significantly depending on
the subset of data used. The other two datasets do not show such
strong proclivity for change. This may be due to the fact that the
Union2.1 dataset is created as a mixture of several different datasets
observed by various different telescopes at different times, thus it
is the most heterogeneous of the three datasets. This may or may not
have an effect on the cosmological results.
\item
We also note, that, although the Panstarrs dataset has the least
number of data points and goes up to only moderate redshifts, it is
the most consistent of the three datasets and least prone to change
upon subset choice. It also shows the least heteroscedasticity, and
has the least RSD of the three. In the future, a more complete
Panstarrs dataset, with more data points at high redshift, may
actually end up being the most stable.
\item
Finally, we note that, barring Panstarrs, the other two datasets show
some strong heteroscedasticity. This variation in residual variance as
a function of redshift may lead to a larger influence of the lower
error points over the best fitting parameters and their confidence
levels, and may result in a bias in the cosmological results.  Of
course, the analysis will depend both on the errors and the distance
from the fits, and typically, given that the spread in errors are
fairly small, the distance from the fit would drive the results. The
errors may still be capable of changing results, especially when the
degrees of freedom of the system increase, making the fit tighter.
\end{itemize}

\section{Cosmology}\label{sec:cosmo}

We are now in a position to verify our findings from the statistical
blind analysis by reconstructing cosmological models from these
datasets.

\subsection{Methodology}

The data is in the form: 
\beq 
\mu(z) = 5 \ {\rm log}_{10} \left(\frac{c(1+z)}{H_0} \int_0^z \frac{dz_1}{h(z_1)}\right) \,\,, 
\eeq 
with $h(z)$
given by 
\beq 
h^2(z) = \omr (1+z)^4 + \omt (1+z)^3 + (1-\omr-\omt)(1+z)^{3(1+w)} \,\,, 
\eeq
for a flat universe with a constant equation of state dark energy $w$
and luminosity distance measured in $10$ pc. The matter and radiation
densities at present are denoted by $\omt$ and $\omr$ respectively,
and $H_0$ is the Hubble constant. It should be noted that at the
redshifts considered, the radiation density $\omr$ is negligible, and
also that the only effect of the parameter $H_0$ is as an additive
constant. Thus marginalizing over $H_0$ does not affect the SNe
results and indeed $H_0$ can be absorbed into the additive nuisance
parameter $M_B$. The flatness of the Universe can be assumed from CMB,
and is a required assumption to reduce the degeneracies between the
different densities. The radiation density is also calculated from
CMB, \eg from \cite{planck}, even though it has little effect. As
mentioned earlier in sec~\ref{sec:data}, the constant equation of
state is not a very realistic model for dark energy, but the supernova
data alone does not have sufficient accuracy to constrain the rate of
change of equation of state of dark energy, and adding other datasets
to the analysis could potentially add the biases of those datasets to
the reconstruction. As the main objective of this paper is not to
constrain cosmology, but rather to examine the consistencies and
inconsistencies of the SNe data, we therefore use the constant
equation of state model and assume a flat universe for all our
cosmological reconstruction.

We use a standard $\chi^2$ minimization analysis, and a Markov Chain Monte Carlo analysis for the confidence levels, with $\chi^2$ defined as
\beq
\chi^2 = \sum_{i=1}^{N_{data}} \left(\frac{\mu_i(\omt, w, \alpha, \beta, M_B)-\mu_{fit}(z_i)}{\sigma_i(\alpha, \beta, \sigma_{int})}\right)^2 \,\,,
\eeq
where the parameters of interest are the matter density $\omt$ and the
equation of state of dark energy $w$. The nuisance parameters $\alpha,
\beta, M_B$ (which includes $H_0$) are marginalized over and we
present our results in the $\omt-w$ parameter space.  To simplify
matters further, we at first utilize a reduced number of parameters
for this exercise where we fix $\omt=0.3$ (commensurate with the
Planck values \cite{planck}) and only constrain $w$. This gets rid of
the potential degeneracy between $\omt$ and $w$. Later, we also obtain
the results in the $\omt$ v/s $w$ parameter space.

\subsection{Results}

Several subsets of the data may be considered, as evinced by the
analysis of the previous section. After considering the full dataset,
one may consider only the common dataset for low redshift in
conjunction with all the high redshift data. One may also truncate the
high redshift data at $z<0.64$, which is the highest redshift for the
Panstarrs sample. For the JLA sample, considering only the common
low-$z$ SNe, this cut-off results in a sample of 546 SNe, while for
Union2.1 it is a sample of 386 SNe. The logic behind this step is to
assess whether the sizeable difference between Panstarrs and the other
two datasets is simply due to the former reaching only medium redshift
ranges. We may form a tighter data sample by throwing out all SNe that
are outliers in the residual distributions for the SNe, as seen in
fig~\ref{fig:nl_box}. We select these potential outliers by
determining which residuals are $2\sigma$ away from the average
residual, using the standard deviation of the residuals distribution
as described in Section~\ref{sec:lin}.  For the Panstarrs, JLA and
Union2.1 data, this means removing 16, 36 and 22 SNe respectively from
the full datasets.  We may also, rather arbitrarily, look at how the
cosmological results change if only the largest low and high redshift
subsamples are used.  For Panstarrs we use the low redshift CfA3 data
subset and the Panstarrs-1 sample at higher redshifts, giving a sample
of 197 SNe. For JLA this would involve the SDSS at low to medium, and
SNLS at medium to high redshifts, with the CfA3 sample at low
redshifts for anchoring. This results in a dataset with 668 SNe. In
the case of Union2.1, this subset contains the sample 6 at low
redshift, sample 8 at low to medium redshift, and samples 14, 15 at
medium to high redshifts, resulting in a total of 369 SNe. This step
would let us know how important the different subsamples are, if the
results are being driven only by the largest subsets, or if the
smaller subsets with a few SNe each also have a role to play in the
reconstruction. Also, in the case of the Union2.1 data, particularly,
one may redo the cosmological analysis without the subset 6 (from the
CfA Year 3 sample), which, though the largest low redshift dataset,
also shows the most discrepant behaviour in comparison to both other
Union2.1 subsets, and the JLA and Panstarrs samples. We attempt to
study all these various possibilities and compare the cosmological
constraints obtained in each case.

\begin{table*}
\caption{\footnotesize 
Best-fit and $1\sigma$ confidence levels on the cosmological parameter
$w$ (with $\omt=0.3$) for various subsets of data. Clow refers to the
common low redshift data subset, H to the high redshift dataset.}
\label{tab:cosmo_w}      
\centering          
\begin{tabular}{|l|ccc|}   
\hline       
&&$w$&\\
\hline                    
{\footnotesize Subset}&Panstarrs&JLA&Union2.1\\  
\hline                    
&&&\\
{\footnotesize All data}&$-1.334^{+0.076}_{-0.117} $&$-1.090^{+0.056}_{-0.024}$&$-1.147^{+0.045}_{-0.045}$ \\ 
&&&\\
{\footnotesize Clow + H}&$-1.306^{+0.095}_{-0.116}$&$-1.078^{+0.060}_{-0.059}$&$-1.118^{+0.051}_{-0.047}$\\  
&&&\\
{\footnotesize Clow + H $(<0.64)$}&$-1.306^{+0.095}_{-0.116}$&$-1.044^{+0.070}_{-0.071}$&$-1.099^{+0.055}_{-0.055}$\\  
&&&\\
{\footnotesize W/O outliers}&$-1.303^{+0.096}_{-0.113}$&$-1.032^{+0.058}_{-0.050}$&$-1.128^{+0.043}_{-0.044}$\\  
&&&\\
{\footnotesize Large subsets}&$-1.259^{+0.110}_{-0.113}$&$-1.026^{+0.059}_{-0.059}$&$-1.027^{+0.054}_{-0.053}$\\
&&&\\
{\footnotesize W/O subset 6}&--&--&$-1.225^{+0.053}_{-0.053}$\\  
\hline          
\end{tabular}
\end{table*}

\begin{figure}
\centering
$\begin{array}{@{\hspace{-0.5cm}}c@{\hspace{-0.25cm}}c@{\hspace{-0.25cm}}c}
\includegraphics[width=0.37\linewidth]{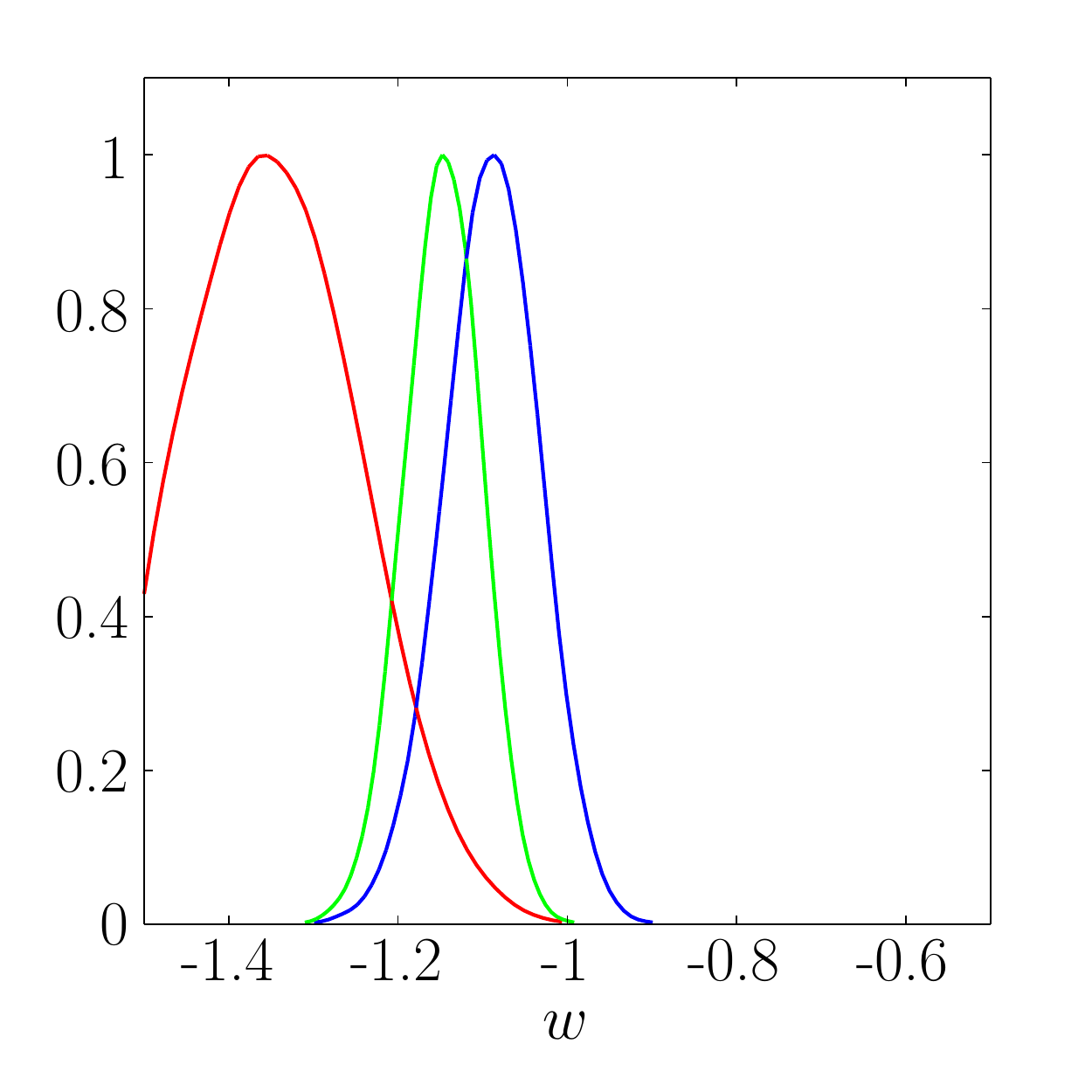} &
\includegraphics[width=0.37\linewidth]{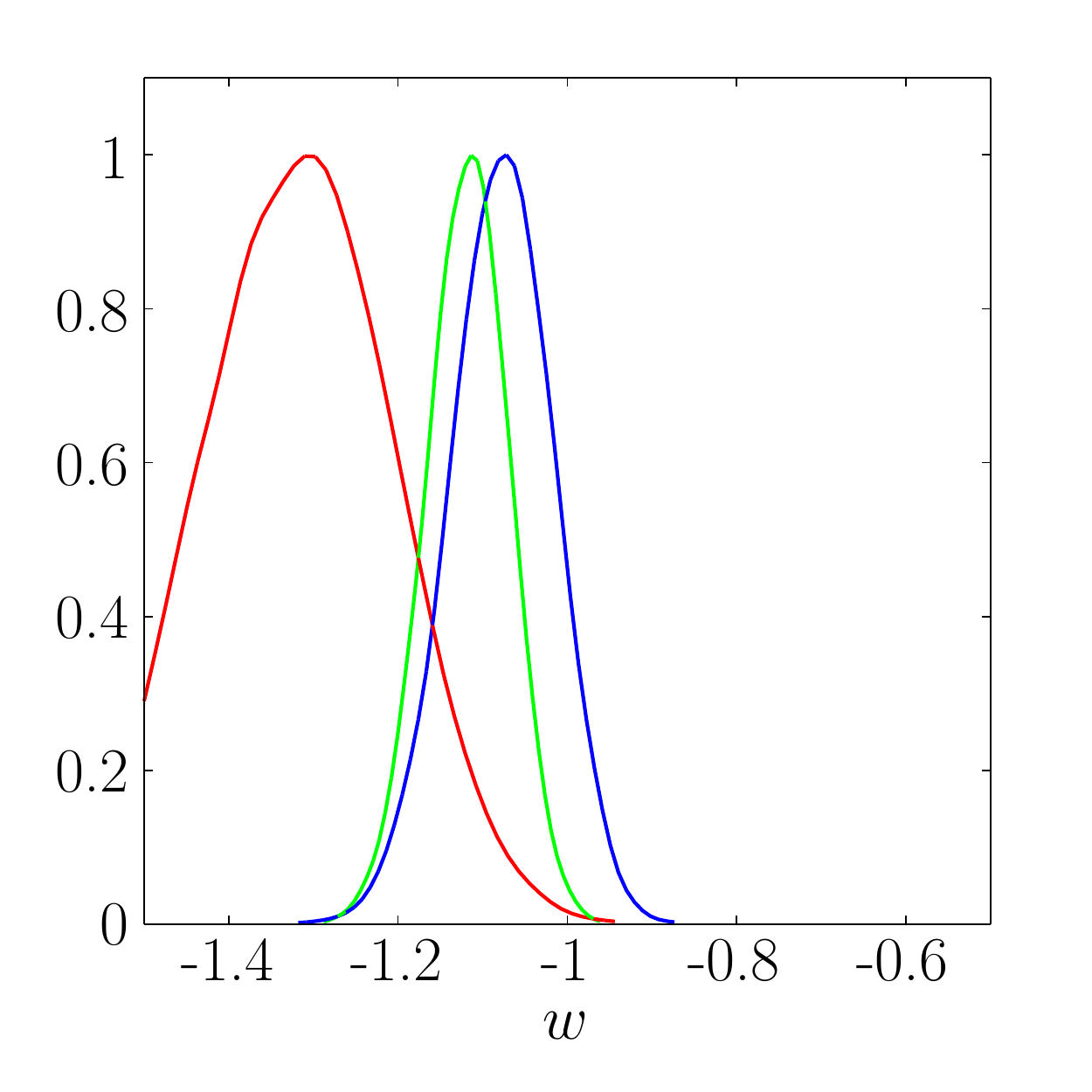} &
\includegraphics[width=0.37\linewidth]{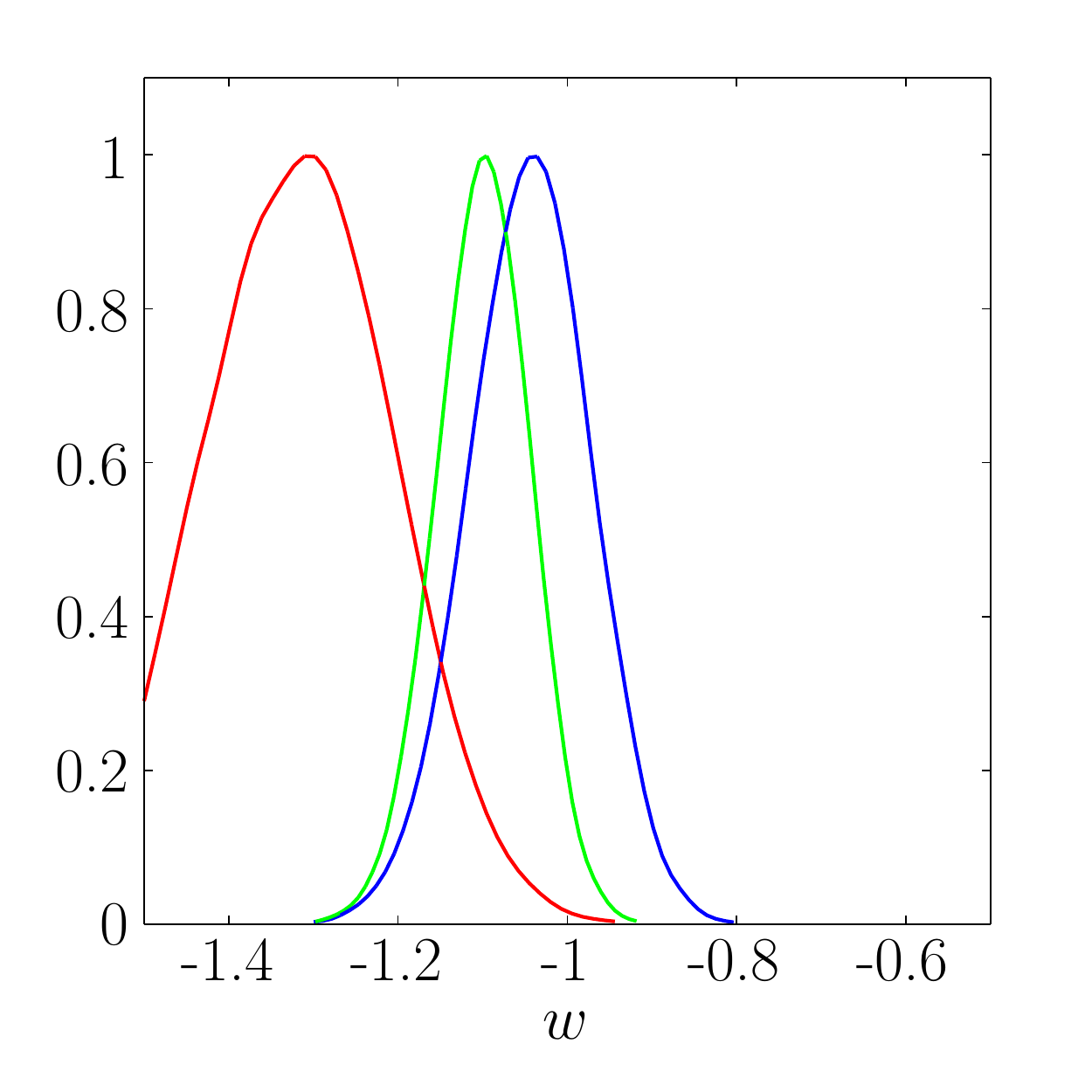} \\
\mbox{(a)} & \mbox{(b)} & \mbox{(c)} \\
\end{array}$
$\begin{array}{@{\hspace{-0.5cm}}c@{\hspace{-0.25cm}}c@{\hspace{-0.25cm}}c}
\includegraphics[width=0.37\linewidth]{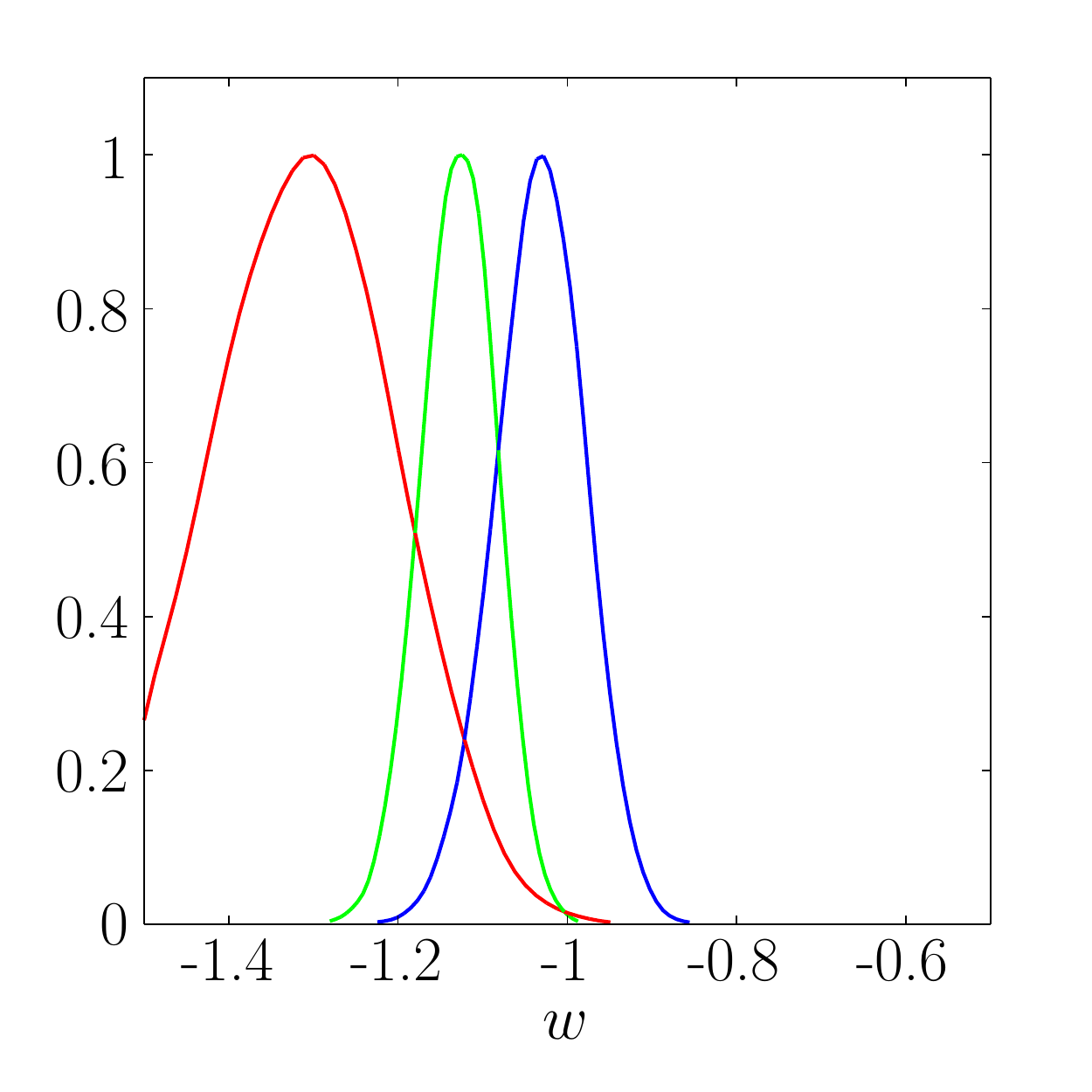} &
\includegraphics[width=0.37\linewidth]{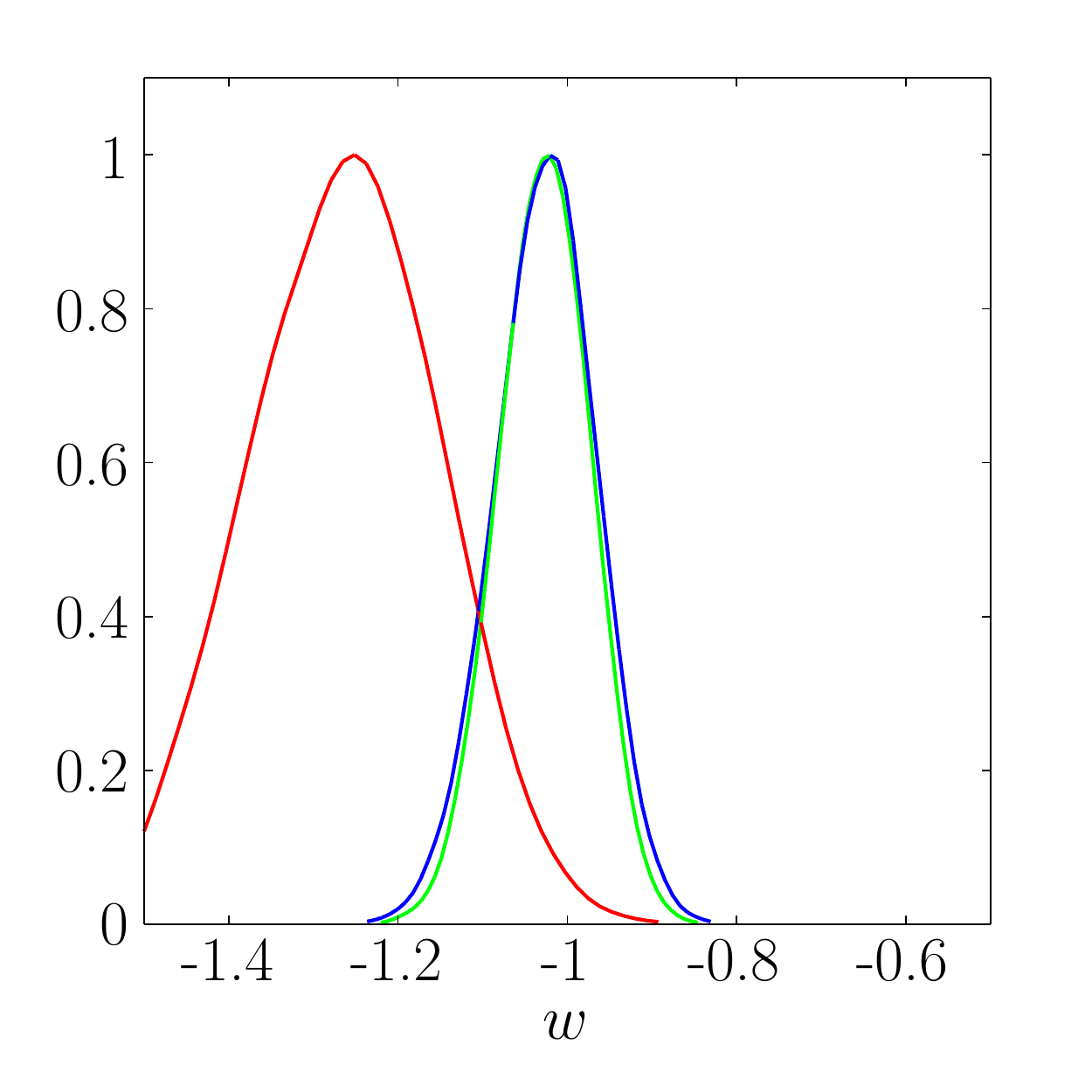} &
\includegraphics[width=0.37\linewidth]{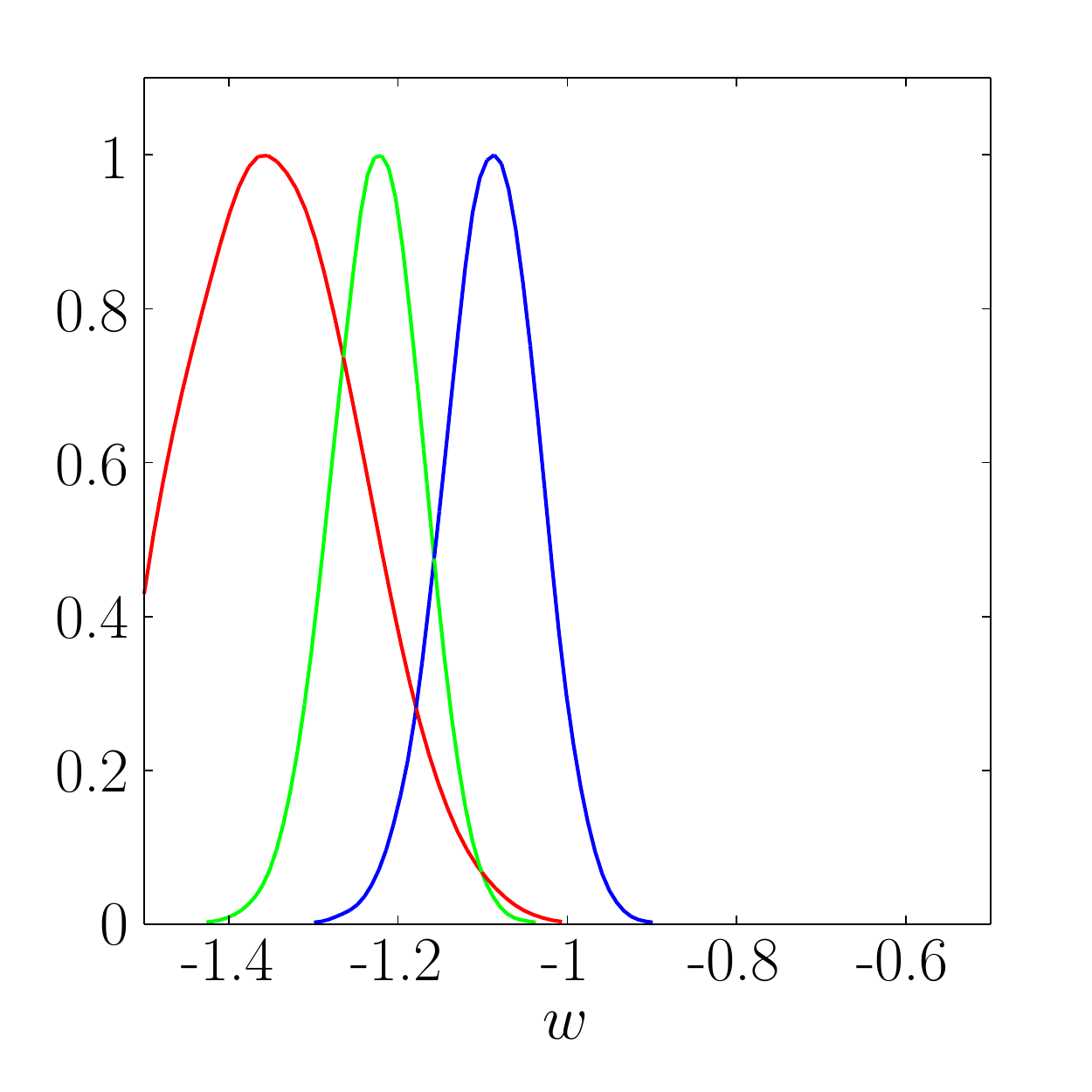} \\
\mbox{(d)} & \mbox{(e)} & \mbox{(f)} \\
\end{array}$
\caption{\footnotesize
Probability distribution of the parameter $w$ (with $\omt=0.3$ fixed)
for Panstarrs (red), JLA (blue) and Union2.1 (green) supernova Type Ia
data. Panel (a) is for the entire datasets, panel (b) is for common
low-$z$ data + high-$z$ data, panel (c) is for common low-$z$ data +
high-$z (<0.64)$ data (with Panstarrs remaining the same as in (b)),
panel (d) shows results for data without the outliers of regression,
panel (e) considers the largest low and high-$z$ subsets for each
sample, panel (f) shows full datasets for JLA and Panstarrs, and
Union2.1 without the data subset 6 which showed anomalous behaviour.}
\label{fig:w_all}
\end{figure}

Table~\ref{tab:cosmo_w} shows the best-fit and $1\sigma$ values of $w$
for the full dataset and all the subsamples outlined above. The
figure~\ref{fig:w_all} shows the likelihood in $w$ for the three
datasets in the different scenarios. We find that, the Panstarrs and
JLA samples are quite insensitive to changes in the dataset. For
various subsamples, the best-fit remains very similar for Panstarrs,
with a slightly larger change for the largest subsets case, which may
well be the effect of too few (197) data points, while the errors
remain nearly the same. This shows that the Panstarrs data, as
expected from the statistical analysis, is very consistent within
itself, even though the results from it are about $2\sigma$ different
from the other two datasets. The JLA data is also very consistent,
with the best-fit values varying only slightly, though the errors show
a slightly bigger variation. Again this result is commensurate with
what we found in the statistical analysis. As expected from the
previous section, the Union2.1 data shows the maximum variation in the
best-fit between the data subsets, about twice as much as the
others. This may be due to the fact that the Union2.1 dataset is
comprised of many different datasets, and therefore it is not as
homogeneous as the other two, in which efforts have been made to keep
at least the high redshift data more homogeneous. This also points to
the fact that if we had a single, high quality source for low redshift
data, this could lead to more stable cosmological constraints, which
would be even less dependent on data selection. We also note that all
three datasets show similar trends for the different subsets. For the
full dataset, Panstarrs has the lowest $w$, and JLA the highest (and
closest to $\Lambda$CDM) while Union2.1 has an intermediate value. For
the common low-$z$ subsample, all three best-fit values for $w$
increase. For the case with the cut-off at $z<0.64$, both the Union2.1
and JLA best-fit increases, thus this cut-off does not make the three
datasets any more consistent. For the case without outliers, the
$w$-values increase as well, but by a lesser amount than in the common
case for Union2.1, and by a greater amount (though by very little) for
JLA and Panstarrs. For the case with the large subsets, the $w$ values
increase further, and the Union2.1 result especially increases by a
larger amount to bring it very close to the JLA result. The Union2.1
dataset gives a lower $w$ when the uncharacteristic subset 6 is taken
out, which makes it slightly more consistent with the Panstarrs
dataset, but still noticeably separated from it in $w$-space. Thus,
while the three datasets are inconsistent with each other, they are
quite consistent in their inconsistency, and changing the selection of
SNe using various criterion does not make them match each other.

\begin{table*}
\caption{\footnotesize
Best-fit and $1\sigma$ confidence levels on the cosmological
parameters $\omt, w$ for various subsets of data. Clow refers to the
common low redshift data subset, H to the high redshift dataset.}
\label{tab:cosmo}      
\centering          
\begin{tabular}{|l|ccc|ccc|}   
\hline       
&&$\omt$&&&$w$&\\
\hline                    
{\footnotesize Subset}&Panstarrs&JLA&Union2.1&Panstarrs&JLA&Union2.1\\  
\hline                    
&&&&&&\\
{\footnotesize All data}&$0.164^{+0.055}_{-0.164}$&$0.250^{+0.104}_{-0.056}$&$0.291^{+0.063}_{-0.042}$&$-1.064^{+0.263}_{-0.134}$&$-0.990^{+0.228}_{-0.182}$&$-1.097^{+0.166}_{-0.147}$ \\ 
&&&&&&\\
{\footnotesize Clow + H}&$0.169^{+0.057}_{-0.169}$&$0.216^{+0.116}_{-0.070}$&$0.280^{+0.068}_{-0.043}$&$-1.047^{+0.271}_{-0.138}$&$-0.908^{+0.246}_{-0.148}$&$-1.080^{+0.178}_{-0.155}$\\  
&&&&&&\\
{\footnotesize Clow H $(<0.64)$}&$0.169^{+0.057}_{-0.169}$&$0.302^{+0.132}_{-0.052}$&$0.331^{+0.088}_{-0.030}$&$-1.047^{+0.271}_{-0.138}$&$-1.096^{+0.177}_{-0.355}$&$-1.211^{+0.098}_{-0.289}$\\  
&&&&&&\\
{\footnotesize W/O outliers}&$0.170^{+0.067}_{-0.159}$&$0.235^{+0.113}_{-0.066}$&$0.305^{+0.057}_{-0.035}$&$-1.044^{+0.267}_{-0.133}$&$-0.914^{+0.232}_{-0.160}$&$-1.159^{+0.154}_{-0.159}$\\  
&&&&&&\\
{\footnotesize Large subsets}&$0.197^{+0.132}_{-0.115}$&$0.202^{+0.124}_{-0.090}$&$0.259^{+0.133}_{-0.071}$&$-1.055^{+0.294}_{-0.163}$&$-0.849^{+0.256}_{-0.111}$&$-0.977^{+0.270}_{-0.186}$\\
&&&&&&\\
{\footnotesize W/O subset 6}&--&--&$0.332^{+0.033}_{-0.018}$&--&--&$-1.347^{+0.040}_{-0.153}$\\  
\hline          
\end{tabular}
\end{table*}

\begin{figure}
\centering
$\begin{array}{@{\hspace{-0.5cm}}c@{\hspace{-0.25cm}}c@{\hspace{-0.25cm}}c}
\includegraphics[width=0.37\linewidth]{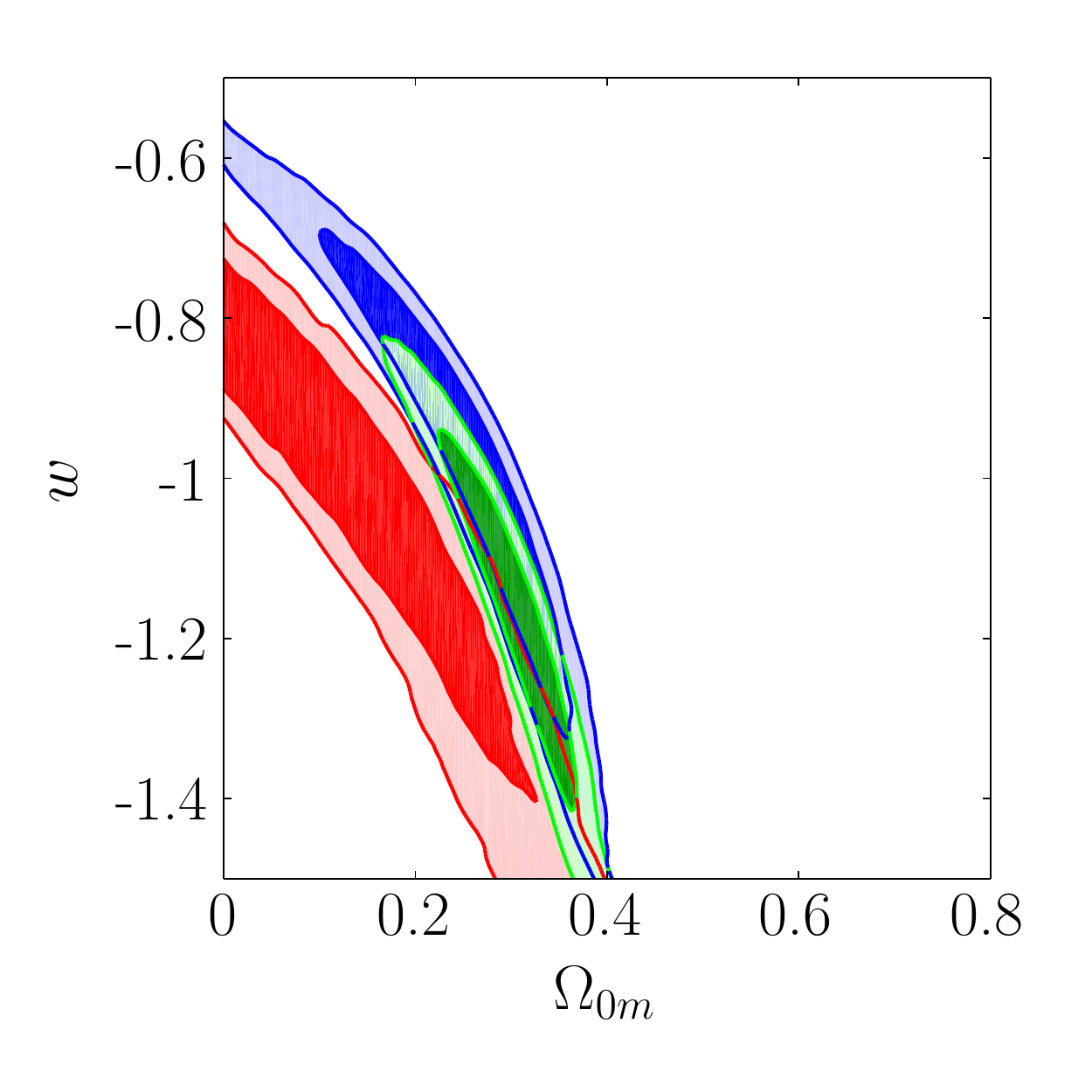} &
\includegraphics[width=0.37\linewidth]{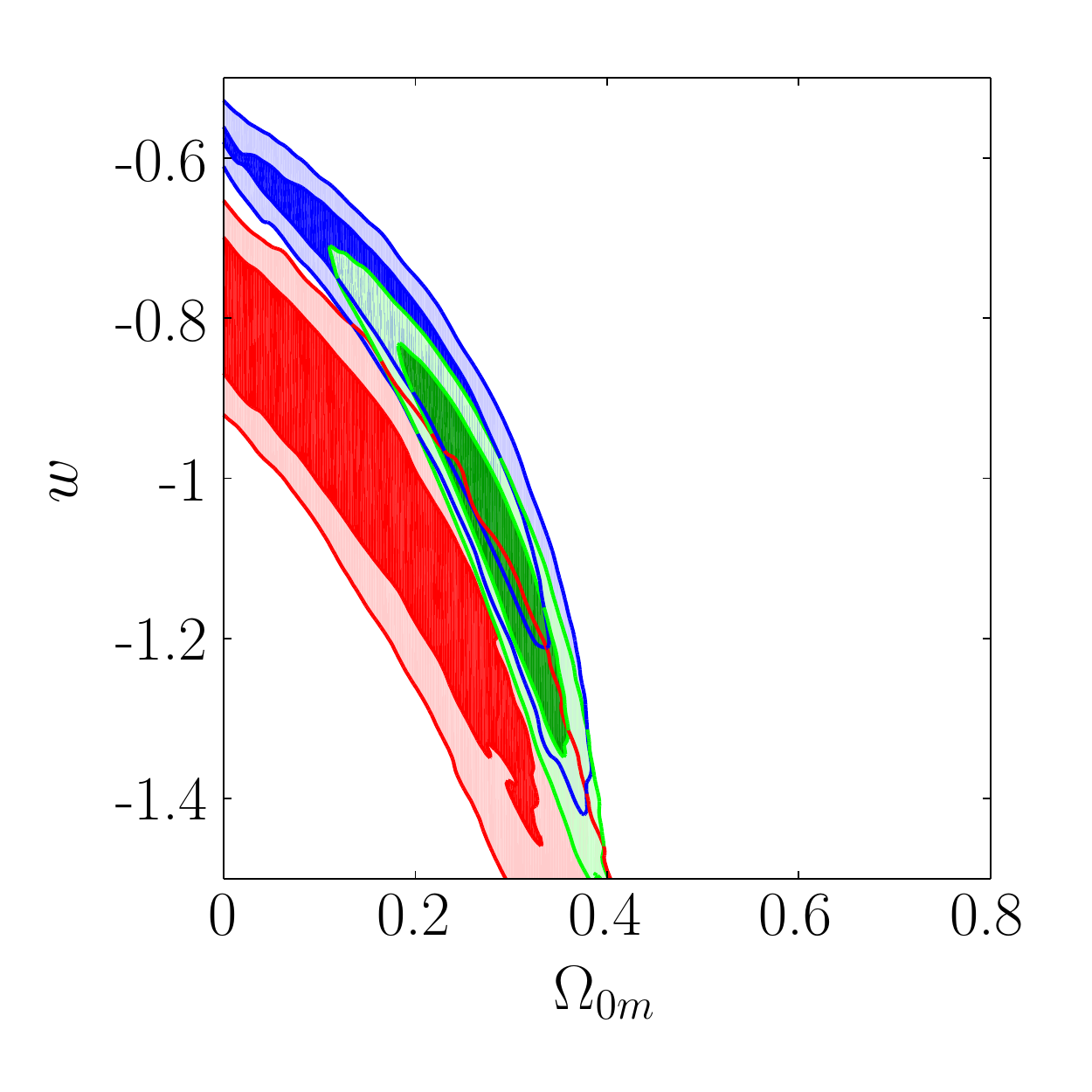} &
\includegraphics[width=0.37\linewidth]{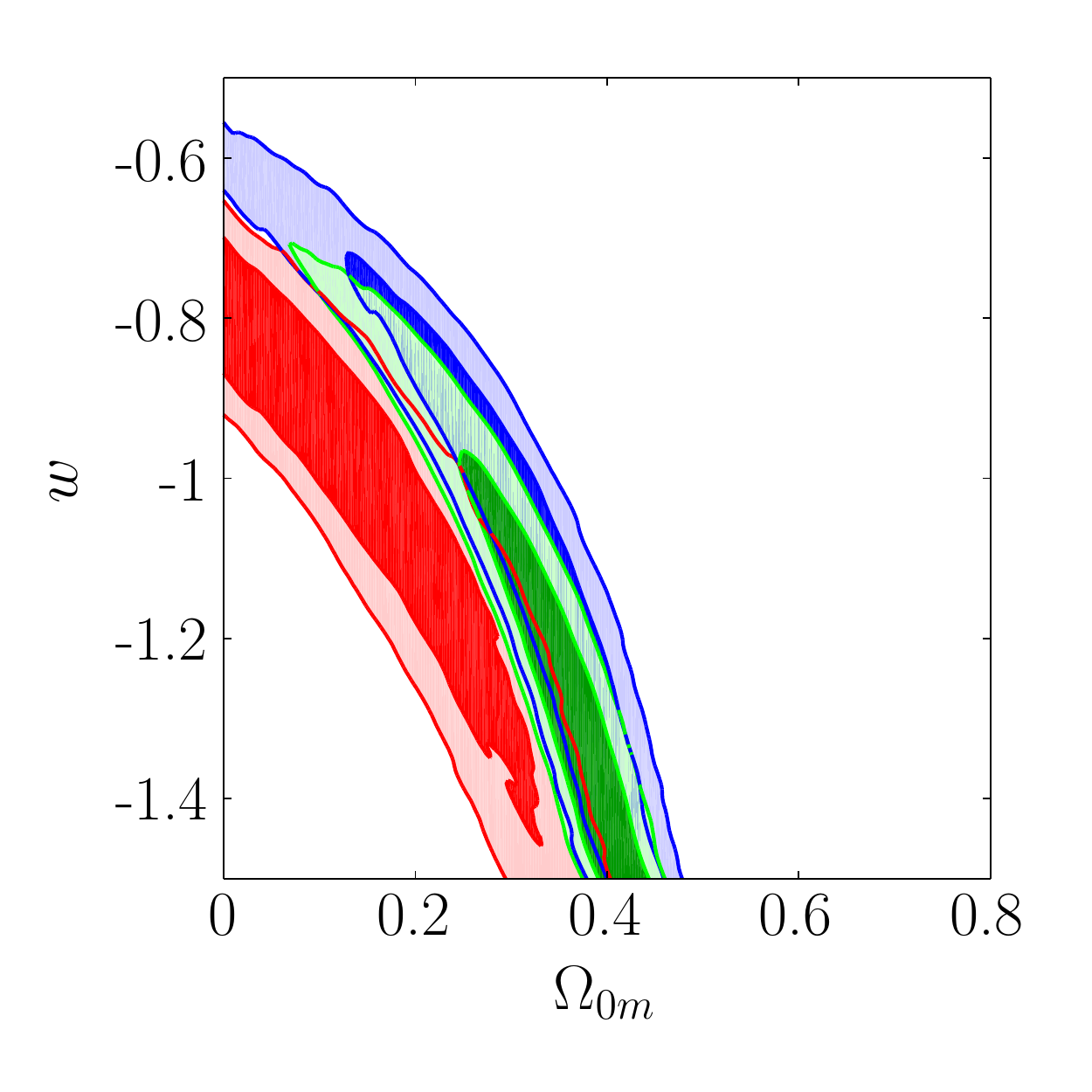} \\
\mbox{(a)} & \mbox{(b)} & \mbox{(c)} \\
\end{array}$
$\begin{array}{@{\hspace{-0.5cm}}c@{\hspace{-0.25cm}}c@{\hspace{-0.25cm}}c}
\includegraphics[width=0.37\linewidth]{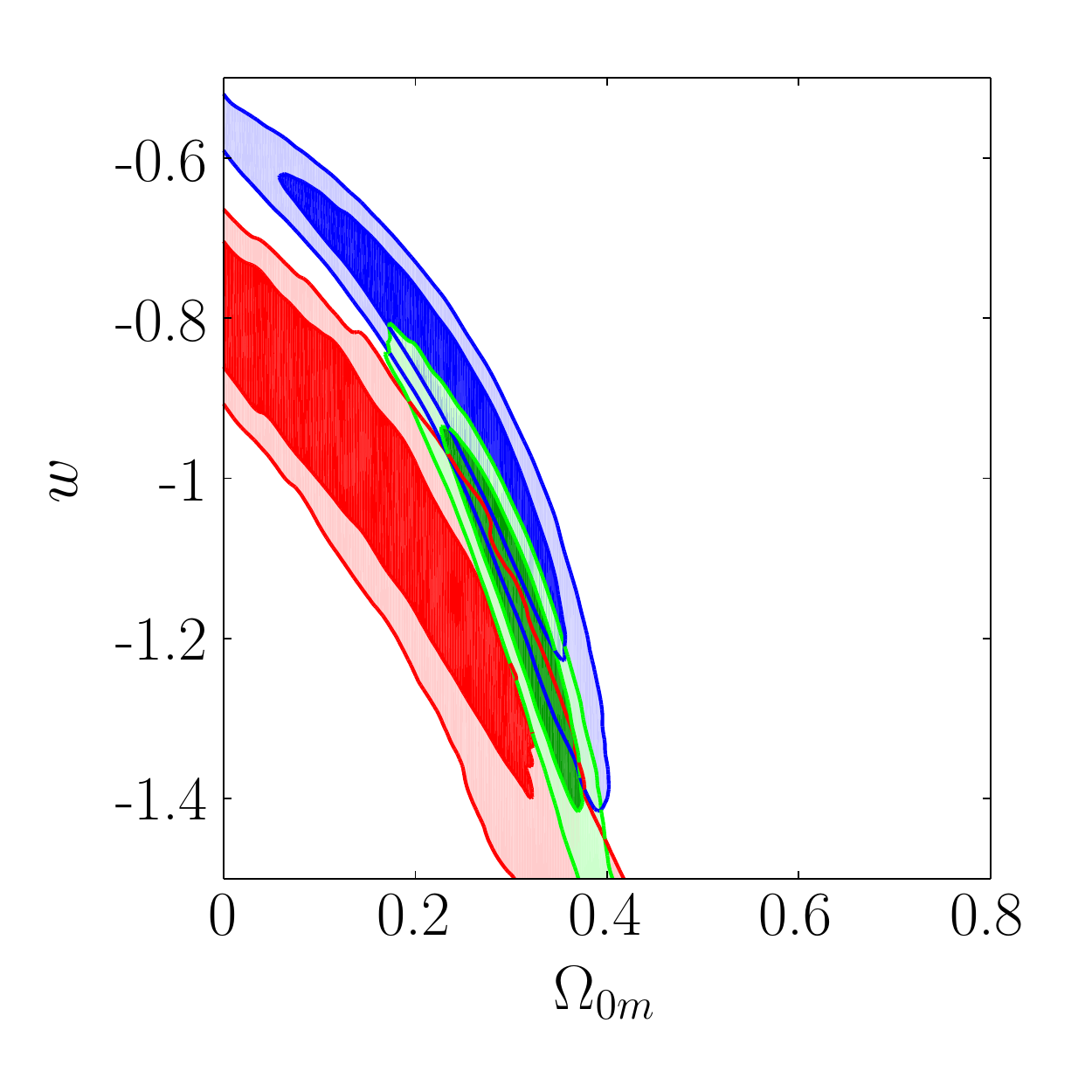} &
\includegraphics[width=0.37\linewidth]{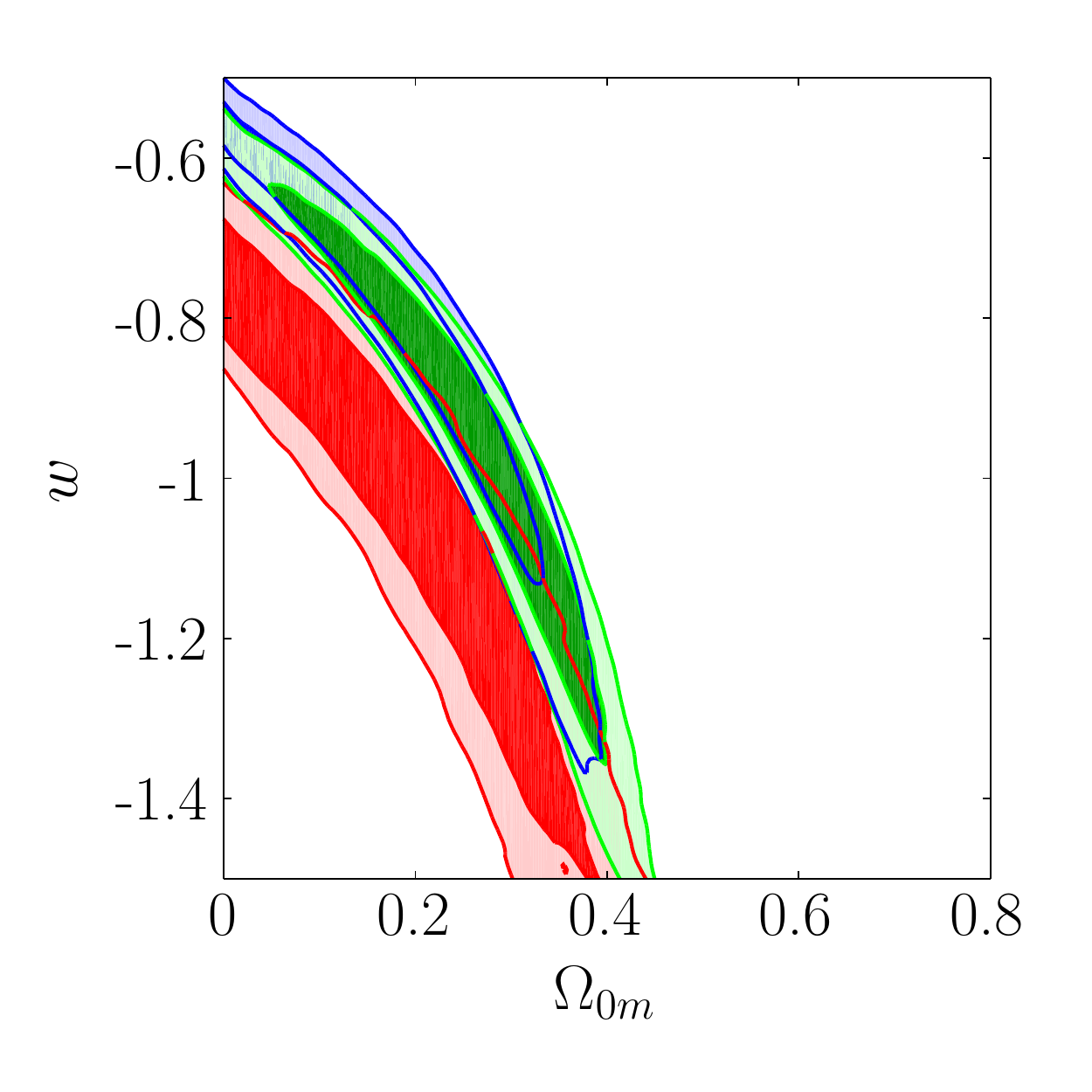} &
\includegraphics[width=0.37\linewidth]{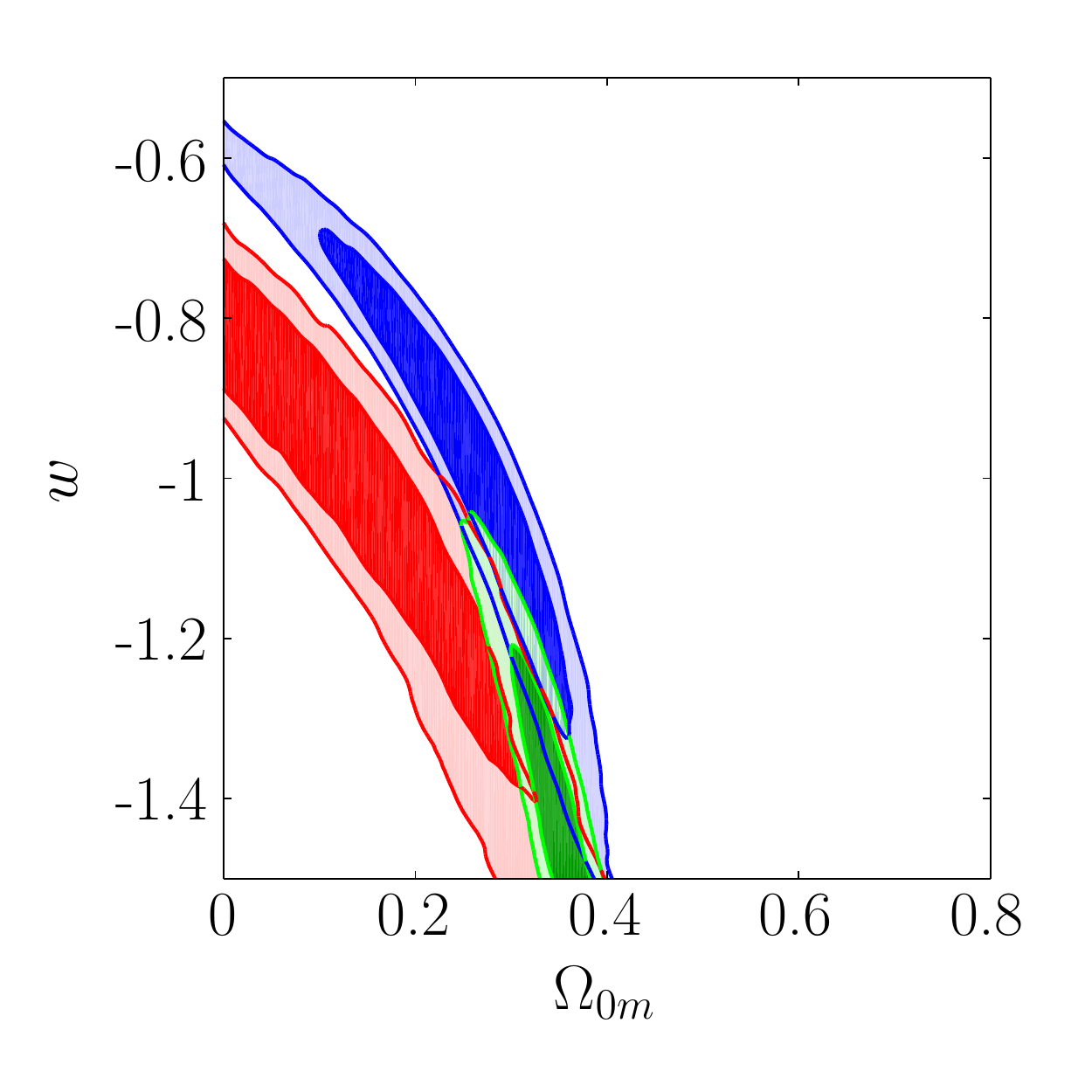} \\
\mbox{d} & \mbox{(e)} & \mbox{(f)} \\
\end{array}$ 
\caption{\footnotesize
$1-2\sigma$ confidence levels in $\omt-w$ parameter space for
Panstarrs (red), JLA (blue) and Union2.1 (green) supernova Type Ia
data. Panel (a) is for the entire datasets, panel (b) is for common
low-$z$ data + high-$z$ data, panel (c) is for common low-$z$ data +
high-$z (<0.64)$ data, panel (d) is for data without the outliers of
regression, panel (e) is for the largest low and high-$z$ subsets for
each sample, and panel (f) shows full datasets for JLA and Panstarrs,
and Union2.1 without the data subset 6 which showed anomalous
behaviour.}
\label{fig:omw_compare}
\end{figure}

We now remove the constraint on $\omt$ and let it be a free
parameter. The resultant analysis will naturally be affected by the
degeneracy between the parameters $\omt$ and $w$ and the constraints
obtained may consequently be poorer. In table~\ref{tab:cosmo}, we show
the best-fit and $1\sigma$ values in the different scenarios for
$\omt$ and $w$. Fig~\ref{fig:omw_compare} shows the 1 and $2\sigma$
confidence levels in the $\omt-w$ parameter space for (a) the entire
Panstarrs, JLA and Union2.1 datasets, (b) the common low-$z$ data +
high-$z$ data, (c) the common low-$z$ data + high-$z$ data cut-off at
$z=0.64$, (d) the data without the $2\sigma$ outliers in the
regression analysis, (e) the case where only the largest datasets in
the low and high redshift ranges are considered for each sample, and
(f) the specific case where the anomalous subset 6 is subtracted from
Union2.1. We have seen in the statistical analysis and also the
cosmological single parameter analysis that the Panstarrs entire
dataset looks quite different than the JLA and Union2.1 data, being
$2\sigma$ away from both, while JLA and Union2.1 also appear to be
about $1\sigma$ away from each other. The same trend is seen in the
two-parameter cosmological results as well, in
figure~\ref{fig:omw_compare}(a), with the Panstarrs confidence levels
$2\sigma$ away, and JLA and Union2.1 about $1\sigma$ away from each
other. We also see that the Union2.1 data appears to give a somewhat
higher $\omt$ and is furthest from the cosmological constant value of
$w=-1$ for the equation of state. The Panstarrs data gives much lower
$\omt$ than is usually expected from other observations such as large
scale structure, and JLA appears to be closest to the standard
cosmological constant model with an amount of matter density that
matches the large scale structure constraints. Trends similar to this
result have been seen in previous works, albeit with additional
information from CMB and BAO data.

We now see the effect of using only the common data subset in the
low-$z$ data, in fig~\ref{fig:omw_compare}(b). Though the results
change in each dataset, the amount by which they change is rather
different. The change to common low-$z$ data appears to affect
Panstarrs least of all and the results are very similar to that for
the full dataset. JLA appears to have a rather large change with
$\omt$ decreasing and $w$ increasing, moving it away from both the
cosmological constant and the other datasets. This is unlike the
results of both the statistical analysis and the single parameter
cosmological analysis. Union2.1 sees a moderate decrease in $\omt$ and
similar increase in $w$. Thus the JLA appears to be most affected by
changing low redshift data, but this does not make it more consistent
with the other datasets.

Next, we cut off the JLA and Union2.1 data at redshift $z=0.64$, based
on the highest redshift SNe available for Panstarrs. We find in
fig~\ref{fig:omw_compare}(c) that the confidence levels for both JLA
and Union2.1 increase, as expected, since the sample size has
shrunk. For JLA, $\omt$ increases while $w$ decreases, thus keeping it
still away from Panstarrs. In the case of Union2.1 as well, a similar
trend is noted. Although Panstarrs and Union2.1 now are slightly more
consistent at $2\sigma$, due to the larger confidence levels for
Union2.1, substantive inconsistency remains.

In the panel (d) of the figure, where we get rid of the outliers of
the regression analysis, we see that once again, Panstarrs results are
quite stable while JLA results change quite majorly and Union2.1
rather less. In any case, the various data selections made on the
datasets do not resolve the inconsistencies of the fitted cosmological
parameters.

From fig~\ref{fig:omw_compare}(e), we see that the Panstarrs data does
show some change in the case where the largest subsets are considered,
possibly because the number of data points has reduced drastically, by
about a third of the total number of SNe. JLA and Union2.1 datasets
also change rather majorly, but the discrepancy between the datasets
remains. We note here, that, unlike in the single parameter case, JLA
and Union2.1 do not become noticeably more consistent with each other.

When the subset 6 is removed from the Union2.1 dataset in the final
panel, this results in somewhat smaller errors, but not more
consistency between datasets, again unlike the single parameter case
where the Union2.1 result had moved towards the Panstarrs values. In
fact, here the cosmological parameters retrieved from Union2.1 become
strongly inconsistent with the other datasets, with a much lower value
of $w$ and much higher value of $\omt$. This may be due to the fact
that subset 6 is the largest low redshift sample, and without it, the
low redshift data may be inadequate in anchoring the full dataset, the
effect of which is felt more strongly for a larger number of
parameters. The smaller errors in this case as opposed to the full
dataset reflects the fact that the subset 6 is rather inconsistent
with the rest of the data, and subtracting it from the dataset reduces
the intra-sample tension. Of course, smaller errors do not necessarily
reflect the accuracy of the result, the result may well become biased
due to lack of anchoring.

We note here that the consistencies in the results should also reflect on the standardization parameters, $\alpha, \beta, M_B$. Among these, $M_B$ being a simple additive, has least effect on the cosmological results, and is the most stable to subset changes. We study the parameters $\alpha, \beta$ for the case where we vary both $\omt, w$, the results are very similar for the other case, since these parameters are practically independent of cosmology. For all three datasets, we find that $\alpha$ is quite stable to changes within the dataset. For JLA, the best-fit varies within $\alpha = 0.137-0.141$, for Union2.1, $\alpha = 0.121-0.128$, and for Panstarrs, $\alpha = 0.146-0.153$, all with $1\sigma$ errors of around $\sim 0.01$. The variation in $\alpha$ within the subsets thus falls within the error bars. For $\beta$, the JLA variation is $\beta = 3.10-3.31$, for Union2.1 it is $\beta = 2.44-2.58$, and for Panstarrs, $\beta = 3.76-4.00$, with errors of the order of $\sim 0.15$. For this parameter, we note that the variation is of the same order as the $1\sigma$ errors for JLA and Union2.1, and somewhat larger for Panstarrs. These results are commensurate with the results obtained in \cite{jla14, union11, panstarrs14}, with the Panstarrs result rather closer to that obtained for the case where $\sigma_{int}$ is dominated by color dispersion. Once again, the values obtained for each dataset are somewhat different to those obtained for the other datasets. The somewhat large variation in $\beta$ may be an effect of host galaxy properties, or of intrinsic scatter, and this difference needs to be studied in more detail, as well as its potential effects on cosmological results. 

The basic finding in this analysis is the same as that of the
statistical analysis and the single parameter cosmological analysis--
we find that using various subsets of the SNe samples does not
necessarily lead to more consistency between the three data samples,
and the inconsistency between them remains. Worryingly, using
different subsets does change the results noticeably for both JLA and
Union2.1 samples, while the Panstarrs dataset, despite being smaller
and at lower redshifts, is extremely stable to such changes. The
Union2.1 data shows this variability in the previous analyses as well,
and its behaviour may be attributed at least partly to the
heterogeneity in this dataset. However, the JLA data had appeared to
be quite well-behaved and stable to such changes in previous analyses,
whereas here large differences are noted. We note here that our
results for the single parameter case is quite similar to those
obtained in \cite{stat_sne2}, and also that our results for the
consistency in the JLA data between the different telescope subsamples
is equivalent with this paper. However, it appears that for the
two-parameter case, the JLA data is sensitive to the degeneracies in
the $\omt-w$ parameter space. Since the SNe data measures the total
energy density, increasing $\omt$ and decreasing $w$ has the same
effect for it as decreasing $\omt$ and increasing $w$, thus data that
look very similar can give rather different results based on this
non-unique behaviour of the parameters. The JLA data appears to be
particularly susceptible to this degeneracy. The reason for this
variability in JLA data in the two-parameter analysis may stem from
the differences explored in sec~\ref{sec:mva}. We saw there that the
Panstarrs data has the most homogeneous error distribution, and that
JLA shows strong trends of heteroscedasticity. The cosmological
analysis depends on a product term, containing the inverse of the
errors of the individual SNe, and the spread in the distribution
around the magnitude fit. The dispersion in magnitude being rather
larger than the spread in errors, it typically drives the
results. However, the spread in errors should also affect the results
to some extent. As we increase the number of parameters, and therefore
the degrees of freedom, the dispersion in magnitude becomes smaller if
the fit becomes tighter, and the spread in the errors now may become
relatively more significant.

To study this effect further, we look at the residuals of the fit for
JLA in the two cases where $\omt$ is fixed and when it is free. We
find that the residuals at best-fit appear to be slightly tighter in
the case with two free parameters. If we now look at the different
subsets of the JLA sample, for the single parameter fit, the variation
of the residuals between datasets is low because of reduced degrees of
freedom. The two-parameter fit has more flexibility and therefore
smaller median residuals for each subset, but larger changes in
residual variance between subsets. This divergence is further enhanced
by the non-linear $z$-dependence of the magnitude errors resulting in
a noticeable difference in the results for different data subsets. For
Panstarrs, a similar variation in the residuals can be seen, but as it
has more homogeneous errors, this behaviour is not enhanced by the
errors, and the fits remain relatively stable. Therefore, the
heterogeneous errors in the JLA data result in a somewhat more
unstable fit when the number of parameters is increased. Adding other
types of data that complement SNe data may be one way to limit this
behaviour, as seen in fig~\ref{fig:w_all} and table~\ref{tab:cosmo_w},
JLA results are very stable if the value of matter density is known,
\ie the degrees of freedom are reduced. Another potential method would
be to model the non-linearity of the magnitude errors and include this
information in the parameter fitting \cite{Davidian90}. Thus the
heteroscedasticity of the data may have a distinct effect on the
cosmological results, and we should be careful when interpreting such
results.

\section{Conclusions}\label{sec:concl}

In this work, we have studied three SNe type Ia datasets available at
present, the Panstarrs, JLA and Union2.1 samples. We have first looked
at the data from a purely statistical point of view, without
considering any cosmological information. We then do a cosmological
analysis, constraining first the single parameter $w$, and then the
two parameters $\omt, w$. We present our findings below.
\begin{itemize}
\item
Simple linear or quadratic regression fits to the data show that the
three SNe samples have divergent behaviours and this can be traced to
the use of different SNe in each analysis. Thus the choice of SNe
appears to make a real difference in the statistical results. These
differences are not easily reconciled, and the Panstarrs, JLA and
Union2.1 datasets remain inconsistent with each other no matter which
data subset is considered. In the cosmological section, these
statistical results are corroborated.
\item
We demonstrate that the Panstarrs data is most homogeneous in its
residuals and has the fewest outliers in the regression analysis. It
also is very stable when considering different subsets. The Union2.1
data is most heterogeneous, it is strongly affected by choosing
different data subsets in the statistical analysis, as well as having
a large anomalous subset (sample 6) at low redshift. The JLA data is
apparently not affected by subset selection in the simple statistical
analysis, but the errors $\sigma_{m_B}$ of JLA do not follow a normal
distribution and show strong clustering.
\item
We find that Panstarrs, the dataset drawn from the most homogeneous
sample, with a single data subsample at high redshift, performs best
in the cosmological fits. The Union2.1 data, drawn from a total of 19
different sources, appears to be less stable than the others. Although
enough surveys now exist at high redshift with sufficiently high
quality data that it is not necessary to use many surveys in
conjunction, the same is not true at low redshift. There we still do
not have a single survey with high quality data that could replace the
various data sources of varying quality currently used. Since
homogeneity of data appears to be important for reliable analysis, we
suggest that the presence of such a homogeneous low redshift sample
would improve the overall quality of cosmological reconstruction.
\item
Significant heteroscedasticity in the magnitude errors, as in the case
of JLA, appears to lead to a less stable cosmological fit which is
dependent on the subset used for analysis, especially for a higher
number of parameters. The Panstarrs dataset, on the other hand, with
its more homogeneous errors, is comparatively impervious to the data
subset selection. If we are to use data with strongly heterogeneous
errors, we should either model the errors along with the cosmological
parameters, or add other datasets to reduce the number of degrees of
freedom. A dataset with homogeneous error residuals, such as
Panstarrs, would be more stable in constraining a larger number of
parameters.
\end{itemize}
In conclusion, we find that the currently available SNe datasets are
somewhat inconsistent with each other (at up to $2\sigma$ confidence
level).  Further data in the future will be able to shed light on this
inconsistency. We note that a single, high quality dataset at low
redshift may go some way towards increasing the robustness of
cosmological reconstruction. Data with homogeneous error residuals
tends to be more immune to changes in the data subsets, thus leading
to more conclusive parameter estimation. Although the Panstarrs
dataset is too small now to draw cosmological conclusions from, it is
stable to small changes and consistent within itself. Therefore future
data releases from Panstarrs may lead to stronger cosmological
constraints. The ideal supernova dataset would be one which is drawn
from the least number of samples, potentially one at high redshift,
and one at low redshift, with a strongly homogeneous distribution of
errors. Cosmological reconstruction from such a supernova sample would
be robust and serve as a good complement to other available
data. Combining disparate supernova samples requires careful
consideration of the systematics involved in the reduction and fitting
techniques for SNe data. We note that certain systematics, such as
the intrinsic dispersion and the effect of host galaxy properties,
could bear closer scrutiny, and we expect to return to the systematics
of supernovae in more detail in future work.

\section{Acknowledgements}
UA was supported in this project by the “DST Young Scientist Program”
of SERB, India. The authors would also like to thank the ChemCam team
at IRAP, Toulouse for the use of the hyperion2 cluster for some of the
calculations of this paper.

\end{document}